\documentclass{aa}
\usepackage{graphicx}
\usepackage{txfonts}
\usepackage{natbib}
\usepackage[caption=false]{subfig}
\usepackage[linktocpage]{hyperref}
\usepackage{xcolor}
\usepackage{threeparttable}
\usepackage{etoolbox}
\makeatletter
\patchcmd\@combinedblfloats{\box\@outputbox}{\unvbox\@outputbox}{}{%
   \errmessage{\noexpand\@combinedblfloats could not be patched}%
}%
 \makeatother

\begin{document}

   \title{Field spheroid-dominated galaxies in a $\Lambda$-CDM Universe}

    \author{M. S. Rosito \inst{1}, S. E. Pedrosa \inst{1}, P.B Tissera \inst{2}, V. Avila-Reese \inst{3}, I. Lacerna \inst{4,5}, L. A. Bignone \inst{2}, H. J. Ibarra-Medel \inst{3}, S. Varela \inst{2}} 

   \institute{Instituto de Astronom\'{\i}a y F\'{\i}sica del Espacio, 
CONICET-UBA, Casilla de Correos 67, Suc. 28, 1428, Buenos Aires, Argentina
         \and
             Departamento de Ciencias F\'{\i}sicas, Universidad Andr\'{e}s 
Bello, Santiago, Chile
        \and 
            Instituto de Astronom\'{\i}a, Universidad Nacional Aut\'{o}noma de M\'{e}xico
        \and 
            Instituto Milenio de Astrof\'isica, Av. Vicu\~na Mackenna 4860, Macul, Santiago, Chile
        \and
            Instituto de Astrof\'isica, Pontificia Universidad Cat\'olica de Chile, Av. Vicuna Mackenna 4860, Santiago, Chile
            }            


\abstract
   {Understanding the formation and evolution of early-type, spheroid-dominated galaxies is an open question
   within the context of the hierarchical clustering scenario,
   particularly in low-density environments.}
   {Our goal is to study the main structural, dynamical, and stellar
     population properties and assembly histories of field
     spheroid-dominated galaxies formed
     in a $\Lambda-$CDM scenario to assess to what extent they  are
     consistent with observations.}
   {We selected spheroid-dominated systems from a $\Lambda$-CDM
     simulation that includes star formation, chemical evolution, and
     supernova feedback.  The sample is made up of 18 field systems with $M_{\rm Star}\lesssim 6\times 10^{10} M_\odot$ that are 
   dominated by the spheroid component. For this sample we estimated
   the fundamental relations of ellipticals and  compared them with current observations. }
   { The simulated spheroid galaxies have sizes that are in good agreement with
     observations. The bulges follow a Sersic law with Sersic indexes
     that   correlate with  the bulge-to-total mass ratios. The
     structural-dynamical properties of the simulated galaxies are consistent 
   with observed Faber--Jackson, fundamental plane, and Tully--Fisher
   relations. However, the simulated galaxies are 
   bluer and with higher star formation rates than the observed isolated early-type galaxies. The archaeological mass growth histories 
   show a slightly delayed formation and more prominent inside-out growth mode than observational inferences based on
   the fossil record method.}
   {The main structural and dynamical properties of the simulated spheroid-dominated galaxies are consistent with observations. This is remarkable 
   since our simulation has not been calibrated to match them. However, the
   simulated galaxies are blue and star-forming, and with later stellar mass
   growth histories  compared to observational inferences. This is mainly due to the persistence of extended discs in the simulations.   The need
   for more efficient quenching mechanisms able to avoid
   further disc growth and star formation is required in order to
   reproduce current observational trends.}

   \keywords{galaxies: formation – galaxies: elliptical and lenticular, cD – 
galaxies: fundamental parameters}

  \authorrunning{M. S. Rosito et al.}
  \maketitle

\section{Introduction}

The formation of galaxies supported mostly by velocity dispersion is currently
thought to encompass different physical processes. 
The monolithic collapse model \citep{Eggen1962} is in disagreement with compelling
globular cluster observations. The scenario proposed by
\citet{SearleZinn78} agrees partially with a hierarchical clustering scenario
in the context of the current cosmological paradigm. Within this scenario, massive early-type galaxies
(ETGs) are
often assumed to be the results of massive and dry mergers
\citep{Toomre1977,Hernquist1993,Kauffmann1996}. While this
mechanism is efficient at producing  classical ellipticals,
detailed photometric and spectroscopic observations
report a more complex situation \citep{Gerhard99,Rix99}.
Massive ellipticals are reported to be slow rotators, supported by velocity
dispersion, while low- and intermediate-mass ETGs tend to be fast
rotators and show power-law surface brightness profiles \citep{Emsellem2007}.
This has been confirmed by results from ATLAS$^{3\mathrm{D}}$ \citep{AtlasI} where ETGs
supported by rotation are seven times more frequent \citep{Emsellem2011}.
The presence of discs in elliptical galaxies varies from embedded to
more intermediate-scale systems \citep{Graham2016}. 
As ETGs tend to populate high-density regions, the interaction of galaxies with their environments is a key process which might prevent the growth of extended
discs via ram pressure stripping \citep{GunnGott1972} or
`strangulation' \citep{Larson1980}, contributing to the quenching of
the star formation (SF) activity. In low-density environments these effects
are not expected to be efficient, and galaxies may thus follow different evolutionary paths. 

As described 
in \cite{Kormendy2016}, there are different formation scenarios to
explain the observed differences among ETGs. Dry mergers are
efficient at forming dispersion-dominated galaxies.  \cite{Naab2013} stress that this scenario may be incomplete and 
that observations require a two-phase assembly: dissipative processes with
in situ  SF at high redshifts and the accretion of stars formed in 
other galaxies.
Minor mergers with mass ratios of $\sim 1\mathord{:}5$ may lead to
the formation of fast rotator ETGs
\citep{Oser12,Gabor12,Lackner12}. Both major and minor mergers have
also been  proposed as possible quenching mechanisms in cosmological
simulations  \citep[e.g.][]{Hopkins2008}.
Another possible scenario is secular evolution,
driven by internal dynamical instabilities or
by interactions/mergers. These mechanisms lead to gas inflows and  the formation of pseudobulges due
to angular momentum redistribution.    
This process requires that a disc  be  previously in place \citep[e.g.][]{Tissera2001,Pedrosa2015}.
Indeed, several works have shown that bulges and ellipticals (spheroids in general) could
have formed in several phases by mergers and
secular processes in such a way that they are currently composed by
multiple stellar populations \citep[e.g.][and 
references therein]{Zavala+2012,Tissera2012form,perez2013,AvilaResse2014}. 
These studies find that the fraction of  ex situ and in situ stars in ETGs correlates with mass: the most massive 
ETGs are dominated by ex situ stars from major mergers, intermediate-mass  ETGs have similar fractions of ex situ and 
in situ stars, and less massive ETGs are dominated by in situ
stars. 

While the ETGs are mostly red and quiescent, there is also a fraction of
blue, star-forming ETGs that host some young stellar populations.
The blue, star-forming fraction increases for smaller galaxies and
decreases for denser
environments \citep[e.g.][]{Schawinski2009, Kannappan2009,Thomas+2010,McIntosh2014,Schawinski2014,Vulcani+2015, Lacerna2016}. 
 \cite{kaviraj2007} studied the UV colours of $\sim 2100$ galaxies from the Sloan Digital Sky Survey (SDSS) at very 
low redshift and found that at least $\sim 30 \%$ are consistent with recent 
SF. They found that many ETGs at $z<0.11$ have $1-3 \%$ of 
their stellar mass younger than $1 \ \mathrm{Gyr}$. 
The origin of  these younger stellar populations is still
under debate. Physical mechanisms such as galaxy interactions or
secular evolution could be an explanation if there is remnant gas in the galaxies.
Recently, \cite{Lacerna2016} have discussed the main photometric, SF, and structural properties of elliptical galaxies from 
a complete SDSS subsample of isolated galaxies \citep[the UNAM-KIAS Catalog;][]{Hernandez-Toledo+2010} and compared them to those 
of cluster ellipticals. They find
that the fraction of blue, star-forming, isolated ellipticals is only slightly higher than that  found in clusters
\citep[see also][]{ Schawinski2009, Kannappan2009, Thomas+2010, McIntosh2014, Schawinski2014}. 
The fractions increase
at lower masses, but they  are never as high as predicted 
in $\Lambda$CDM-based semi-analytical models
\citep{Kauffmann1996,Niemi+2010}.

On  the other hand, ETGs follow clear scaling relations.
The Faber--Jackson relation (FJR)  \citep{Faber1976}  relates structural (photometric) 
and kinematic properties so that the  luminosity increases with
increasing velocity dispersion, $  L\ \propto \ \sigma^{4}$.  However,  
the exponent may depend on the galaxy type and luminosity band \citep{Kormendy13}.
The most notable relation followed by ETGs is the called the fundamental 
plane  (FP) \citep{Faber1987, Dressler1987, Davis1987}, which links the effective radius $R_{\rm eff}$, the stellar velocity 
dispersion $\sigma_{\rm e}$, and the average surface brightness
$\Sigma_{e}$. An unsolved issue regarding this observed relation 
is a tilt from the predictions of the virial theorem \citep{Cappellari2013}. It may be related to 
a systematic variation in the stellar population or initial mass
function (IMF) \citep{Prugniel1996, 
Forbes1998} or the non-homology in the surface brightness distribution
\citep[e.g.][]{Prugniel1997, Graham1997, Bertin2002, Trujillo2004} or  the variation in the amount of dark 
matter \citep[e.g.][]{Renzini1993, Ciotti1996, Borriello2003}, among
others. However, the FP  is  in agreement with the virial predictions 
if dynamical mass is used instead \citep{cappellarireview2016}.
Recent studies also  find that ETGs with a gaseous disc  follow the
Tully--Fisher relation \citep[TFR;][]{TullyFisher1977}.
   \cite{Heijer2015} measured HI rotation velocities for a subsample of ETGs from the ATLAS$^{3\mathrm{D}}$ sample and determined
the  TFR using magnitudes in the K band and stellar masses.

In this paper we analyse the main structural-dynamical relations and the stellar population properties of spheroid-dominated 
galaxies (SDGs) in a $\Lambda$CDM-based cosmological simulation  \citep{Pedrosa2015}. These SDGs can be
related to low- and intermediate-mass ETGs in the field. 
Furthermore, we compare the properties
of our simulated SDGs with those of isolated ETGs (including lenticular galaxies) from the UNAM-KIAS
Catalog. We also compare the global and radial stellar mass growth histories calculated from the archaeological analysis of the
stellar populations with the corresponding histories inferred for ETGs from the MaNGA/SDSS-IV survey \citep{Bundy+2015} by means
of the fossil record method \citep{IbarraMedel2016}. Our aim is to explore whether the simulated SDGs within the context of the $\Lambda$-CDM scenario
are consistent with the observed field ETGs.

The results for SDGs discussed in this paper are complementary to those
reported by \citet{Pedrosa2015}, \citet{Tissera2015}, and \cite{Tissera2016} regarding
disc-dominated galaxies (DDGs). Simulated galaxies are identified from the
same simulation and by following similar criteria. In the mentioned
papers, the authors report that the simulated disc galaxies follow the
size-stellar mass relation, the TFR, 
and that the chemical gradients are in agreement with observations.  It is then
relevant to analyse whether the simulated galaxies with a dominating velocity-dispersion component also satisfy observational constraints. 
Our findings will also be  important in order to study the effects of
environment on the preprocessing of physical properties as galaxies move to
higher density regions. New observations are starting to provide information
on dispersion-dominated galaxies in low-density environments \citep{Ashley2017} which will be available for comparison in the near future.

This paper is organised as follows. We present our numerical simulations in Section 2. In Section 3 we characterise simulated galaxies, and  mention how the morphological decomposition was made and describe the surface brightness profile of the selected ETGs. Our results are presented in Sects. 4 and 5 for the main scaling relations and   colour and specific SF rate (sSFR), respectively,  and in Sect. 6   we analyse stellar mass growth histories. In all cases, we compare our results with observations.
Finally, a summary of our results is presented in Section 7. Table \ref{T_acron}  lists most of the acronyms and definitions used in this paper.

\begin{table}[h]
        \centering
        \caption{List of acronyms and definitions used in this paper.} 
         \resizebox{9cm}{!} {
                \begin{threeparttable}
                        \begin{tabular}{cc}
                                \hline
                                \hline
                                 AGN &  active galactic nucleus \\
                                 ATLAS$^{3\mathrm{D}}$ & a volume-limited survey of local ETGs\\                           
                                 DDG & disc-dominated galaxy (simulation)\\
                                 ETG & early-type galaxy (observations) \\
                                 FJR & the Faber--Jackson relation\\
                                 FP & fundamental plane\\
                                 IMF & initial mass function\\
                                 ISM & interstellar medium\\
                                 LSST & Large Synoptic Survey Telescope \\
                                 LTG & late-type galaxy (observations)\\
                                 MaNGA & Mapping Nearby Galaxies at the APO \\
                                 MGH & (Stellar) mass growth history \\
                                 SDG & spheroid-dominated galaxy (simulation)\\
                                 SF & star formation\\
                                 sSFR & specific SF rate \\
                                 SN & supernova\\
                                 TFR  & the Tully--Fisher relation\\
                                 UNAM-KIAS & a catalogue of SDSS very isolated galaxies\\
                                \hline
                                $B/T$ & dynamical bulge-to-total mass ratio\\
                                $M_{\rm Bar}$ & galaxy baryonic mass \\
                                $M_{\rm Star}$ & galaxy stellar mass \\
                                $n$ & Sersic index \\
                                $R_d$ & disc scale length \\
                                $R_{\mathrm{hm}}$ & radius which contains half of the total stellar mass\\
                                 $R_{\mathrm{eff}}$  & radius which contains half of the mass/luminosity obtained from a Sersic \\
                                  &  fit to the surface density/brightness profile (B, D, or T) and using Eq. (\ref{eq:reff})\\
                                 $\sigma$ & velocity dispersion\\
                                $V$ & rotation velocity\\
                                \hline
                        \end{tabular}
                \end{threeparttable}
                }
        \label{T_acron}
\end{table}

\section{Numerical simulations}

In  this work we use  the cosmological  simulation S230D from the Fenix set analysed
first by \citet{Pedrosa2015}, which is 
consistent with a $\Lambda$-CDM universe with $\Omega_{m}=0.3$, 
$\Omega_{\Lambda}=0.7$,  $\Omega_{b}=0.04$, and $H_{0}=100 \ h \mathrm{ \ km \ 
s^{-1}  \ Mpc^{-1}} $ with $h=0.7$, and a normalisation of the power spectrum of $\sigma_{8} = 0.9$. 
The size of the simulated box is $14 \mathrm{\ Mpc}$ per side. The
initial condition has  $2 \times 230^{3}$ total particles with a mass
resolution of  $5.9 \
\times{}10^{6} \ h^{-1} \ \mathrm{M}_{\odot}$ and $9.1 \times{}10^{5} \ h^{-1} 
\ \mathrm{M}_{\odot}$ for the dark matter particle and initial gas particle, 
respectively. The maximum gravitational softening is $0.5 \ h^{-1}\mathrm{kpc} $.

The initial conditions
were chosen to describe a typical region of the
universe where no massive group  is present (the largest halos
have virial masses smaller than $\sim 10^{13} \
\ M_{\odot}$).  To check the effects that the small simulated
  volume might have on the growth of the structure, \citet{DeRossi2013} compared the halo mass growth histories of galaxies
in a simulation similar to the one used here with those 
estimated by \citet{Fakhouri2010} for the halos from the Millennium
Simulation. This comparison showed that the growth
of the simulated halos is   accurately described in these simulations in
  the mass range of interest.

The simulation was run using GADGET-3, an updated version of GADGET-2 
\citep{springel2003, springel2005}, optimised for massive parallel simulations of highly inhomogeneous
systems. It includes treatments 
for metal-dependent radiative cooling, stochastic SF, and chemical and 
energetic supernova (SN) feedback \citep{scan2005, scan2006}. The SN feedback model is
capable of triggering galactic mass-loaded winds without
introducing mass-scale parameters. As a consequence, galactic winds naturally
adapt to the potential wells of galaxies where star
formation takes place.
It also includes a multiphase model for the ISM that allows the
coexistence of the hot diffuse phase and the cold dense gas phase
\citep{scan2006, scan2008}. Stars form in dense and cold
gas clouds.  Some of them end their lives as SNe, injecting
energy and chemical elements into the ISM assuming a Salpeter IMF \citep{Salpeter55}. Each
SN event releases $7\ \times \ 10^{50}$ erg, which are distributed equally
between the cold and hot phases surrounding
the stellar progenitor.
Our simulation does not include the effects of feedback from active galactic
  nuclei (AGNs).
Previous results show that AGN feedback is expected to play an important role in the evolution of 
 massive galaxies formed in halos with masses larger than $\sim
 10^{12}\ \mathrm{M}_{\odot}$ \citep[see e.g.][]{Somerville2015,Rosas2016}. 
 Most of our galaxies formed in halos less massive than $10^{12} \mathrm{M}_{\odot}$.

The adopted code uses  the chemical evolution model developed by \citet{mosco2001} and
adapted to GADGET-3  by \citet{scan2005}. This model considers the enrichment
by SNeII and SNeIa adopting the yield prescriptions of
\citet{WW95}  and \citet{Iwamoto1999}, respectively. 
A detailed description of this SN feedback model is given extensively by \cite{scan2008}. 
It is important to stress that the SN feedback scheme does not include parameters that depend on the global properties of the given galaxy (e.g.  total mass, size).
\citet{Pedrosa2015} analysed the angular momentum content of the disc
and spheroid components of galaxies in the S230D
using a higher gas density threshold for star formation and a lower energy per SN event 
than in previous experiments of this project \citep[e.g. ][]{DeRossi2013, Pedrosa2014}.
They found that this combination of star formation and feedback
parameters produces systems that can better reproduce observational trends
such as the size-mass relation and the angular momentum content
\citep{Pedrosa2015}, the metallicity gradients of the disc components
\citep{Tissera2016, Tissera2017}, and the chemical abundances of the circumgalactic
medium \citep{Machado2018}.

The synthesised
chemical elements are distributed  between the cold
and hot phase ($80\%$ and $20\%$, respectively). These values were tuned in order to
provide a better description of  metallicity gradients of the stellar
populations and the gas-phase medium in the disc
components of the galaxies \citep{Tissera2016, Tissera2016b} and the circumgalactic 
medium \citep{Machado2018}.

The lifetimes for SNeIa are randomly selected within the range [0.1, 1] Gyr. This model is found to nicely reproduce
 mean chemical trends \citep{jimenez2015}.

\section{Characterisation of simulated galaxies}

We  use the galaxy catalogue  constructed by \citet{Pedrosa2015} where 
a friends-of-friends algorithm  is applied to identify the  virialised
structures at $z=0$ and then the SUBFIND code
\citep{springel2001} to
select 317 galaxies. For our analysis, only galaxies
resolved with more than $10\,000$ baryonic particles within the
$R_{\mathrm{opt}}$\footnote{The optical radius, $R_{\mathrm{opt}}$, is
defined as the radius that  encloses $ \sim 80 \%$ 
of the baryonic mass (gas and stars) of the galaxy \citep{tissera2000}.} are considered. 
This minimum number of baryonic particles yields a subsample of 39
galaxies, with stellar masses 
in the range $[0.27-11.6] \times 10^{10} \ \mathrm{M}_{\odot}$.  The stellar
masses  are  measured within the
$R_{\mathrm{opt}}$.

To assess the global environment inhabited by simulated galaxies, for each central galaxy we  identify its
neighbours within a distance of $1.5$ times the virial radius and with a minimum stellar mass of $\sim
4\times10^8$ M$_{\odot}$.
We find that the  maximum ratio between stellar mass of the   neighbours
and that of the 
central galaxy is $\sim 0.2$. Hence, the simulated central galaxies have no close massive
companions,  and are consequently  classified as field galaxies. Satellite galaxies
will not be studied in this paper because they are expected to follow different evolutionary paths than central galaxies.
In any case, there are only three satellite galaxies with more than $10\,000$ baryonic particles.

\subsection{Morphological decomposition}

We classify galaxy morphology by resorting to a {dynamical}
decomposition, applying the
method and criteria described by \cite{tissera2012}. We calculate $\epsilon = 
J_{z} / J_{z,max}(E)$ for each particle, where $J_{z}$ is the angular momentum 
component in the direction of the total angular momentum and $J_{z,max}(E)$ is the 
maximun $J_{z}$ over all particles at a given binding energy ($E$). We adopt
the criterion that those particles with
$\epsilon > 0.5$ are associated with the disc component and the rest of them with the spheroid component. In order to
discriminate between the  bulge (hereafter also called  the spheroid) and the stellar halo, 
we consider the particle binding energy so that the most bounded particles 
are taken to belong to the spheroid.

\begin{figure}
  \centering
\includegraphics[width=0.45\textwidth]{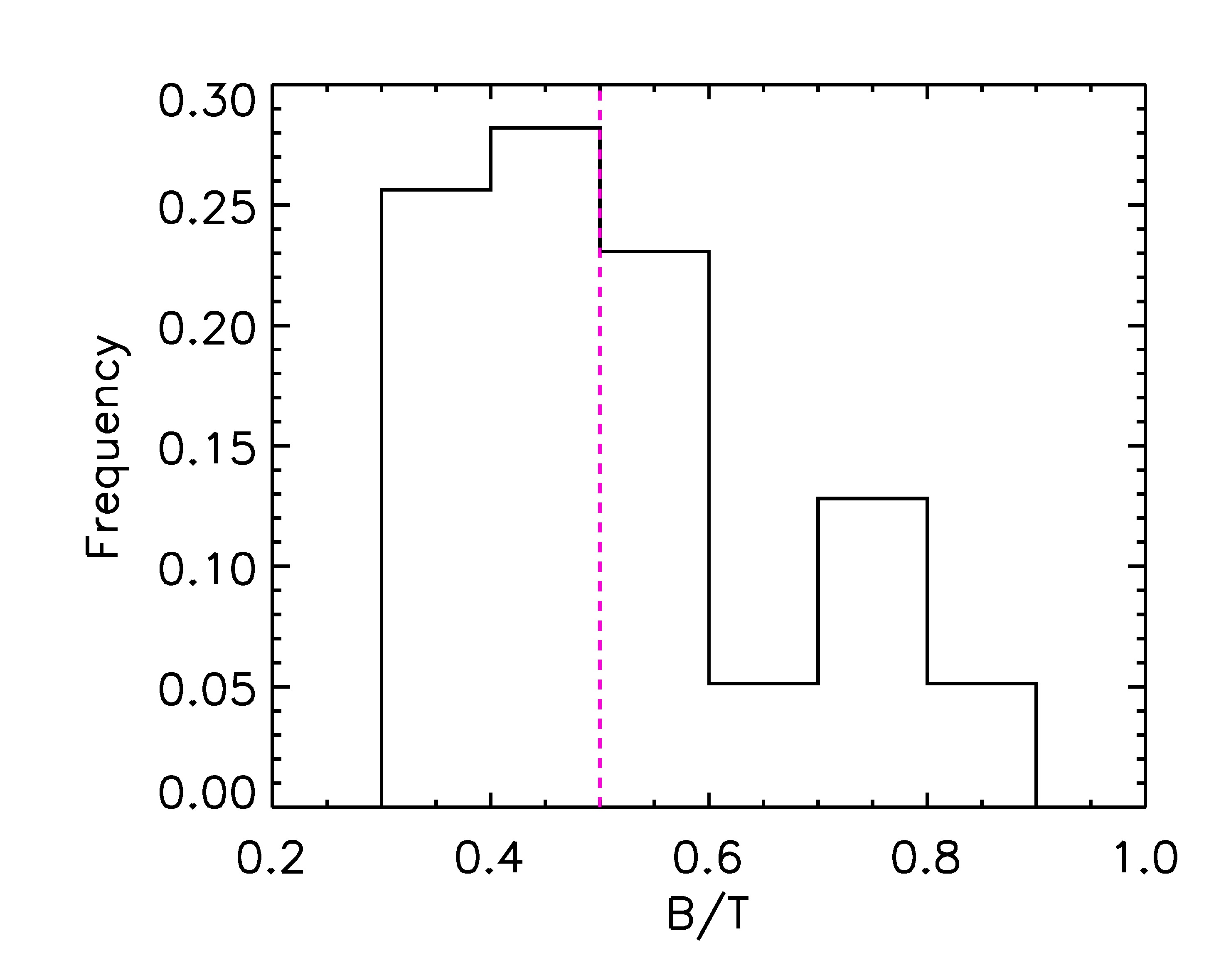}
  \caption{Distribution of the dynamical $B/T$ ratios for the 39 well-resolved
    simulated galaxies. Only those with $B/T > 0.5$ (red dashed line)
    are classified
    as SDGs.}
   \label{fig:BT}
\end{figure}

To classify the simulated galaxies according to morphology, the bulge-to-total
stellar mass ratios ($B/T$) are estimated using the stellar masses of
the bulge (central spheroid) and disc, defined as mentioned above. We adopt a threshold of
$B/T = 0.5$ to
separate between spheroid-dominated (SDG) and disc-dominated (DDG) galaxies.
In Fig. \ref{fig:BT} we show a histogram of the  $B/T$ ratios of the
subsample. Those with $B/T>0.5$ (i.e. the SDGs) are analysed here. Central galaxies with $B/T<0.5$ have been
studied in previous papers \citep{Pedrosa2015,
 Tissera2015,Tissera2016}. From Fig. \ref{fig:BT} we 
can appreciate that all of the  SGDs have a disc component,
and hence that there are  no pure ellipticals in this sample.
This is a very important aspect to bear in mind for the comparison  with observations, as
discussed in Section 5. 

After applying the morphological selection, the final sample of SDGs contains 18 objects, 
with stellar masses in the range $[0.27-6.33] \times 10^{10} \ 
\ \mathrm{M_{\odot}}$, i.e. all of them are sub-Milky Way mass
galaxies. The total stellar mass of the simulated galaxies is
obtained by adding the stellar masses of the  spheroid and disc components within $R_{\mathrm{opt}}$.
We also estimate the stellar half-mass
radius, $R_{\mathrm{hm}}$, as the radius that encloses 50\% of the total stellar mass. 
Table \ref{table:table1} summarises the properties of the 18 analysed
central spheroid-dominated galaxies.
Although our sample is small, we carry out a detailed analysis of the
dynamical and astrophysical properties which contribute to understanding
the complex history of formation of these galaxies and to set
constraints on the subgrid physics. This is of utmost importance, since
the interpretation of the observations relies on the comparison with
numerical models.

\subsection{Spheroid and disc surface mass densities}
\label{structure}

For the disc and spheroid components, the projected stellar-mass surface
density distributions are computed.
As we have dynamically separated spheroid and disc components, it
is straightforward to fit a Sersic profile \citep{Sersic1968} to the
projected surface distributions, obtaining the 
central surface brightness  $I_{0}$, the scale radius $R_{\mathrm{b}}$, and the 
 Sersic index $n$: 
\begin{equation}
 I(R)=I_{0}\exp{\big({-(R/R_{b})^{(1/n)}}\big)}
 \label{eq:Sersic}
\end{equation} 
For our analysis the projected stellar-mass
surface density is considered a proxy of the luminosity surface
brightness (equivalent to adopting  a mass-to-light ratio $M/L=1$, which is
close to observations for optical-infrared bands). 
When $n=1$, the Eq. (\ref{eq:Sersic}) recovers the exponential
profile that is fitted to the stellar-mass surface density of the
disc components, obtaining in this way the scale length $R_{\mathrm{d}}$ (Table \ref{table:table1}).

For the spheroid component, the surface density profiles are fitted within
the radial range defined by the gravitational softening and the radius that encloses $90 \%$ of the total spheroid mass.
In the case of the disc component, the fit is performed within the latter and $R_{\mathrm{opt}}$.

In Appendix \ref{app:image1}, Fig. \ref{fig:image1} shows  the
synthetic images of the 18 SDGs, the distributions of
$\epsilon$,
and the projected surface density  for the spheroid and disc  components, and the corresponding best-fitted
profiles.  
We also include the projected density profiles of those particles supported by rotation but coexisting with the spheroid.
As can be seen, these particles determine
a variety of surface density profiles: some SDGs have
discs which continue exponentially to the central part (e.g. SDG 897), while
others get flatter  (e.g. SDG 925) or change the profiles to
merge with that of the spheroid components  (e.g. SDG 790). 
As mentioned before, with different degrees of importance, {all the SDGs
have a disc components}. In this figure we also include the
observability radius.\footnote{From the post-processed galaxies we find the radius where the cumulative surface brightness in the $r$ band attains a value of 23 mag arcsec$^{-2}$. 
This is roughly the detection limit for the SDSS galaxies, see Appendix \ref{app:image1}.}
 As can be seen
in all cases, the disc components are below the observability threshold.
This is also seen in the right panels of Fig. \ref{fig:image1}, where the horizontal dotted lines indicate the stellar surface density of the given galaxy corresponding to the observability radius. 
Most of the external discs of the simulated galaxies would not be observed in the SDSS galaxy images.
In particular, as can be seen
from Fig. \ref{fig:image1}, SDG 288 also has spiral arms. According to
the morphological classification of \citet{sandage1961}, this galaxy
  might not be an ETG. However, according to the dynamical
  $B/T$ ratio (0.74 in this case), the disc components represents a small fraction of the
  total stellar mass (i.e. the disc is a tenuous extended rotating system).

As mentioned above, the inner discs that coexist with the spheroids have
stellar-mass surface density profiles that behave differently. Some of them
follow the exponential profiles of the discs determining a unique
exponential profile, while others  break  off  either to follow
the spheroid profiles or to set a new flatter trend. 
We detect  that 3 out
of the 18 analysed SDGs have discs with stellar-mass surface
density profiles following those of the spheroid components. These spheroids
have $n \sim 2-3$.

\begin{figure}[!ht]     
  \centering
  \includegraphics[width=0.45\textwidth]{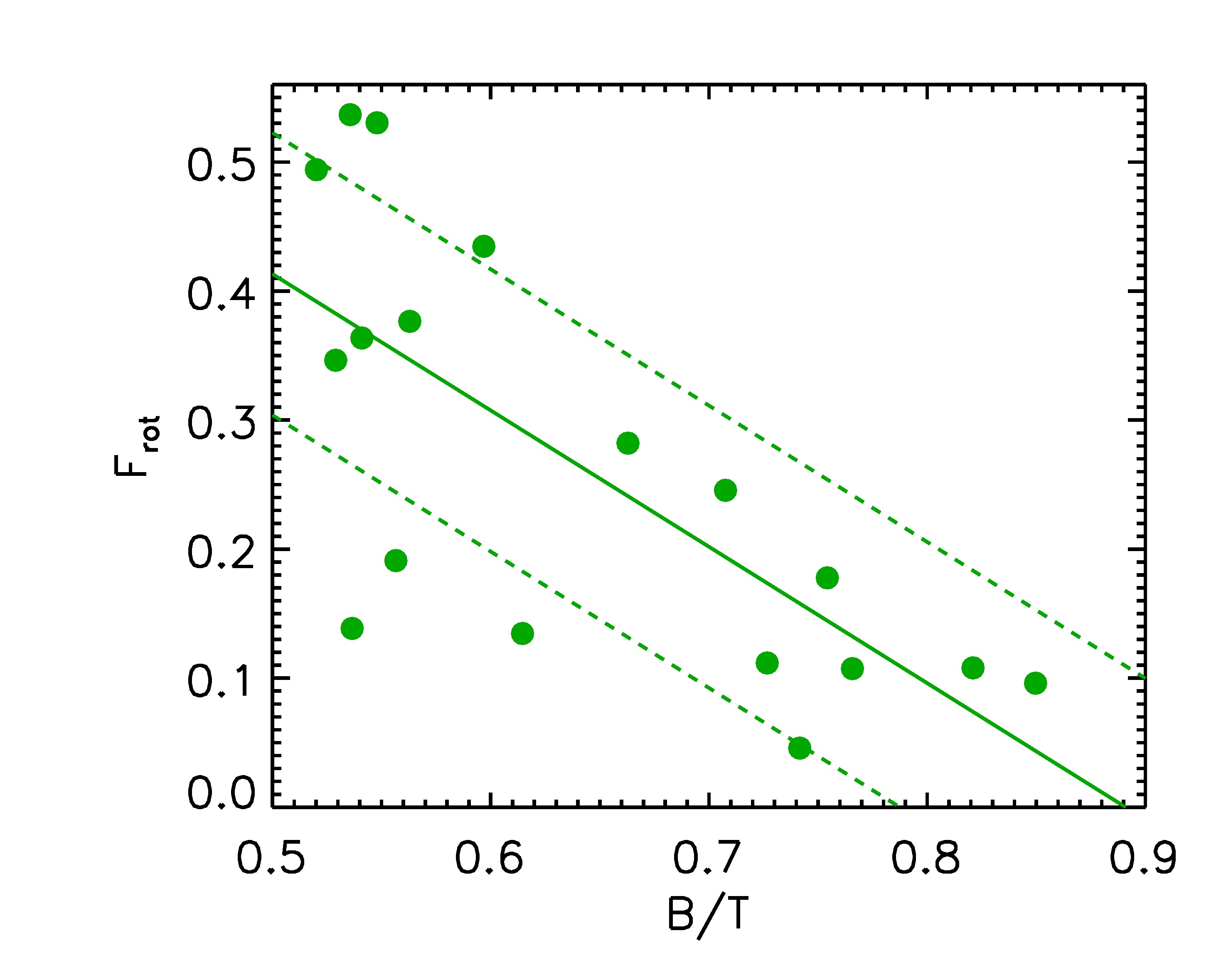}
  \caption{Stellar-mass fractions of the discs that coexist with
      the spheroidal components as a function of the fraction $B/T$. A linear
    regression fit is included (green solid line) along with its 1$\sigma$ dispersion (green dashed lines).}
  \label{fig:fracfrac}
\end{figure}

In order to quantify the relative importance of the inner discs with
respect to the spheroid components (i.e. defined  by the dispersion-dominated stars), 
we calculate the stellar mass fraction, $F_{\mathrm{rot}}$,
of  stars with $\epsilon >0.5$  and with binding energies low enough to be classified
as part of the spheroid component. This fraction is a measure of the rotating
component embedded in the spheroid region. As can be seen from
Fig.~\ref{fig:fracfrac}, a clear correlation between
$F_{\mathrm{rot}}$ and  $B/T$  is found with a Spearman correlation
coefficient of  $-0.77$ ($p=0.0002$). 
Figure~\ref{fig:fracfrac} shows that those systems which  overall are more
dominated by the spheroid component have a smaller contribution of
an inner rotating disc. We note that the dispersion is significant,
particularly for those SDGs with $B/T$ $\sim 0.5-0.6$, where
$F_{\mathrm{rot}}$ can vary between $\sim 0.2$ to $\sim 0.5$,
reflecting very different dynamical assembly histories.

The significant variations in the morphologies and the spheroid (bulge)--disc coexistence are the
result of the different formation histories. In the following
sections, we  study these features and to what extent they are able to reproduce
observed properties of SDGs.

\subsection{Spheroid Sersic index versus  $B/T$}
\label{n-vsB/T}

\begin{figure}[!ht]     
  \centering
  \includegraphics[width=0.45\textwidth]{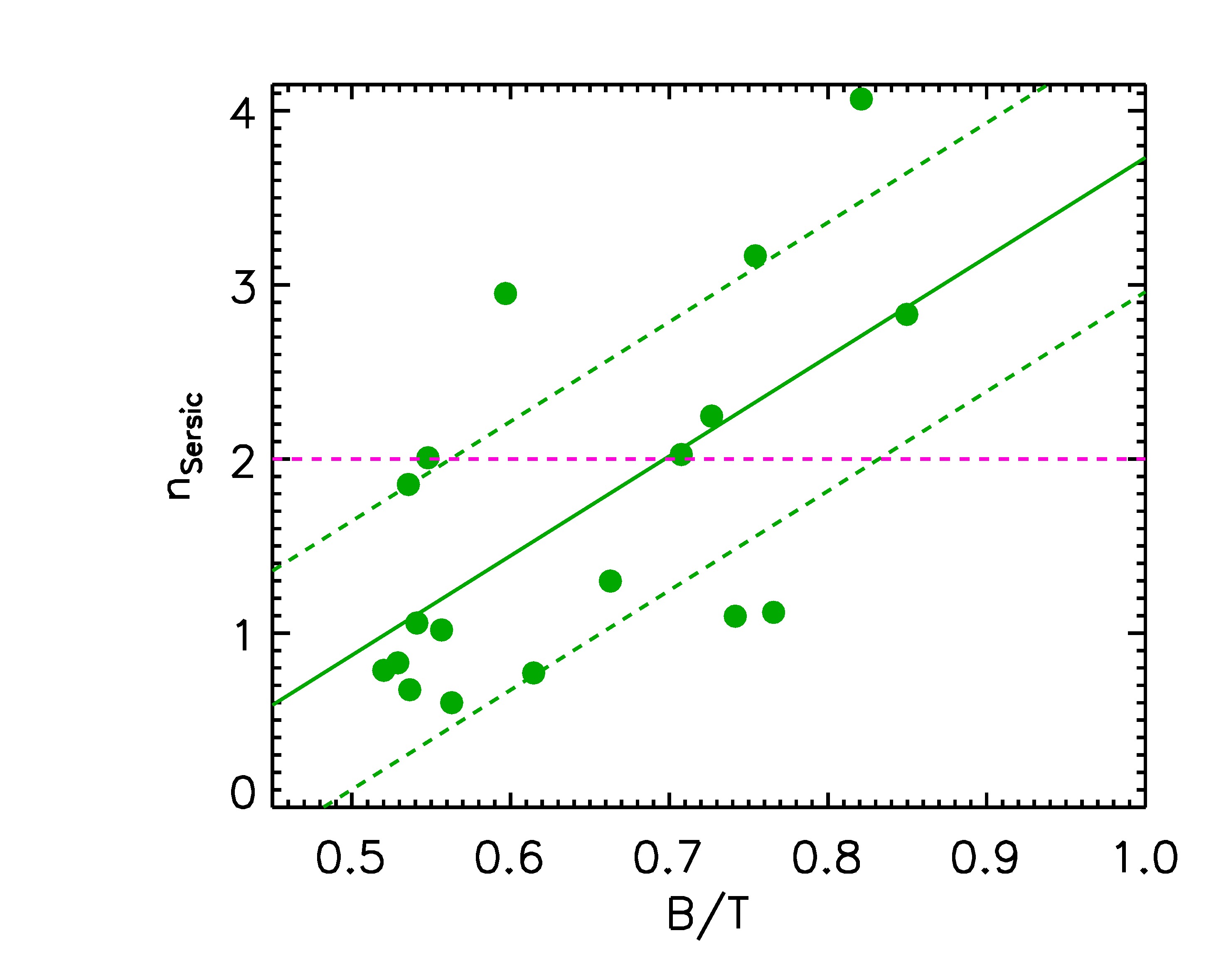}
  \caption{Spheroid Sersic index obtained for the simulated 
    SDGs as a function of their dynamical B/T mass ratios. The green
    solid and dashed lines represent the best linear regression fit and the
    1$\sigma$ dispersion. The magenta
    dashed line denotes $n=2$, which is often assumed to be the limit
    between classical and pseudo-bulges.}
   \label{fig:sersic}
\end{figure}

As shown in Fig. \ref{fig:sersic}, a correlation is detected between the dynamical $B/T$ ratio and the Sersic index. 
The Spearman correlation 
coefficient obtained is $0.63$  ($p =0.01$), implying that the trend is statistically significant, albeit with a large dispersion. 
A linear regression yields a slope of $5.72 \pm 1.89$ and an intercept of $-1.99 \pm 1.13$.
The errors are calculated by a bootstrap method.
The simulated B/T ratios are in a range of  values associated with
elliptical and lenticular (S0) galaxies. In particular, $\sim 50$\% could be classsified as S0 galaxies.  Hereafter, we refer to
 them as spheroidal-dominated systems for the sake of
simplicity and because the spheroidal component is
always the more massive one. 
On the other hand,  we note that if the stellar mass of the disc coexisting with the spheroid ($F_{\rm rot}$, see Fig. \ref{fig:fracfrac}) is assigned to the spheroid (as probably done in the photometric bulge/disc decompositions), then the B/T ratios of the simulated galaxies is larger than shown in Figs. \ref{fig:BT} and \ref{fig:fracfrac}.

It has also been claimed that the  structural Sersic index, $n$, is able  to distinguish between classical bulges and pseudo-bulges,
being $n\sim 2$ the value that separates  these two
types \citep[e.g.][]{Tonini2016, Fisher2008, 
Combes2009}. Not only does this parameter change
for the classical and pseudo-bulges, but also  many other galaxy properties such as colour, sSFR, rotational support, and
kinematics \citep[see][for a recent review]{Kormendy2016}. 
From  Fig. \ref{fig:sersic}, we can see that  half of our SDGs
have $n\leq 2$. These changes   and the presence of
a disc component inside the spheroids (see above) suggest
that the simulated spheroids (bulges) are composite stellar systems formed by the action of different 
formation channels such as mergers, interactions, or local
instabilities \citep[see also Tissera et al. 2017
submitted]{DeLucia2010,Zavala+2012,perez2013}. The extended discs
could grow because  our SDGs
inhabit  low-density environments. In higher density regions, these discs would be prevented from
growing or  surviving by a higher impact of  ram pressure stripping or strangulation.

\section{Scaling relations}

In this section, we analyse the size-mass relation, the FJR, the FP, and the
TFR   determined for the simulated SDGs.
It is important to analyse these relations since none of them has
been used to set the parameters of the subgrid physics modelling, and  the
degree of agreement (or disagreement) thus provides important clues for
improving the models. In all figures in this section, simulated
  galaxies are distinguished according to the B/T ratio. However, the
  global relations are calculated over the whole sample in order to
  have a better statistical estimation.

Since our simulated galaxies are dynamically dominated by the velocity dispersion
components, we  resort to the samples of observed ETGs from the
ATLAS$^{3\mathrm{D}}$ Project \citep{AtlasI, Cappellari2013}
and ETGs with $\mathrm{H_{I}}$ disc
components identified from the ATLAS$^{3\mathrm{D}}$ survey by \cite{Heijer2015}. 
The presence of these discs allows  rotation velocities to be measured accurately. 
We compile a clean ATLAS$^{3\mathrm{D}}$ sample by  excluding  those
galaxies that belong to the Virgo cluster since our SDGs are field
systems. The clean ATLAS$^{3\mathrm{D}}$ sample is compared with simulated trends when appropiate.

\subsection{Size-mass relation}
\label{mass-size}

\begin{figure}
  \centering
  \includegraphics[width=0.45\textwidth]{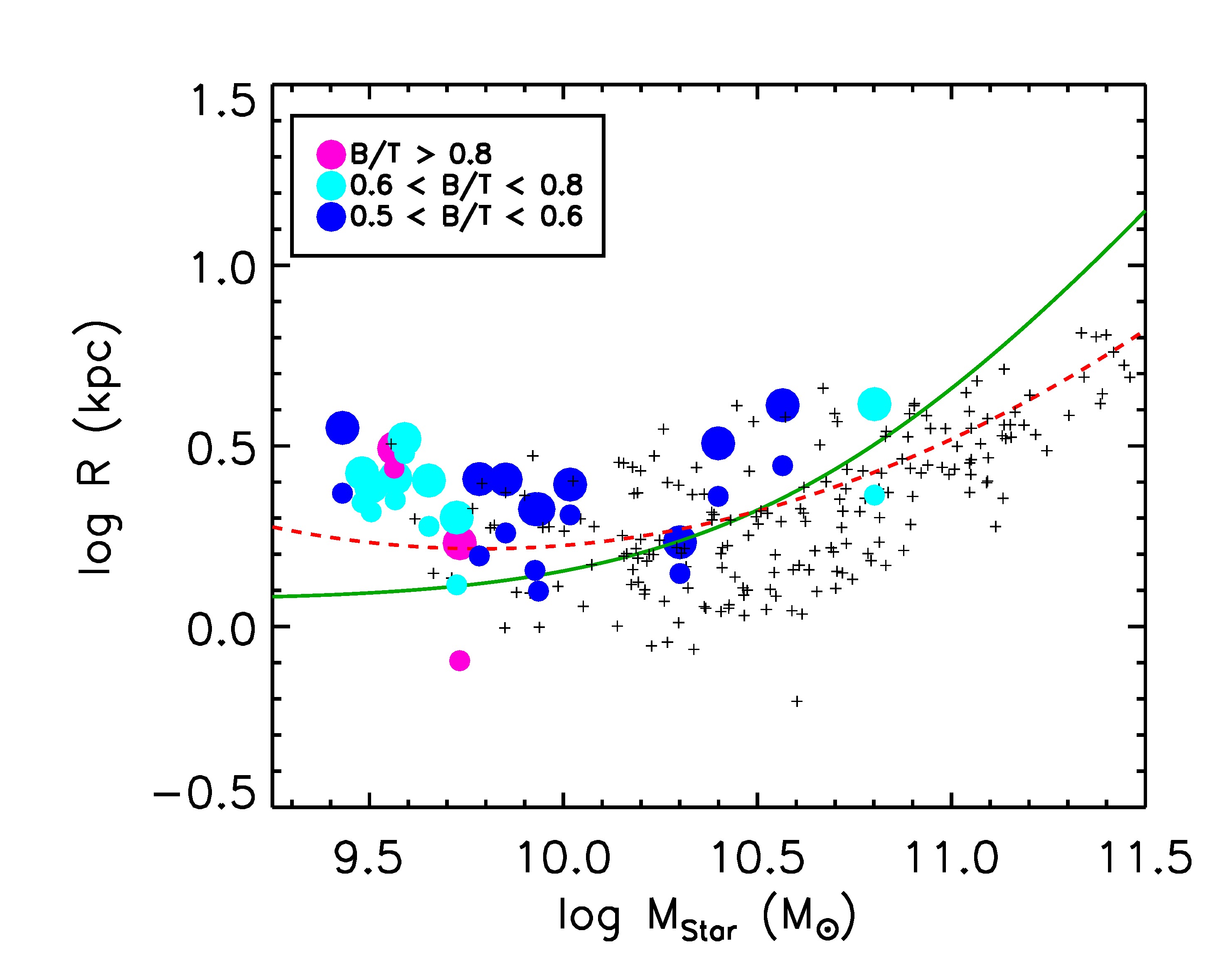}
  \caption{Mass-size relation estimated for  the simulated SDGs
    ($R_{\mathrm{hm}}$,  big  circles, and
    $R_{\mathrm{eff}}$, small circles, both coloured according to the dynamical $B/T$ ratios),  ETGs from the clean ATLAS$^{3\mathrm{D}}$ sample (black
    crosses), and the observed relations reported by \citet{Mosleh13}
   (green solid line) and \cite{Bernardi2014} (red dashed line) for
   ETGs. }
   \label{fig:MS}
\end{figure}

\begin{figure*}[!ht]    
  \centering
  \includegraphics[width=0.45\textwidth]{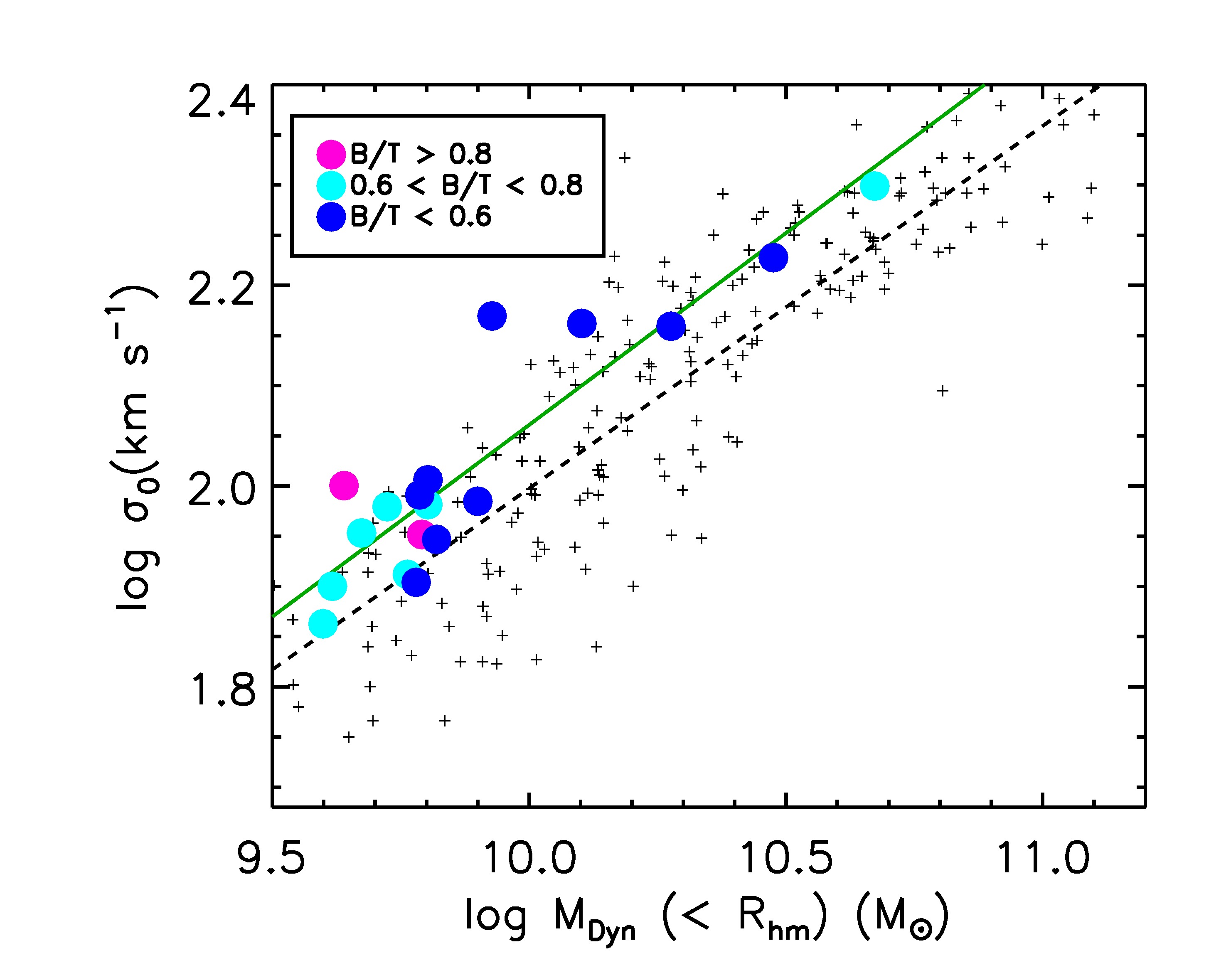}
  \includegraphics[width=0.45\textwidth]{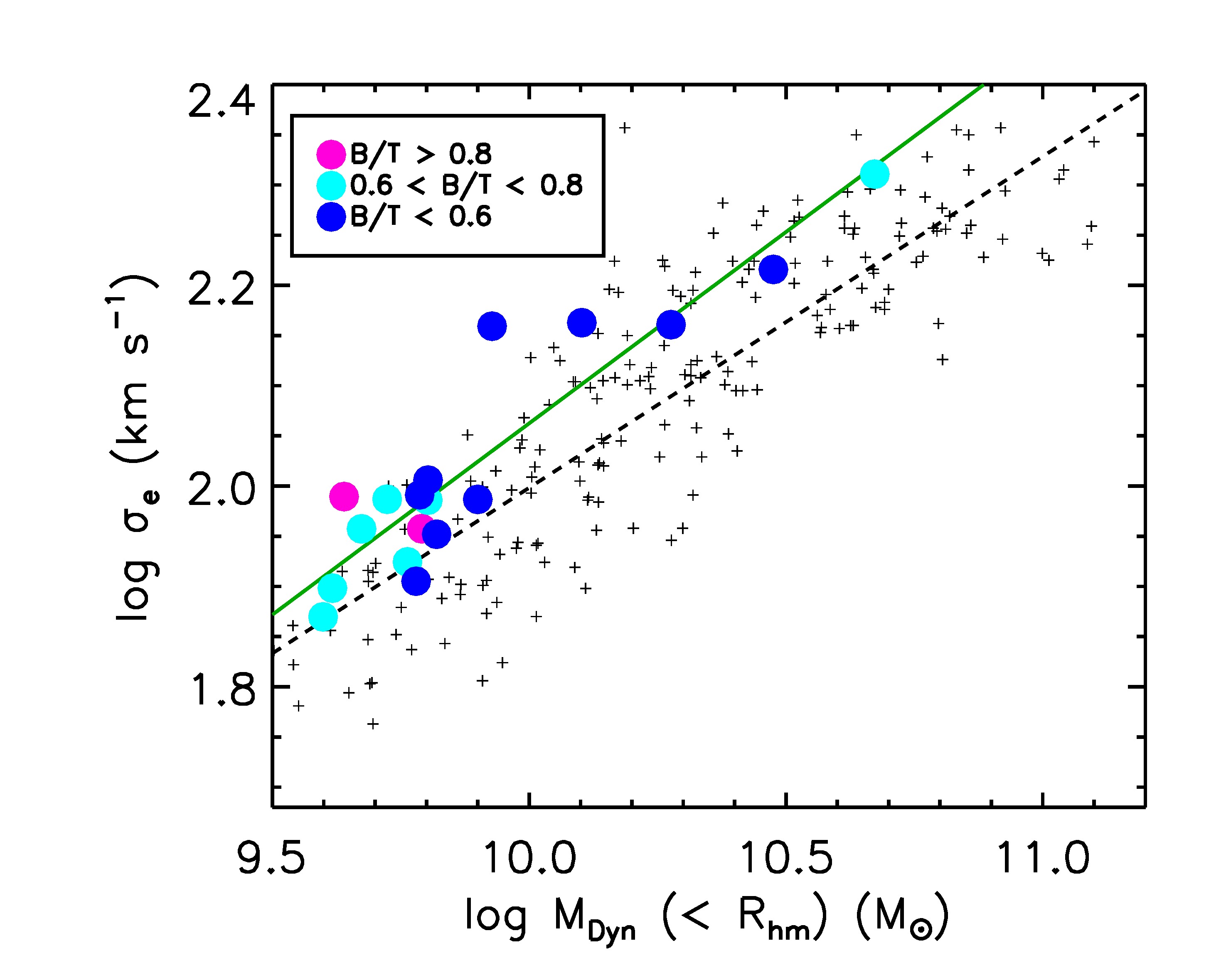}
   \caption{Faber-Jackson relation for the simulated SDGs (filled
    circles, coloured according to the dynamical $B/T$ ratios) estimated by using $\sigma_{0}$ (left panel) or
$\sigma_{e}$ (right panel). 
    For comparison, the corresponding relations obtained from the
    clean ATLAS$^{3\mathrm{D}}$ sample are included (see Section 4).
   The linear regression to the simulated data (green solid lines) and observational
   data (black dashed lines) are depicted for comparison. }
   \label{fig:FJ}
\end{figure*}

One of the fundamental scaling relations is the size-mass
relation \citep[e.g.][]{Mosleh13,Bernardi2014}. 
To check that our SDGs satisfy this observed relation, we first use the $R_{\mathrm{hm}}$, which in simulations
is straightforward to measure (see above). 
Observationally, this radius is commonly estimated from an infrared surface brightness profile 
down to a given aperture (in the infrared the mass-to-light ratio is close to 1) or by fitting a given law to the profile
and extrapolating this law to calculate the total galaxy light.
Applying a similar procedure, we then define  the effective
radius  $R_{\mathrm{eff}}$.
To better compare these data with  observations, we use the Sersic profiles fitted to the
total surface density obtained by combining the disc and spheroid
components within $R_{\mathrm{opt}}$. Then,
$R_{\mathrm{eff}}$ is calculated by
applying the well-known relation \citep[e.g.][]{Saiz01}
\begin{equation}
  \frac{R_{\mathrm{eff}}}{R_{0}}=(2n-0.324)^n
  \label{eq:reff}
,\end{equation}
which links scale  lengths and the Sersic index;
$R_0$ is the scale length of the Sersic profile fitted as mentioned above. 

To compare the simulations  with observations, we use the results reported by
\citet{Mosleh13}  and \citet{Bernardi2014} for ETGs.
\citet{Mosleh13} have adopted the functional form to relate stellar mass ($M_{\mathrm{Star}}$) and size for spiral galaxies
given by \citet{Shen03}
\begin{equation}
  \bar{R}_{\mathrm{eff}}(\mathrm{kpc}) \ = \ \gamma \ \bigg( \frac{M_{\mathrm{Star}}}{M_{\odot}} \bigg)^{\alpha} \ \bigg(1+\frac{M_{\mathrm{Star}}}{M_{0}} \bigg)^{(\beta - \alpha)}
  \label{eq:shen}
,\end{equation}
where  $\bar{R}$ is the median of the log-normal distribution of
$R_{\mathrm{eff}}$ (in kpc)  in stellar mass bins.
\citet{Mosleh13} study a sample selected from
the Max-Planck-Institute for Astrophysics (MPA)--Johns Hopkins
University (JHU) SDSS
\citep{Kauffmann03,Salim07}. 
The fitted parameters ($\alpha$, $\beta$, $\gamma$, $M_0$) vary with morphology, colour, 
or sSFR. We take those corresponding to ETGs (table 1 in \citet{Mosleh13}).

\citet{Bernardi2014} also find that the size-mass 
relation depends on morphology using the SDSS DR7.
They calculate $R_{\mathrm{eff}}$ from different fits to the surface
brightness profiles finding 
\begin{equation}
 \mathrm{log}(R_{\mathrm{eff}} {\mathrm{(kpc)}}) \ = \ p_0 \ + \  p_1 \mathrm{log}( M_{\mathrm{Star}}-0.24) \ + \ p_2 (\mathrm{log}(M_{\mathrm{Star}}-0.24))^2
 \label{eq:bernardi}
.\end{equation}
From \citet{Bernardi2014}  we take the case of a single-Sersic
profile fit for ETGs (their table 4). A correction from Chabrier
\citep{Chabrier2003} IMF to Salpeter \citep{Salpeter55} is applied.

Finally, we also compare the simulated trends with the ETGs from the clean ATLAS$^{3\mathrm{D}}$ sample.
The stellar masses are calculated from the luminosities given in table 1 in
\cite{Cappellari2013}, using the mass-to-light ratios 
from  table 1 in \citet{CappellariAtlasXX} for a Salpeter IMF.

In Fig. \ref{fig:MS}, we show the comparison between the simulated SDGs
and the above-described observational estimates. As can be seen the simulated SDGs have $R_{\mathrm{eff}}$ in
reasonably good agreement with observations, whereas the $R_{\mathrm{hm}}$  are slightly
larger ($\sim 0.2\ \mathrm{dex}$), especially at lower masses. 
It is possible that these  characteristic radii of our simulated SDGs
are indeed slightly larger than those of observed ETGs, due to the presence of extended discs in all of the
simulated galaxies.

Hence, overall the simulated mass-size relations of the ETGs are in good agreement with
observations. We note that \citet{Pedrosa2015} reported that the sizes of
the DDGs are also in reasonable agreement with observations. These
findings suggest that the adopted SN feedback model is able to
reproduce the mass-size relations of both types of galaxies without
resorting to any fine-tuning.

\begin{figure*}[!ht]
  \centering
 \includegraphics[width=0.45\textwidth]{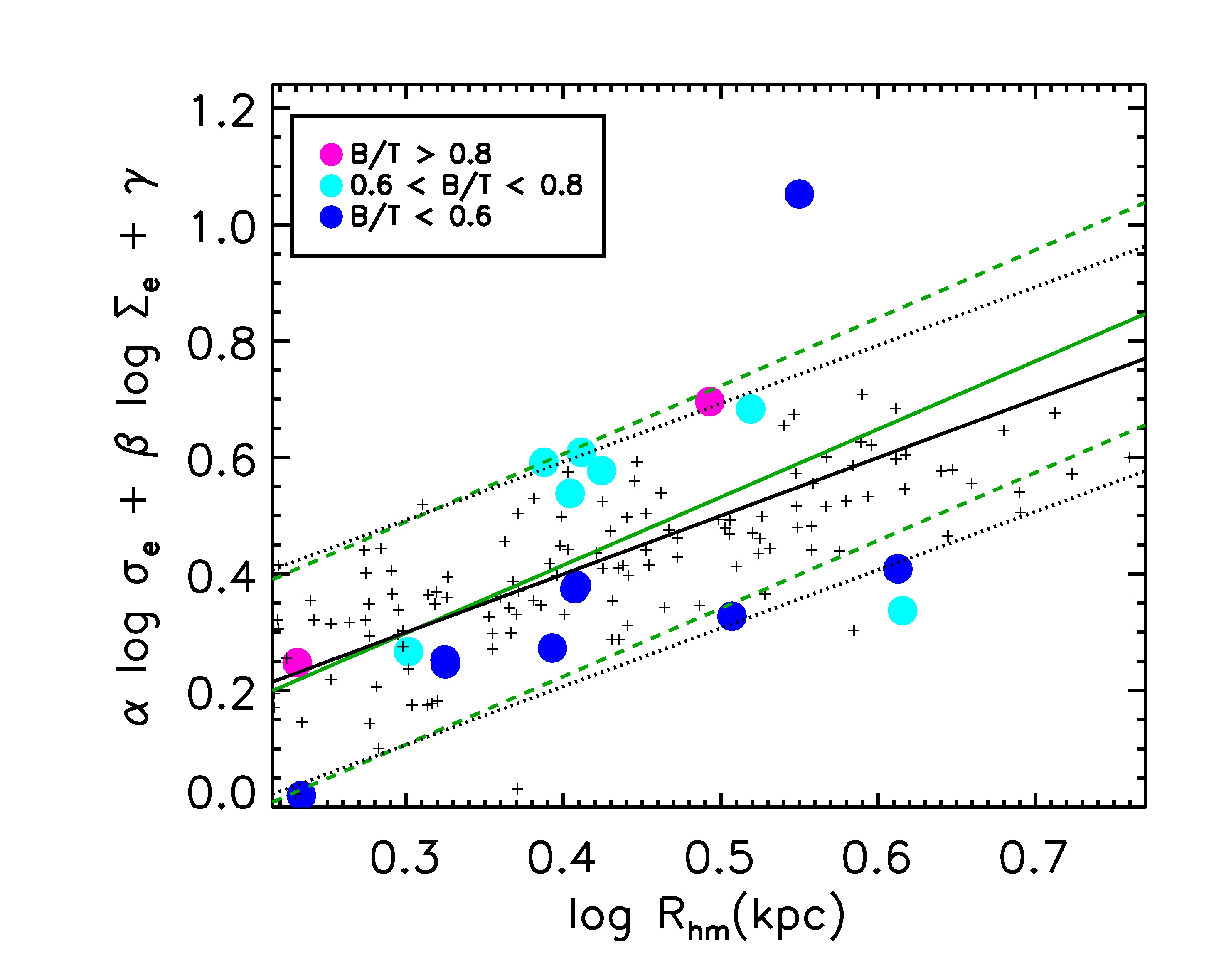} 
  \includegraphics[width=0.45\textwidth]{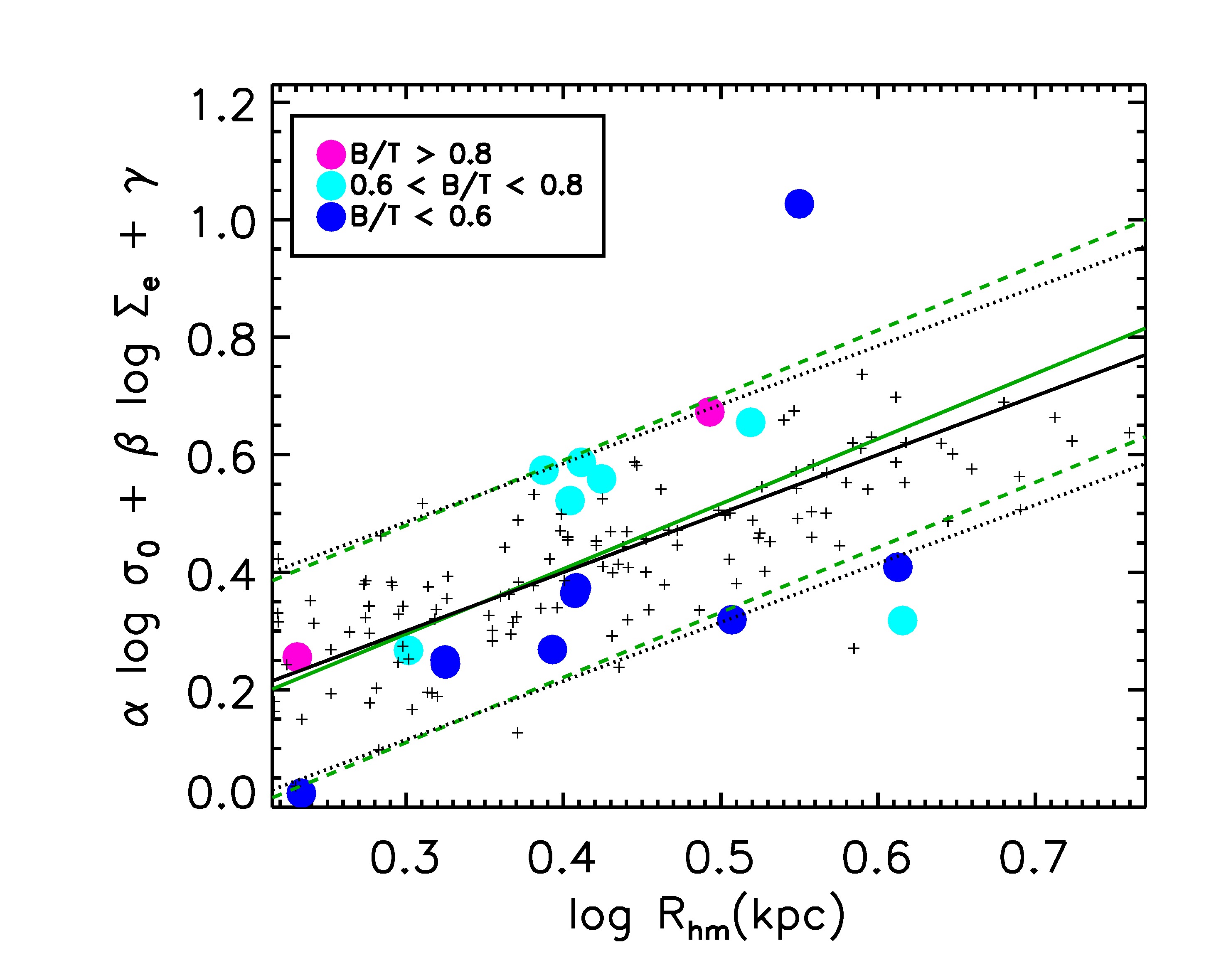}
  \caption{Fundamental plane for the simulated SDGs calculated with the 
parameters estimated for our clean ATLAS$^{3\mathrm{D}}$ sample. The black line denotes
the one-to-one relation and the green line shows the 
best fit to our simulated SDGs (solid circles, coloured according to
the dynamical $B/T$ ratios). The {\it rms} is $0.19$ and $0.18$ for the least
squares regression (dashed green lines) when considering
velocity dispersion within $R_{\mathrm{eff}}$  and central velocity dispersion, respectively,
and  the corresponding regression for 
the one-to-one relation (${\it rms}=0.19$ for both cases) is shown by dotted black lines.}
   \label{fig:FPsimu}
\end{figure*}

\subsection{Faber--Jackson relation}

The FJR \citep{Faber1976} relates the luminosity and the central velocity dispersion
so that $ L\ \propto \ \sigma_{0}^{\gamma}$,
where $\gamma$ is a constant. We use the dynamical masses
instead of luminosities.  For the simulated SDGs,   the  central velocity
dispersion  ($\sigma_{0}$) is calculated  within $1.5 \ h^{-1}$ kpc
(corresponding to three gravitational softening radii) and the 
dynamical masses at  $R_{\mathrm{hm}}$.
For galaxies in the clean ATLAS$^{3\mathrm{D}}$ sample, the dynamical masses are estimated by following
\cite{Cappellari2013}: $M_{\rm JAM} \approx 
2 \times M_{1/2}$, where $M_{1/2}$ is the total mass within a sphere enclosing
half of the galaxy light. To make the conversion to mass, the mass-to-light ratios provided by 
\cite{Cappellari2013} are adopted. 
We also explore whether the same relation holds when the central
velocity dispersion is replaced by the velocity 
dispersion calculated within  $R_{\mathrm{eff}}$ ($\sigma_{e}$).

Figure \ref{fig:FJ} shows both the observational and the  simulated
FJRs obtained by using  $\sigma_{0}$ (left panel) or
$\sigma_{e}$ (right panel). The linear regression for the simulated SDGs 
yields $0.40  \pm  0.05$ for both slopes. The
errors are calculated via a bootstrap method.
Similar linear regression fits to relations defined by the clean
ATLAS$^{3\mathrm{D}}$ galaxies  yield $0.36  \pm 0.01$ and $0.33 \pm 0.01$ respectively. 
Hence, the simulated SDGs follow a FJR in agreement with
observations within one standard deviation. 
No clear dependence is found on the dynamical $B/T$ ratio 
as can be seen from this figure.
We note, however, that our galaxy sample is too small to be able to see 
a robust trend on this point.

\subsection{Fundamental plane}

The  FP  \citep[][]{Dressler1987, Davis1987, Faber1987}
relates the size ($R_{\mathrm{eff}}$) with the surface density ($\Sigma_{e}$) and
velocity dispersion ($\sigma$):
\begin{equation*}
 R_{\mathrm{eff}} \propto \sigma^{\alpha} \ \Sigma_{e}^{\beta}
 \label{fig:FP}
\end{equation*}

The    surface brightness given by the data available in Table 1 in  \citet{Cappellari2013} is transformed into mass surface densities by adopting
$M/L=1$  ($ \Sigma_{e}= \frac{L}{2 \pi R_{\mathrm{eff}}} $).
Similarly to the FJR, the FP is estimated for  $\sigma_{e}$ and
$\sigma_{0}$. 

In Fig. \ref{fig:FPsimu} we compare the simulated FP obtained
using our clean  ATLAS$^{3\mathrm{D}}$ sample. 
For this sample, using $\sigma_e$ for the fit, $\alpha=0.92$ and $\beta=-0.69$ (left panel) for the observed data.
When fitting using $\sigma_0$, $\alpha=0.87$, and $\beta=-0.66$ (right panel).
The black line corresponds to the one-to-one relation ($y=x$).
As can  be seen, the simulated FP (green lines) is within one standard
deviation of the observed relation. 

As  is well-known,  ETGs tend to show a
tilt in the FP with respect to the value derived
assuming virialisation \citep{Binney1987}. Its origin is still under
debate and there are  a variety of possible causes  such  as a
variable or non-homologous IMF \citep{Prugniel1996, Forbes1998},
dependence on dark matter halo features caused by the non-linear
assembly of the structure, among others. However, the FP obtained by
using dynamical masses is consistent with the predictions from the
virial theorem \citep{cappellarireview2016}.

\subsection{Tully-Fisher relation}

\begin{figure*}[!ht]    
  \centering
  \includegraphics[width=0.45\textwidth]{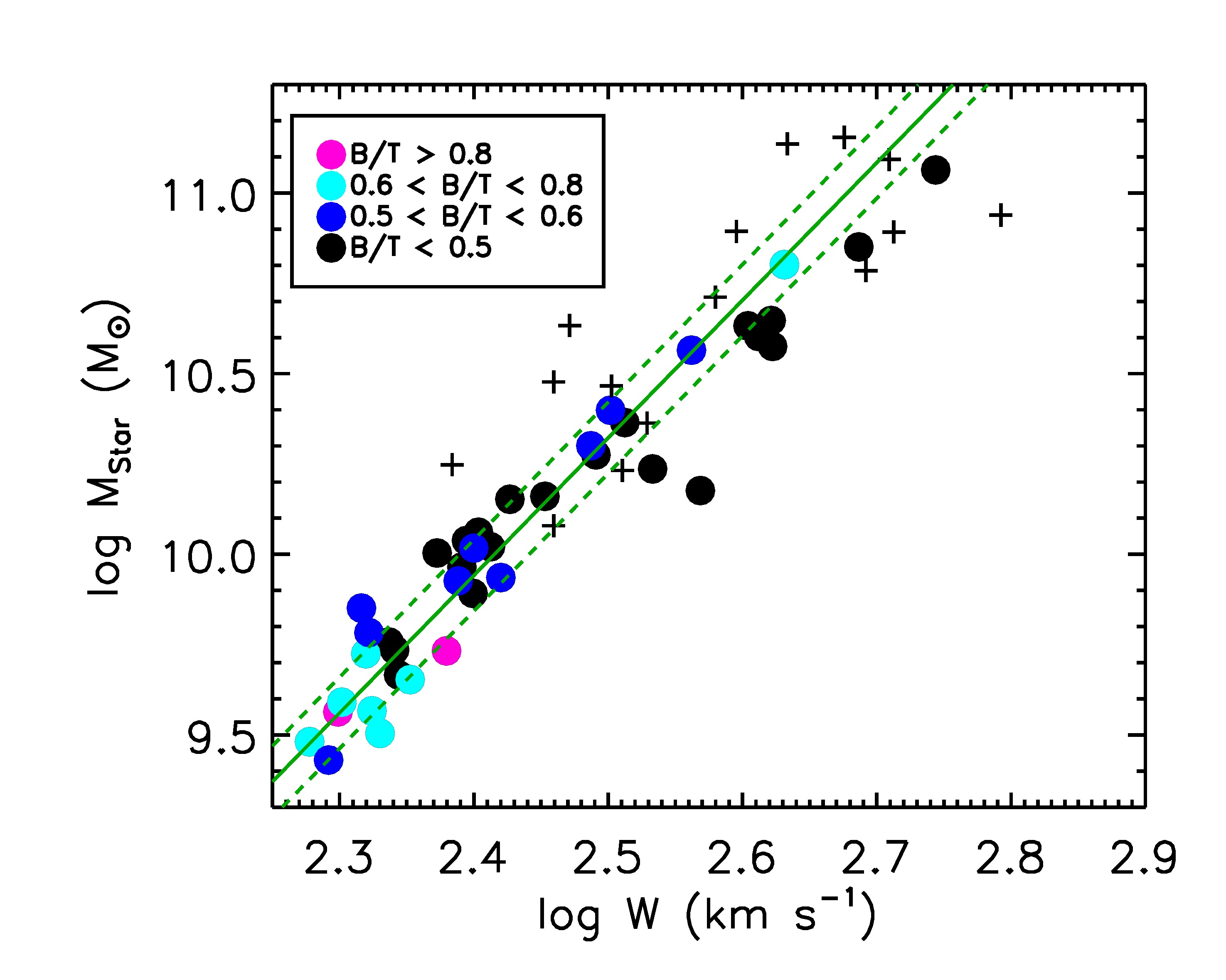} 
  \includegraphics[width=0.45\textwidth]{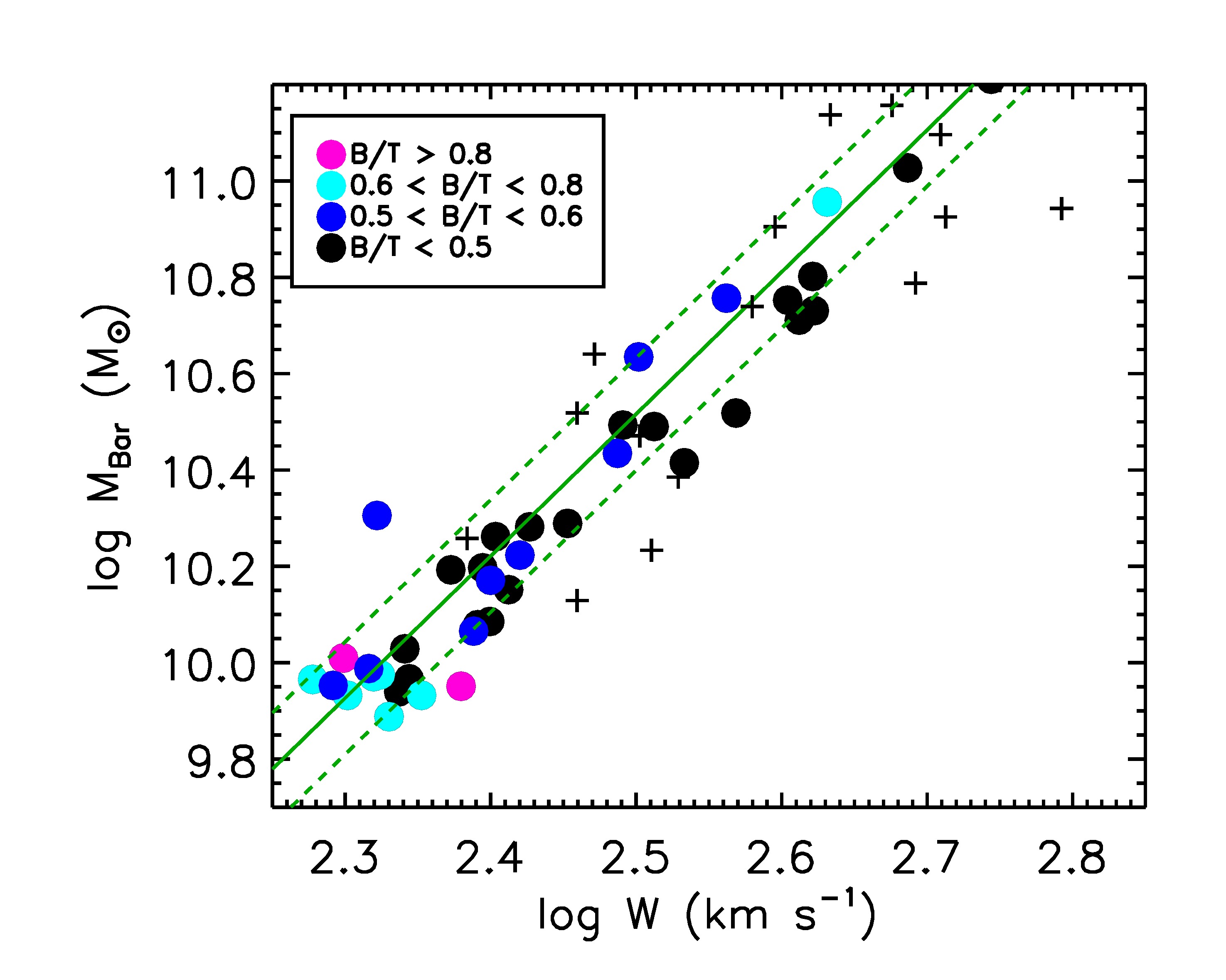}
  \caption{Stellar (left panel) and baryonic (right panel) TFRs for the simulated SDGs (filled circles coloured according to the dynamical $B/T$ ratio) and DDGs (black solid circles),
    and the observed values reported by \citet{Heijer2015} for ETGs
    with extended disc components  (black
    crosses). The linear regression for simulated SDGs (green lines) 
    are shown; the \textit{rms} dispersions are also shown (dashed lines). }
   \label{fig:tf}
\end{figure*}

Although ETGs are dominated by a velocity dispersion component,
 a rotating disc  is also frequently  detected through cold or ionised
gas observations.  \citet{Emsellem2011} report most of the ETGs in the
ATLAS$^{3\mathrm{D}}$ sample as fast rotators.
In particular, \cite{Heijer2015} study the TFR for
16 ETGs from the ATLAS$^{3\mathrm{D}}$ survey using an extended
$\mathrm{H_{I}}$-component which determines a disc component. In this work, the
rotation velocity is measured at very large radius (on
average, within $R/R_{\mathrm{eff}} = 7.3$). We compare this sample
 with the simulated TFR.
For consistency with our previous comparison, a galaxy that belongs to the
Virgo cluster has been excluded.

We determine the stellar and baryonic TFRs for our
simulated SDGs  and fit a relation of the form $\mathrm{log} (M_{i})=a \ (\mathrm{log} (W)-2.6)+b$,
where $i$ represents the stellar or the baryonic mass and  $W$ is
twice the  rotational velocity $V$ calculated at  twice $R_{\mathrm{opt}}$. The stellar and baryonic masses 
are determined  within $R_{\mathrm{opt}}$.
The estimation of stellar masses for the sample of
\citet[][Table 1]{Heijer2015} is done in the same way  as in
Section 4.1 by using $r$-band mass-to-light ratios, i.e. their star formation history (SFH) case.
The baryonic masses are calculated as the sum of stellar mass and $\mathrm{H_{I}}$ mass multiplied by a factor of 
$1.4$ to take into account  helium and metals \citep{Heijer2015}.

In Fig. \ref{fig:tf} (left panel) we show the stellar
TFR. In this case, the linear regression yields $a=3.81 \pm 0.20$ and
$b=10.70 \pm 0.04$ for the simulated SDGs. This is steeper than that
reported by  \citet{Heijer2015} ($a=2.40 \pm 0.50$ and $10.72 \pm 0.06$), but within the scatter and uncertainties involved in the stellar mass determination of the observations and in agreement with the TFR for
spiral galaxies. For disc galaxies, it is known that the slope of stellar TFR is steeper than that of the baryonic TFR, and the
scatter of the former is smaller than that of the latter \citep[e.g.][]{Avila-Reese+2008}. These trends are followed by our
simulated SDGs.  Our simulated stellar TFR is also in agreement with results
from the cosmological EAGLE simulation \citep{Ferrero2017}. 

Similarly, Fig.~\ref{fig:tf} (right panel) shows the baryonic TFR  for the simulated SDGs and that
reported in \citet{Heijer2015}. 
The best fitting parameters for the
simulated relation are $a=2.95 \pm 0.31$ and $b=10.81 \pm 0.08$, which
are in agreement with the observed values within the estimated
errors. In particular, the unconstrained observed  TFR 
(i.e. when they calculate the parameters without fixing any of them),
where the  baryonic mass is calculated with SFH, yields a slope of $a=2.51 \pm 0.42$ and zero-point of $b=10.71 \pm 0.05$. 

For comparison,  in Fig. \ref{fig:tf} we include the subsample of simulated 
DDGs with more than 10,000 baryonic particles (black filled circles). Both the SDG and DDGs
follow roughly similar stellar and baryonic TFRs, although disc-dominated galaxies
seem to determine a slightly flatter relation in the former case. This might be caused by
the action of the  SN feedback  which is more efficient for lower
stellar mass galaxies in the velocity range where most of the SDGs are \citep{deRossi2010}.

Overall, the main structural and dynamical relations of our simulated SDGs 
are in reasonably good  agreement with those reported for 
observed ETGs. This is an  encouraging result, considering that none of these relations has
been fine tuned to be reproduced as mentioned before.

\subsection{Dark matter fraction as a function of stellar mass}

A key prediction of the $\Lambda$-CDM scenario is that galaxies are
embedded in  dark matter halos. However, it is not yet clear what
the fraction of dark matter within ETGs is.  Thanks to integral field spectroscopy studies, like those performed in the ATLAS$^{3\mathrm{D}}$ survey,
estimations of $F_{\rm dm}= M_{\rm dm}/M_{\rm dyn}$ are now possible for ETGs \citep{CappellariAtlasXX}. 

\begin{figure}
  \centering
  \includegraphics[width=0.45\textwidth]{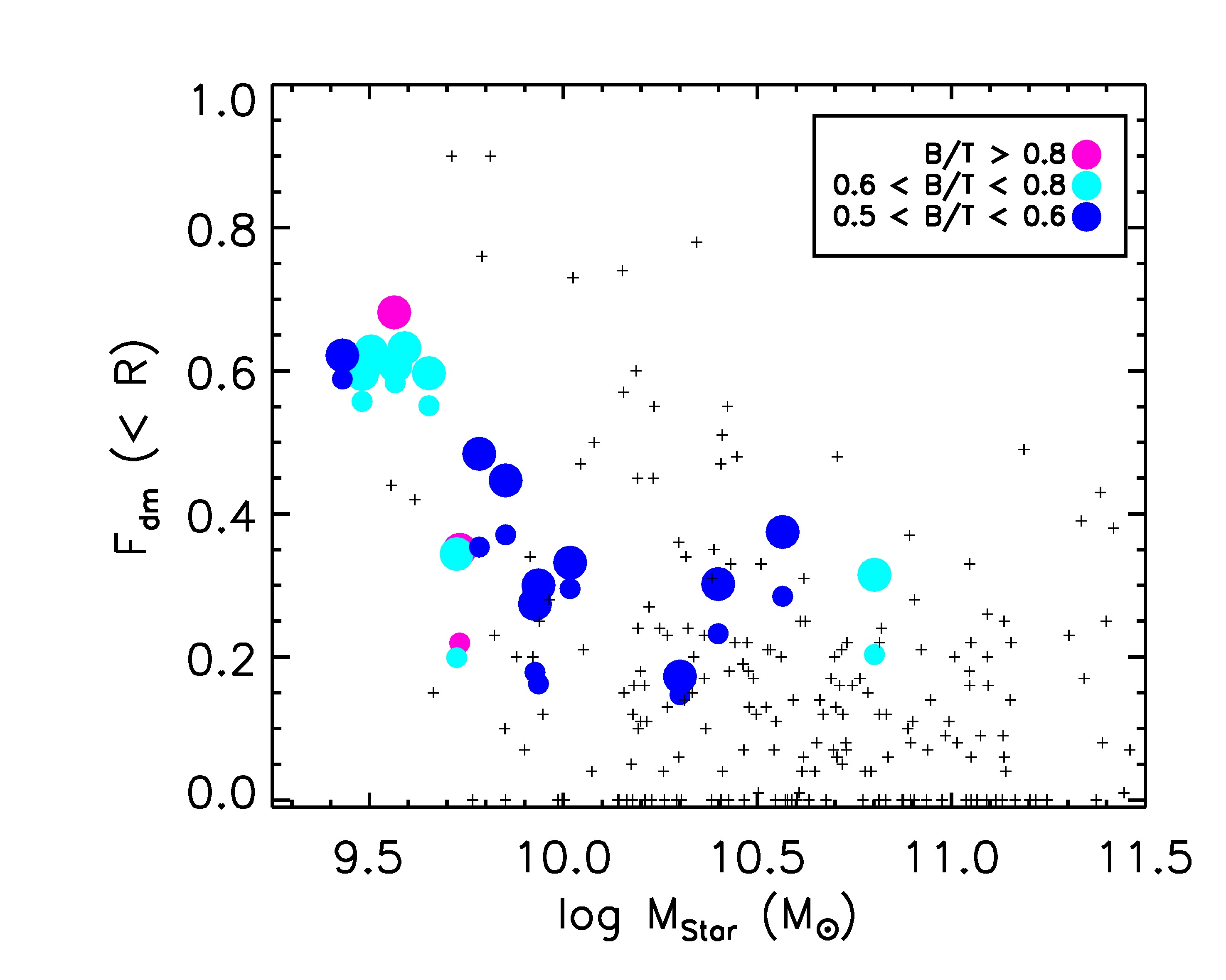}
  \caption{Dark matter fraction,  $F_{\rm dm}= M_{\rm dm}/M_{\rm dyn}$, within $R_{\mathrm{hm}}$ (big solid circles) and $R_{\mathrm{eff}}$ 
  (small solid circles), both coloured according to the dynamical $B/T$ ratios as a function of stellar mass for the simulated SDGs. The crosses are dynamical estimates for the 
   ATLAS$^{3\mathrm{D}}$ ETGs \citep{Cappellari2013, CappellariAtlasXX}. }
   \label{DMfraction}
\end{figure}

In Fig. \ref{DMfraction}, the $F_{\rm dm} $  measured for our SDGs at $R_{\mathrm{hm}}$ and $R_{\mathrm{eff}}$ 
are plotted. Since $F_{\rm dm}$ increases with the radius, the values of $F_{\rm dm}$ are higher for $R_{\mathrm{hm}}$ than
for $R_{\mathrm{eff}}$ . For both definitions of
$F_{\rm dm}$ the smaller the galaxy, the larger the dark matter
fraction. This trend is in agreement
with the dynamical inference for the ATLAS$^{3\mathrm{D}}$ survey
\citep{Cappellari2013}, as shown  in  Fig. \ref{DMfraction} (for the observed values we have used the stellar masses calculated in
Section 4.1). 
 The agreement with the observational inferences is
reasonable. However, none of the simulated SDGs has values of $F_{\rm dm}(<R_{\mathrm{eff}})$ smaller than 0.15, while 
many of the ATLAS$^{3\mathrm{D}}$ ETGs have these values. These are galaxies dominated by baryons and should be
very compact in the centre, with high surface densities. The presence of discs in all of our SDGs make them likely less compact.

\begin{figure*}[!ht]    
  \centering
\includegraphics[width=0.45\textwidth]{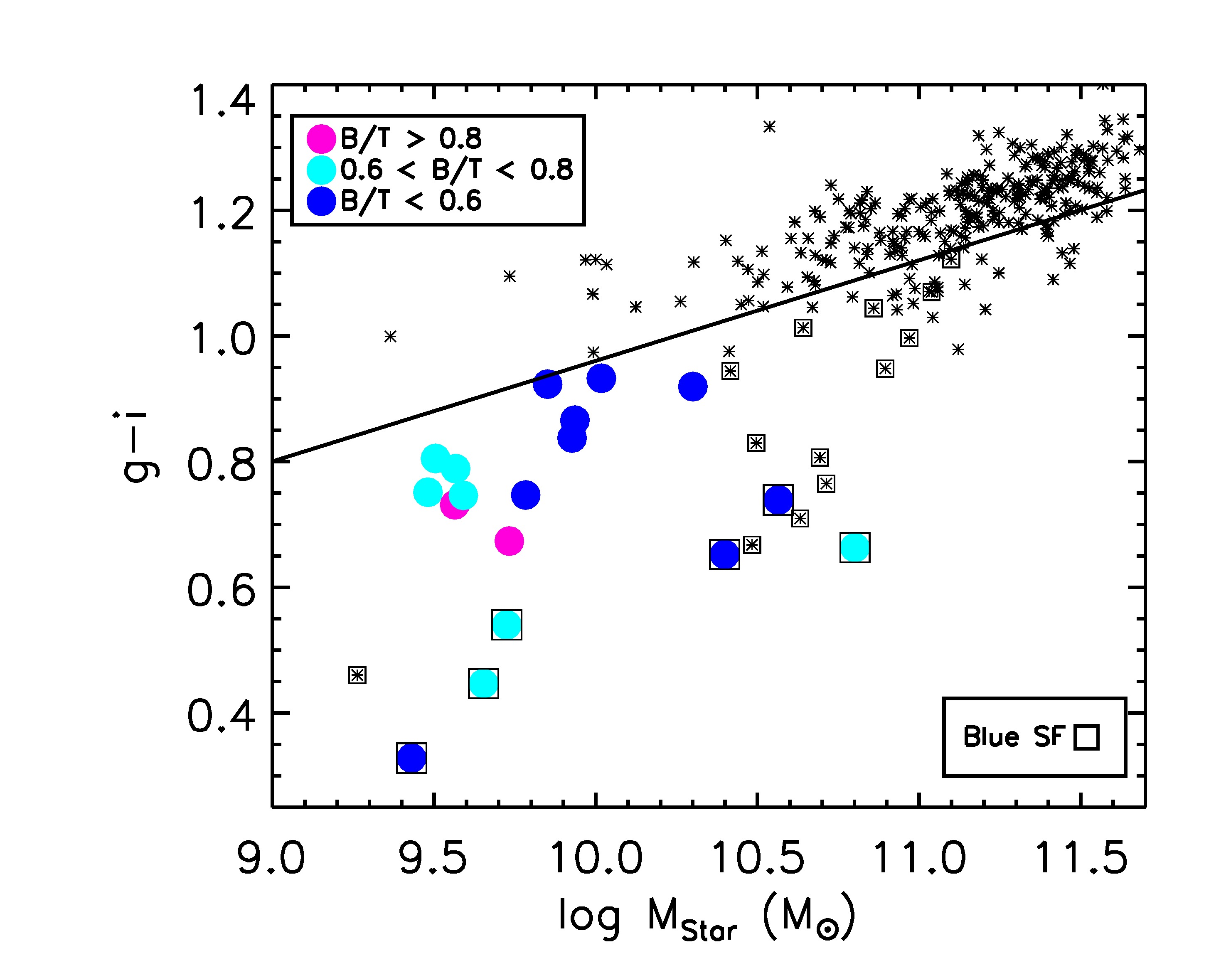}
  \includegraphics[width=0.45\textwidth]{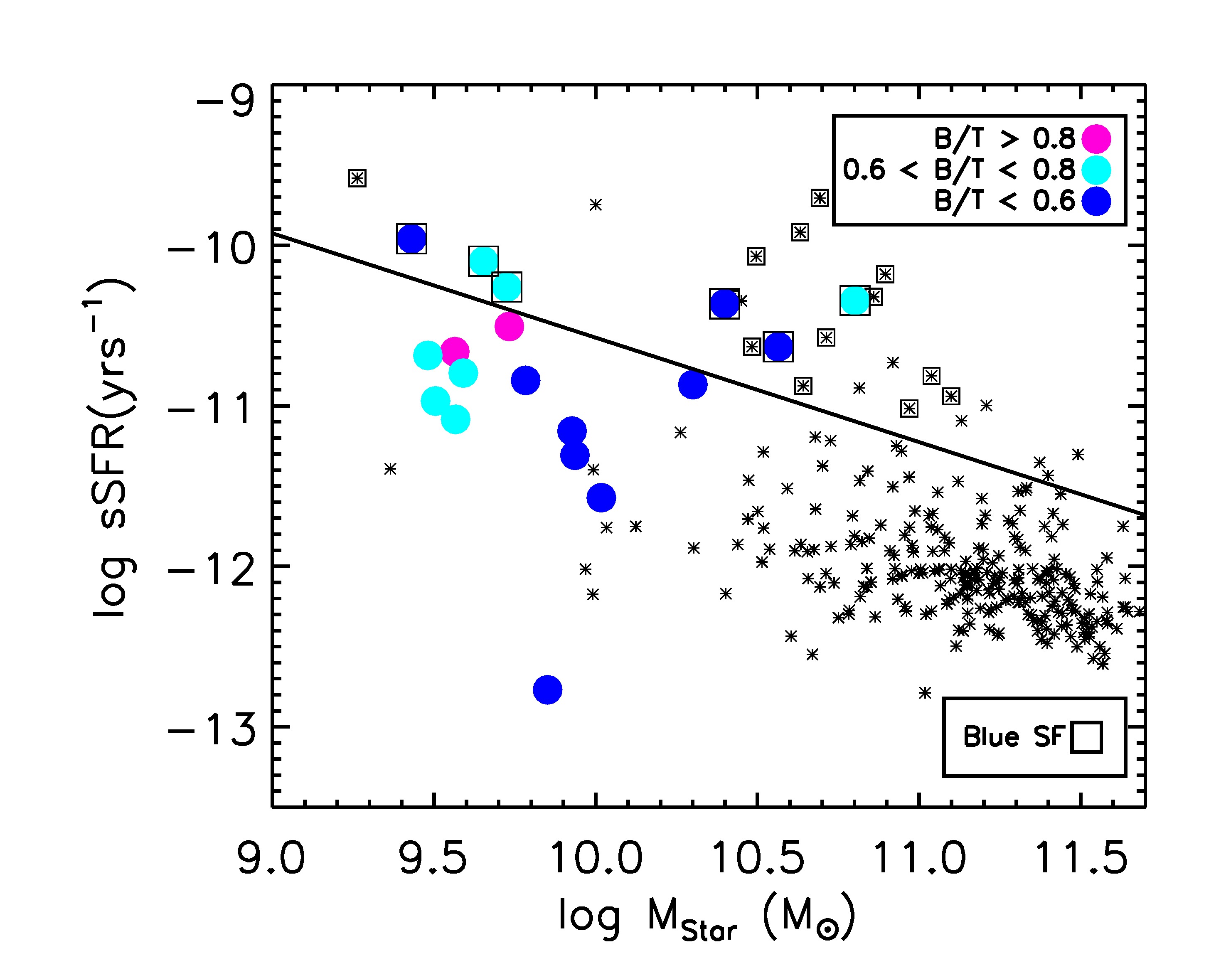}
  \caption{Colours and sSFRs as a function of the $\mathrm{M_{Star}}$ (left and right panels, respectively) for the simulated SDGs 
  (filled circles coloured according to the dynamical $B/T$ ratios). 
    Observational results for isolated ETGs \citep[black asterisks;][]{Hernandez-Toledo+2010} are included. 
    In the left panel, blue and red galaxies are separated by the
    black line reported by \cite{Lacerna2014},
while in the right panel the black line depicts the limit between 
    star-forming and passive galaxies \citep{Lacerna2014}.
    Galaxies that are blue and star-forming at the same time are marked with black open squares.
    }
   \label{fig:Fig3}
\end{figure*}

\begin{sidewaystable*}
 \caption{Main properties of simulated SDGs.}
 \label{table:table1}
 \centering
 \begin{tabular}{ccccccccccccccc} 
 \hline\hline             
 ID & $M_{\mathrm{vir}}$ &N baryons  & $M_{\mathrm{Star}}$ & $R_{\mathrm{opt}}$ & $R_{\mathrm{hm}}$ & $B/T$ & $F_{\mathrm{rot}}$ & $n_{\mathrm{Sersic}}$ & $R_{\mathrm{eff, Sersic}}$ &  $R_{\mathrm{d}}$  & $n$  & $R_{\mathrm{eff}}$ & sSFR & $g-i$ \\
  & $(10^{11} \mathrm{M_{\odot}})$ &  & $(10^{9} \mathrm{M_{\odot}})$ & (kpc) & (kpc) &  &   &  & (kpc) &(kpc) &  & (kpc) & ($10^{-11}\mathrm{yr}^{-1}$)  & \\
 (1) & (2) & (3) & (4) & (5) & (6) & (7) & (8) & (9) & (10) & (11) & (12) & (13) & (14) & (15)\\
 \hline
 288  &  20.74 & 133719  & 63.34 & 26.58 & 4.13 & 0.74 & 0.05 & 1.10 & 2.26 & 18.48 & 1.07 & 2.31 & 4.52 & 0.66 \\
 579  &  10.59 & 78666   & 36.73 & 30.88 & 4.10 & 0.54 & 0.36 & 1.06 & 2.20 & 7.32  & 1.07 & 2.79 & 2.33 & 0.74 \\
 613  &  7.80  & 61831   & 25.05 & 22.61 & 3.21 & 0.53 & 0.35 & 0.83 & 1.63 & 5.44  & 0.96 & 2.29 & 4.32 & 0.65 \\
 735  &  4.30  & 40504   & 19.97 & 10.09 & 1.71 & 0.56 & 0.38 & 0.60 & 1.06 & 2.23  & 0.50 & 1.40 & 1.35 & 0.92 \\
 881  &  2.79  & 21211   & 10.41 & 15.38 & 2.47 & 0.52 & 0.49 & 0.79 & 1.47 & 3.18  & 0.81 & 2.04 & 0.27 & 0.93 \\
 790  &  3.98  & 23397   &  8.62 & 17.47 & 2.11 & 0.60 & 0.43 & 2.95 & 0.96 & 3.73  & 3.10 & 1.25 & 0.49 & 0.87 \\
 897  &  2.19  & 17297   &  8.45 & 9.99  & 2.11 & 0.54 & 0.14 & 0.67 & 1.19 & 3.83  & 1.10 & 1.43 & 0.70 & 0.84 \\
 885  &  2.89  & 14635   &  7.10 & 22.08 & 2.55 & 0.55 & 0.53 & 2.01 & 1.41 & 5.21  & 1.44 & 1.82 & 0.02 & 0.92 \\
 746  &  2.75  & 21649   &  6.07 & 32.66 & 2.56 & 0.54 & 0.54 & 1.85 & 1.03 & 4.32  & 2.09 & 1.57 & 1.44 & 0.75 \\
 823  &  3.63  & 13422   &  5.40 & 18.09 & 1.70 & 0.85 & 0.10 & 2.83 & 0.79 & 3.93  & 2.84 & 0.80 & 3.12 & 0.67 \\
 946  &  2.08  & 11990   &  5.31 & 18.37 & 2.00 & 0.66 & 0.28 & 1.30 & 1.12 & 3.42  & 1.73 & 1.30 & 5.49 & 0.54 \\
 868  &  2.53  & 12661   &  4.50 & 11.51 & 2.54 & 0.61 & 0.13 & 0.77 & 1.49 & 3.29  & 1.09 & 1.89 & 7.95 & 0.45 \\
 925  &  2.16  & 11439   &  3.89 & 20.74 & 3.30 & 0.71 & 0.25 & 2.03 & 2.54 & 4.82  & 1.79 & 3.01 & 1.60 & 0.75 \\
 904  &  2.39  & 12963   &  3.69 & 16.00 & 2.58 & 0.73 & 0.11 & 2.25 & 2.43 & 4.02  & 1.58 & 2.24 & 0.82 & 0.79 \\
 1005 &  1.78  & 12609   &  3.66 & 20.51 & 3.11 & 0.82 & 0.11 & 4.07 & 5.09 & 5.08  & 1.71 & 2.74 & 2.17 & 0.73 \\
 969  &  2.25  & 10405   &  3.20 & 14.85 & 2.44 & 0.77 & 0.11 & 1.12 & 2.08 & 3.81  & 0.86 & 2.08 & 1.07 & 0.81 \\
 979  &  1.71  & 10713   &  3.03 & 23.89 & 2.66 & 0.75 & 0.18 & 3.17 & 1.82 & 5.32  & 2.59 & 2.20 & 2.06 & 0.75 \\
 917  &  2.76  & 12325   &  2.69 & 19.20 & 3.55 & 0.56 & 0.19 & 1.02 & 1.73 & 4.47  & 1.13 & 2.34 & 10.96 & 0.33 \\

\hline

\end{tabular}
     \tablefoot{Column (1): Galaxy ID. Column (2): Virial mass. Column
       (3): Number of baryon particles within optical radius. Column
       (4): Stellar mass within optical radius. Column (5): Optical
       radius. Column (6): Stellar half-mass radius. Column (7): Bulge-to-total stellar mass ratio. Column (8): Fraction of rotating
       low-energy component and spheroid mass. Column (9): Sersic
       index of the spheroid component. Column (10): Spheroid
       effective radius calculated from equation \ref{eq:reff}. Column
       (11): The disc scale length estimated from  the exponential fit. Column (12): Total Sersic index. Column (13): Total effective radius calculated from equation \ref{eq:reff}. Column (14): Specific star formation rate. Column (15): $g$ -- $i$ colour. Galaxies are listed in order of descending stellar mass.}

\end{sidewaystable*}

\section{Galaxy colours and specific star formation activity}
\label{sec5}

In this section, we focus on the analysis of the SF activity and the
galaxy colours of the simulated SDGs. 
Integrated magnitudes and
colours are computed from the resulting fully integrated spectral energy distribution (SED)
of each galaxy, based on its age, mass, and metallicity (for more details see Appendix \ref{app:image1}).
We compare our data with the sample of ETGs (elliptical and
lenticular) from the UNAM-KIAS Catalog of Isolated Galaxies
\citep{Hernandez-Toledo+2010} selected from DR5 SDSS under strict isolation
criteria. \citet{Lacerna2016} have studied in  detail  a subsample of pure elliptical galaxies 
from the UNAM-KIAS Catalog (see more details therein).

In Fig. \ref{fig:Fig3} (left panel) the $g-i$ colours are plotted against $M_{\mathrm{Star}}$.
As can be seen, the simulated SDGs occupy a range of stellar masses shifted to lower masses 
with respect  to the bulk of the observations, but in the mass interval where they overlap the simulated
galaxies are clearly bluer than the observed ones. 
The black line separates blue and red galaxies according to a relation presented
in \citet{Lacerna2014},
\begin{equation} \label{eq:colour}
 g-i= 0.16 \ [\mathrm{log}(M_{\mathrm{Star}})-10.56]+1.05 ,
\end{equation}
where $M_{\mathrm{Star}}$ is in units of M$_{\odot}$, and the masses were corrected to a  Salpeter IMF \citep{Salpeter55}.
All the simulated SDGs lie on the blue side of the $g-i$  versus $M_{\mathrm{Star}}$ relation, while
most of  the observed isolated pure ETGs are in the red sequence.
In particular, only $20 \%$ of these observed galaxies in the same mass range (approximately $[10^{9.4},10^{10.8}] \ \mathrm{M_{\odot}}$) are blue. 
The blue colours of simulated SDGs are consistent with a more important contribution of
young stellar populations. 

\begin{figure}[!ht]     
  \centering
\includegraphics[height=0.675\textwidth]{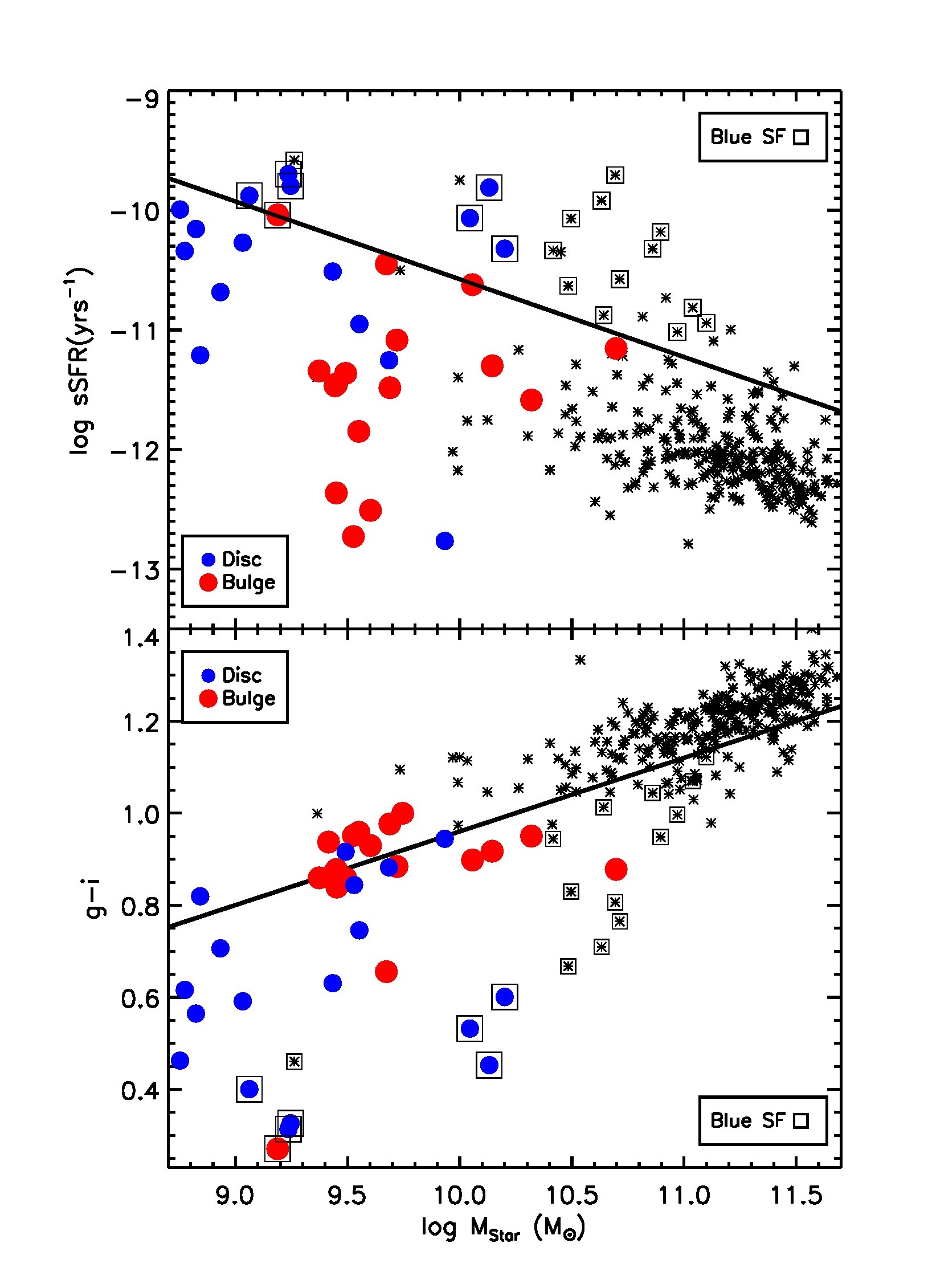}
  \caption{Properties of the spheroid (red circles) and disc (blue circles) components: $g$--$i$ (top panel) and sSFR (bottom panel) as a function of $\mathrm{M_{Star}}$.  For comparison, we include the same observations as in Fig. \ref{fig:Fig3} (black asterisks).
  We depict blue star-forming galactic objects
  as black open squares,
  and the limiting lines reported by \cite{Lacerna2014} between red and blue
    galaxies in the top panel, and between star-forming and non-star-forming galaxies in the bottom panel.
    }
   \label{fig:varias}
\end{figure}

To   investigate the SF activity of the simulated galaxies
and the relation with the colour distributions, we calculated the
SFR by using stars younger than 1 Gyr.
This SFR is not an `instantaneous' measure of the activity, but we choose it because
it is less affected by numerical noise than measures using lower ages given that most of the 
SDGs have low values of SFR at $z=0$. 

In Fig. \ref{fig:Fig3} (right panel), we plot sSFR (=SFR/$M_{\mathrm{Star}}$), as a function of $M_{\mathrm{Star}}$. The black line 
separates passive and star-forming galaxies according to \citet{Lacerna2014},
\begin{equation} \label{eq:ssfr}
 \mathrm{log}(\mathrm{sSFR})=-0.65 \ [\mathrm{log}(M_{\mathrm{Star}})-10.56]-10.94 ,
\end{equation}
where $M_{\mathrm{Star}}$ is in units of $\mathrm{M_{\odot}}$ and  sSFR in units of  yr$^{-1}$ (corrected for Salpeter IMF).
There are 6 out of 18 SDGs ($\sim 33$ \%) in the star-forming region,
while the rest are below the demarcation line and  follow the same trend. With respect to the isolated ETGs from the observations
\citep{Hernandez-Toledo+2010}, our simulated SDGs
are about $\sim 0.5-1$ dex more active. 
The six star-forming SDGs are highlighted with a blue open square in
this figure  and in the following ones. As can be seen in both panels of Fig. \ref{fig:Fig3}, the fraction of simulated blue star-forming 
SDGs is  higher than the observed fraction for isolated ETGs ($33$ \% vs. $14$ \% in the same mass range).

\subsection{Properties of spheroids and discs}

Because each simulated SDG has been decomposed dynamically into a
spheroid and a disc component, similar estimations of colours and
sSFRs as presented above can be performed for each of these components. 
In particular, we explore whether  the colours and sSFR  of the spheroids would agree more closely with observations.

In Fig. \ref{fig:varias}  we show the sSFR and $g$--$i$ colour as a function
of the stellar mass for the disc and spheroid components.
The differences between the two components are evident.
The disc components of the simulated SDGs are nearly
all blue according to \citet{Lacerna2014}  and  with high sSFR (there are only two red discs and both of them are passive), whereas the
spheroid components  show redder colours and much lower sSFR values, as we
expected. 
We find  $\sim 39$ \% of the spheroids are red and $\sim 94$ \%\  are quiescent.
In the mass range in common with the observations,
$[10^{9.4},10^{10.8}] \ \mathrm{M_{\odot}}$, $\sim 50$ \%\  of the spheroids 
are red and all are quiescent  (except only one which is just above the
limit). 
 Therefore, comparing these  fractions with  those of observed
 isolated ETGs in the mentioned mass range, we find that the star
 formation activity of the spheroids are in rough
 agreement. However the colours of the simulated spheroids are still
 bluer than observations ($\sim 80$ \%).

\begin{table*}
\caption{Fraction of young stars in the spheroid and disc components.} 
\label{table:table2}   
\centering              
\begin{tabular}{ccccccc}   
\hline\hline   
  ID & \multicolumn{3}{c}{Spheroid} & \multicolumn{3}{c}{Disc}  \\
   &  < 2 Gyr &  < 3 Gyr & < 4Gyr & < 2 Gyr &  < 3 Gyr &< 4Gyr\\
  \hline                   
288 & 0.01 & 0.03 & 0.04 & 0.31 & 0.50 & 0.64 \\
579 & 0.01 & 0.01 & 0.01 & 0.09 & 0.14 & 0.16 \\
613 & 0.01 & 0.02 & 0.12 & 0.19 & 0.38 & 0.54 \\
735 & 0.03 & 0.05 & 0.08 & $\sim$ 0 &  $\sim$ 0 & 0.02 \\
881 & $\sim$ 0 & $\sim$ 0 & $\sim$ 0 & 0.02 & 0.03 & 0.08 \\
790 & 0.02 & 0.02 & 0.06 & 0.04 & 0.19 & 0.35 \\
897 & 0.01 & 0.01 & 0.01 & 0.02 & 0.04 & 0.08 \\
885 & $\sim$ 0 & $\sim$ 0 & $\sim$ 0 & 0 & 0 & 0.01 \\
746 & $\sim$ 0 & $\sim$ 0 & $\sim$ 0 & 0.07 & 0.14 & 0.21 \\
823 & 0.09 & 0.16 & 0.22 & 0.02 & 0.04 & 0.09 \\
946 & $\sim$ 0 & $\sim$ 0 & 0.01 & 0.19 & 0.20 & 0.26 \\
868 & 0.01 & 0.01 & 0.07 & 0.39 & 0.61 & 0.69 \\
925 & $\sim$ 0 & 0.02 & 0.05 & 0.14 & 0.32 & 0.43 \\
904 & 0.02 & 0.03 & 0.05 & 0.10 & 0.20 & 0.23 \\
1005 & 0.02 & 0.02 & 0.03 & 0.22 & 0.24 & 0.24 \\
969 & $\sim$ 0 & $\sim$ 0 & $\sim$ 0 & 0.09 & 0.14 & 0.17 \\
979 & 0.01 & 0.02 & 0.03 & 0.14 & 0.18 & 0.20\\
917 & 0.16 & 0.21 & 0.28 & 0.42 & 0.47 & 0.51 \\
  \hline 
  
\end{tabular} \\

\end{table*}

\begin{figure}[!ht]     
  \centering
  \includegraphics[width=0.45\textwidth]{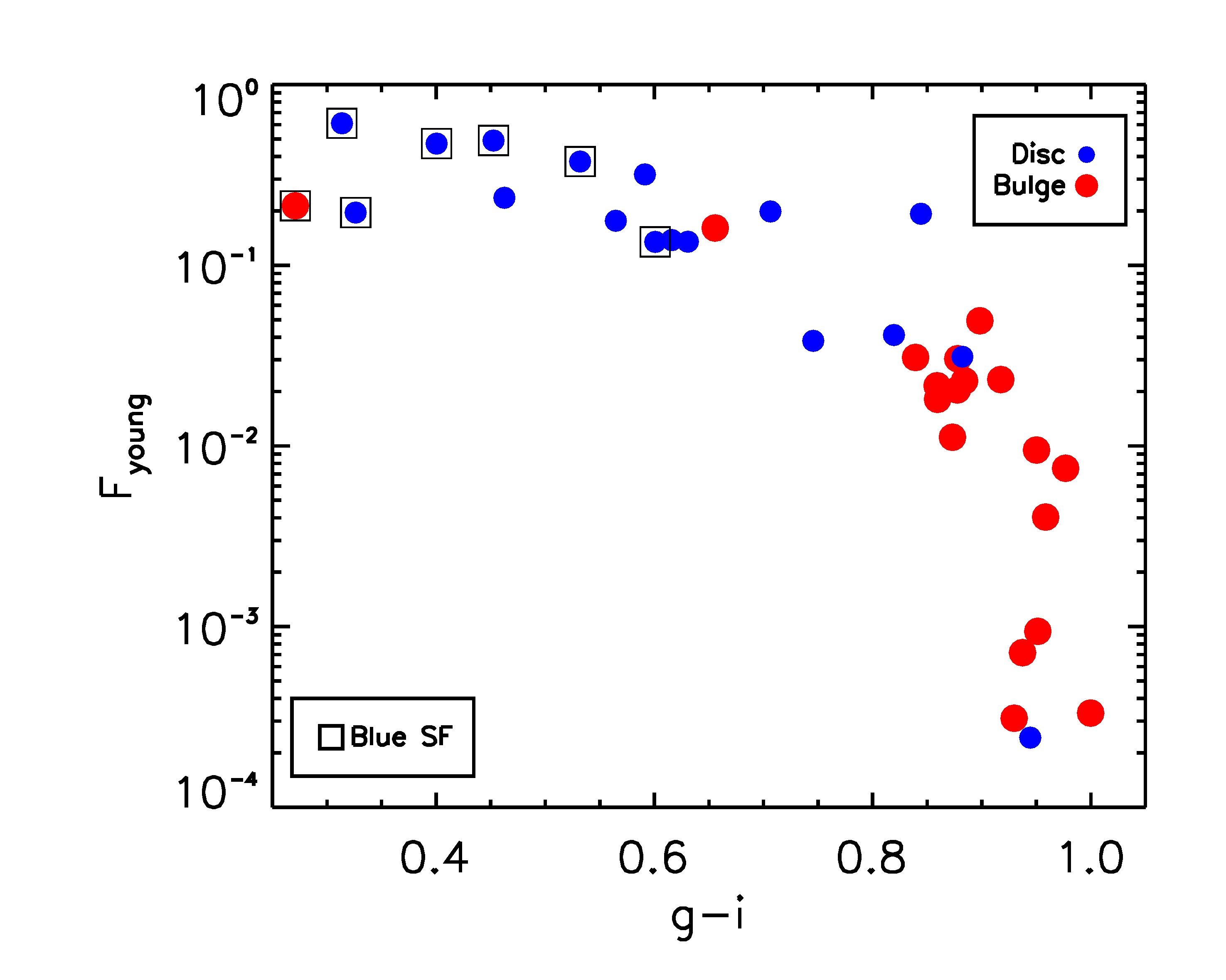}
  \caption{Fraction of stars younger than 3 Gyr as a function of the
    $g$--$i$ for the  spheroid (large red   circles) and disc (small blue 
     circles) components. We also depict blue star-forming galaxies with open black squares.
    }
  \label{fig:frac1}
\end{figure}

\begin{figure}[!ht]
  \centering
   \includegraphics[width=0.45\textwidth]{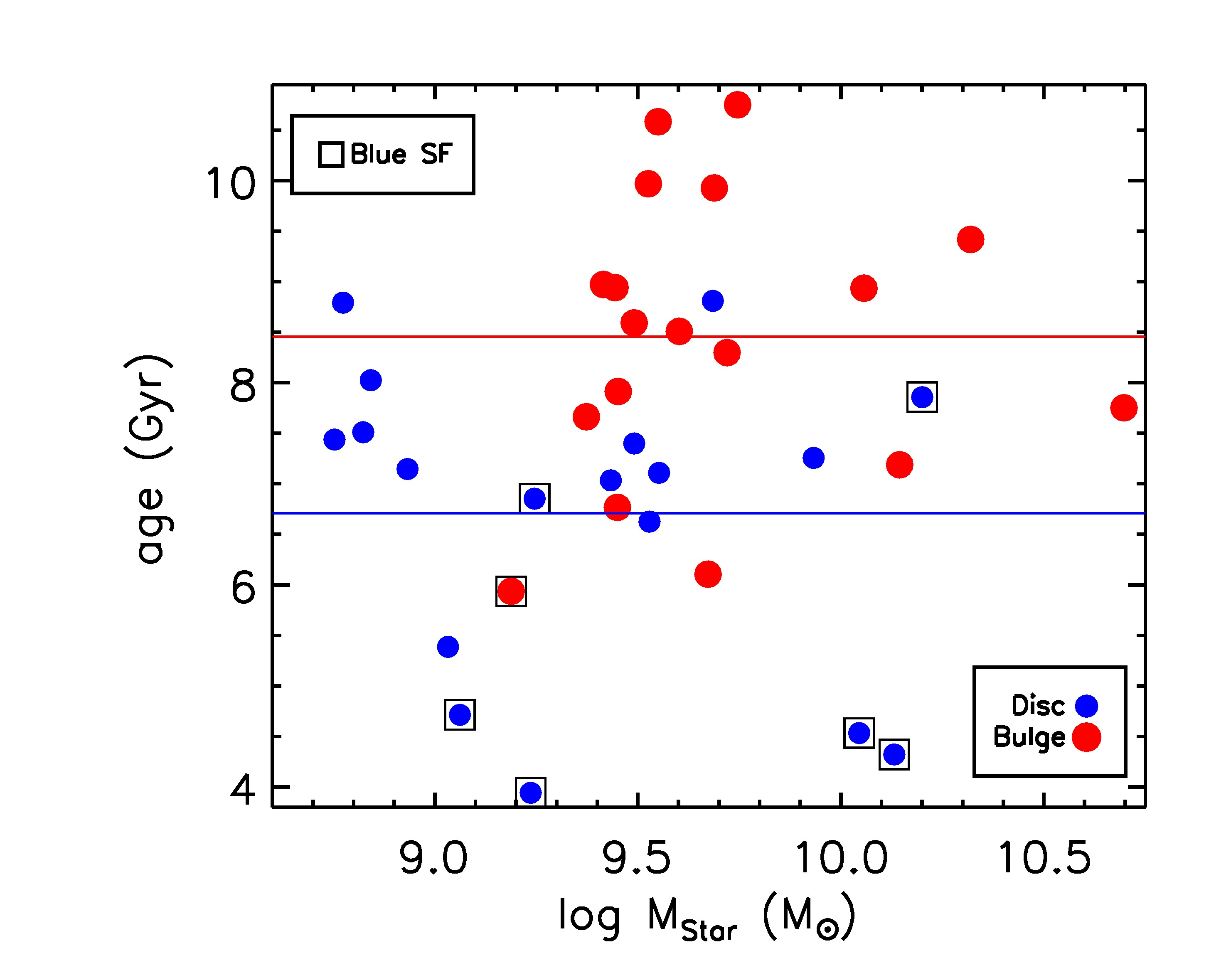}
  \caption{Average mass-weighted ages of the stellar populations in
    the  spheroid (large red
circles) and disc (small blue  circles) components of the SDGs.  
The horizontal lines represent the average ages of all spheroids (red)
and discs (blue).}
  \label{fig:fig6_comp}
\end{figure}

As expected, the disc components  form stars more actively than the spheroids.
This can be quantified by estimating the fraction of young stars. 
In Fig.\ref{fig:frac1} we show the distribution of the fraction of
stars younger than 3 Gyr as a function of $g-i$ for the spheroid and disc components. 
It is clear that most spheroids have no recent SF, while the discs show the
opposite, i.e. that most of them have experienced more recent active episodes, except 
a single disc which does not have stars younger than
3 Gyr (not shown in Fig.~\ref{fig:frac1}). 
A similar trend is found when a 2 Gyr threshold is adopted instead. In Table
\ref{table:table2} the fractions of young stars using different age
thresholds are given.
The colours of the spheroids are not as red as
expected and this is due to the presence of an intermediate-age stellar
populations of about 3-4 Gyr old that, although small, it is enough to make colours bluish.

We can also estimate the average ages of the stars in the spheroid and disc
components of the simulated SDGs. Figure \ref{fig:fig6_comp} shows these
distributions. As
expected, the disc components are younger (horizontal lines correspond to
each group average age) than the spheroids: 
$\sim 6.7 $ Gyr compared to
$\sim 8.5 $ Gyr. 
There is not clear dependence of the mean ages on the stellar mass in
agreement with the results reported by \citet{Lacerna2016}.

Our  findings show that the simulation produces field SDGs  which are
globally bluer and more star-forming than observed isolated ETGs,
mainly due to the persistence of a disc component in all of  them. 
 We note, however,  that even for the spheroid components the simulations show a trend where  an excess of blue systems is formed, 
 with a significant fraction of intermediate-age stellar populations. Some extra mechanisms for avoiding further disc growth and/or
 efficiently quenching SF seem to be necessary.
For our most massive simulated galaxies, the inclusion of AGN feedback could work in this direction.
Simulations of massive galaxies with AGN feedback have shown that the SF rate is 
reduced whilst this feedback is active \citep[e.g.][]{Khalatyan+2008,Grand+2017}. 
As a result, the galaxies end with a lower fraction of intermediate-age stars that are redder than when  AGN feedback is not included. 
\citet{Dubois+2013} have shown that in the absence of AGN feedback, a large number of stars accumulate in the central galaxies to form overly massive, blue, compact, and rotation-dominated galaxies; instead,  when AGN feedback is 
included, these blue massive LTGs turn into red ETGs. However, \citet{Newton+2013} and \citet{Park+2017} have explicitly shown 
that the reduction in SF is significant only for merging galaxies (which is common for massive halos), mainly because the 
AGN feedback heats the gas and prevents  the formation of a new disc, eliminating the possibility of a SF 
burst that would otherwise  happen. For Milky Way-sized galaxies without mergers or with minor mergers, the effect of AGN feedback is minor
for the SF history. All analysed galaxies are sub-Milky Way systems, so the
effects of AGN feedback is expected to be minor.

\begin{figure*}
  \centering
  \includegraphics[width=0.935\textwidth]{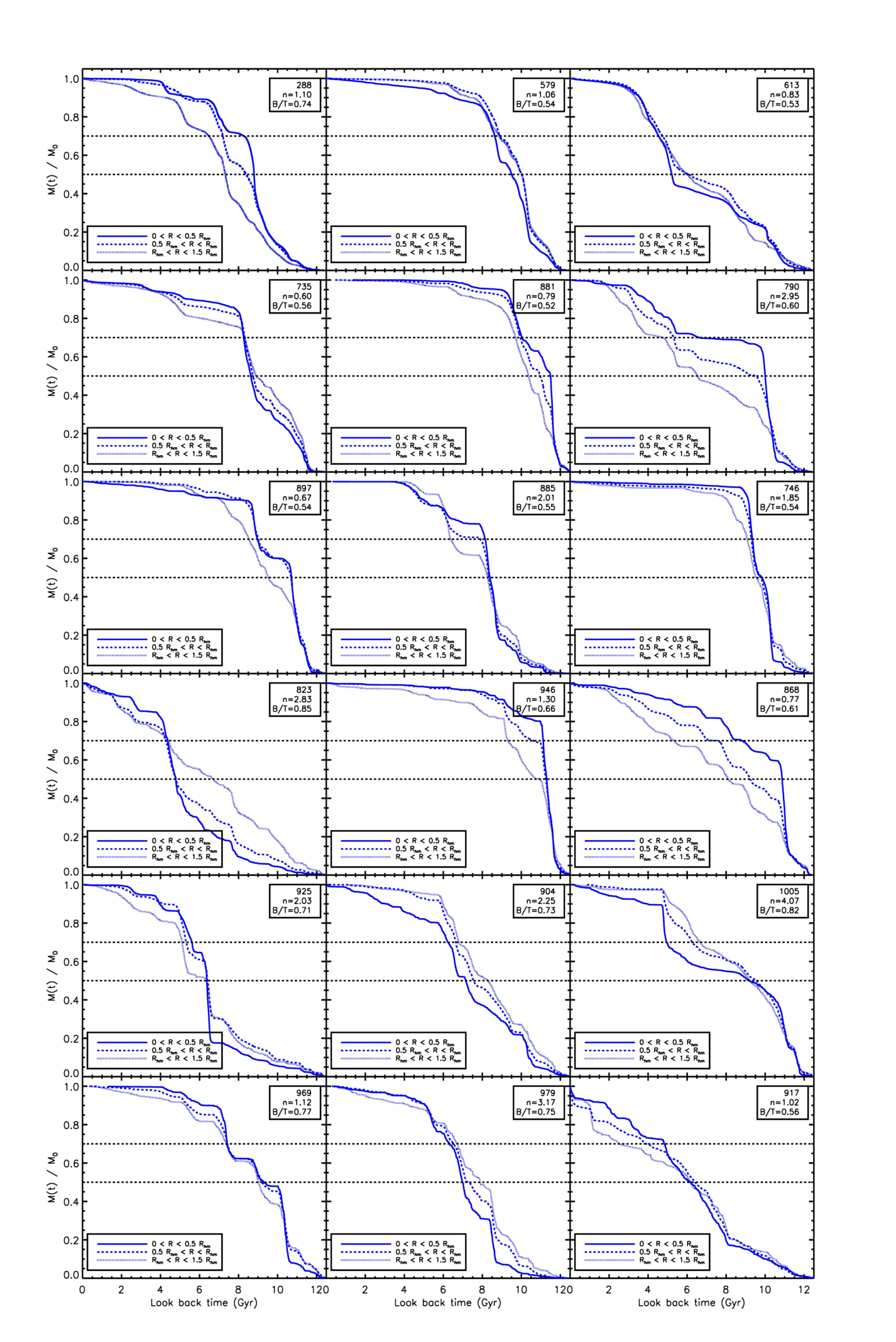}
  \caption{Normalised MGHs estimated in three radial intervals of our 18 SDGs, in order of descending stellar masses (from left
to right and from top to bottom). In each panel, the $n_{\mathrm{Sersic}}$ parameter and
the dynamical $B/T$ ratio are also included.}
  \label{fig:MGHs}
\end{figure*}

\begin{figure*}
  \centering
  \includegraphics[width=0.7\textwidth]{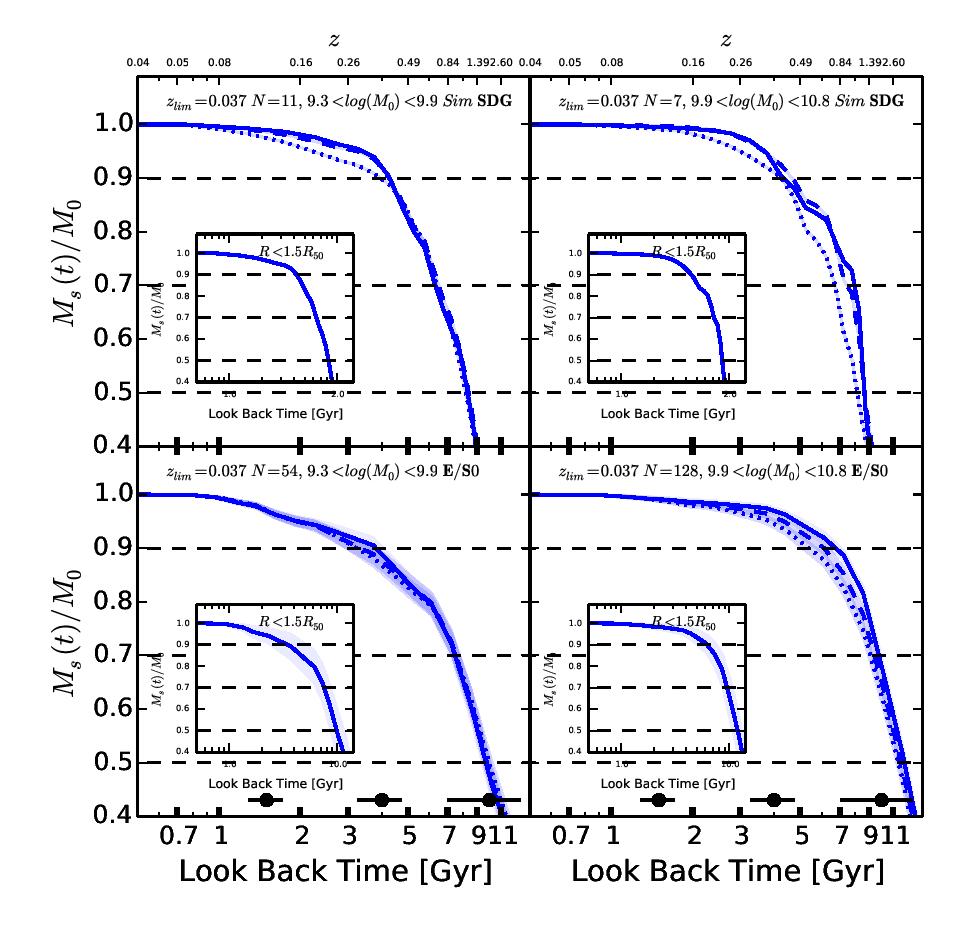}
  \caption{Average global and radial normalised MGH for our simulations (top) and MaNGA ETGs at $z<0.037$ (bottom). We make the division between low (left panels) and high (right panels) stellar mass, adopting  
  $\mathrm{log}(M_{\mathrm{Star}}) = 9.9$ as a threshold. Radial bins of [0--0.5], [0.5--1.0], and [1.0--1.5] $R_{\mathrm{hm}}$ are represented by solid, dashed, and dotted lines, respectively. }
  \label{fig:ES0-MGHs}
\end{figure*}

\section{Stellar mass growth histories}
\label{sec_ah}

In this section, we analyse the global and radial `archaeological' mass growth histories (MGHs) of the simulated
SDGs. This study provides clues to the   galaxy assembly in a
cosmological framework and  allows a direct comparison with observational findings obtained
from integral field spectroscopy. \citet{IbarraMedel2016} have analysed galaxies over a wide mass range from the first data release of the 
Mapping Near Galaxies at APO   \citep[MaNGA;][]{Bundy+2015, SDSS+2016}
survey from SDSS-IV \citep{Blanton+2017}. These authors confirm that the global MGHs 
significantly depend on stellar mass: the more massive the
system, the earlier  their stars formed on average (downsizing).
 They have also  found that the way galaxies
radially grow  their stellar masses depends significantly on galaxy morphology, sSFR, or colour. 
Regarding   ETGs,  \citet{IbarraMedel2016} have found that for a given mass they assemble
earlier than late-type galaxies, but also follow a clear downsizing trend. The radial MGHs of
ETGs are more homogeneous than those of late-type galaxies (LTGs). On average, the radial MGHs tend to follow
a weak inside-out behaviour, but individually there are many cases where the outer regions
can be younger than the innermost ones, suggesting either an outside-in assembly at some 
epochs or processes of global stellar migration from inside to outside. Paths involving
inside-out or outside-in scenarios have been  reported \citep[e.g.][]{SB2007}. Currently, the
formation of spheroidal systems is expected to be a complex process as
different mechanisms can contribute, such as  gas collapse and infall,
mergers, and internal dynamical processes \citep[for recent reviews, see e.g.][]{Brooks+2016, Kormendy2016}.

\cite{IbarraMedel2016} analysed 454 galaxies at $z < 0.037$. The global
MGHs and those in radial bins were normalised to their corresponding
final  masses at $z = 0.037$
(look-back time of $\sim 0.5$ Gyr); fixing the same final epoch for
all galaxies is necessary for calculating the mean MGHs. They explore the dominant direction of mass growth as a function of
mass or morphology. 

We perform a similar archaeological analysis of our SDGs to that of  \cite{IbarraMedel2016}. The age distribution of the stellar
particles at $z=0$ is used to construct the MGHs. In Fig. \ref{fig:MGHs} the
normalised MGHs in three radial bins for
the analysed  galaxies are plotted.
The radial bins are defined at [0--0.5], [0.5--1.0], and [1.0--1.5]
$R_{\rm hm}$.
The panels are in order of decreasing total stellar masses (from left
to right and from top to bottom).

As can be seen in Fig.~\ref{fig:MGHs},  most of systems experience periods of
fast growth in the cumulative mass distributions associated with mergers and starbursts.
The rates of growth are quite different among the SDGs, with weak dependence on mass. 
In general, our SDGs show an inside-out formation mode although there is a large variety of behaviours when looked at in detail. Some systems 
exhibit a combination of modes in their radial MGHs (e.g. SDG 735, SDG
885, SDG 925), while a few galaxies show a clear outside-in mode (e.g. SDG 904).

In Fig. \ref{fig:ES0-MGHs} we plot the average global and radial MGHs of the simulated SDGs (upper panels) and
the MaNGA ETGs (E/S0; lower panels) in two mass ranges. In this case, the MGHs are normalised and start at a look-back time
of 0.5 Gyr (see above). The average radial MGHs of the more massive SDGs attain 70\% of their masses
at look-back times $\approx 8, 7,$ and 6 Gyr for the inner, intermediate, middle and outer radial bins, respectively, while for the
less massive SDGs this happens at 6.2, 6.4, and 6.1 Gyr, respectively. 
The radial MGHs show an average trend 
of inside-out growth mode, though very moderate. For the massive SDGs, the outer MGHs (dotted
line) are shifted to later times with respect to the inner MGHs. For the less massive SDGs, this happens only
at late times. 

The global average MGHs (insets in Fig. \ref{fig:ES0-MGHs}) of the simulated SDGs are shifted to later times with respect to
those inferred from observations, in particular at the earliest epochs. As discussed in \citet{IbarraMedel2016}, the fossil record 
determinations for the oldest ages are very uncertain (see horizontal error bars in the lower panels) and likely bias the 
stellar population ages to even older values. Both observations and simulations show evidence of mass downsizing, but for the latter
the trend is weaker. Regarding the radial MGHs, the simulated galaxies show evidence of a more
pronounced inside-out growth mode than observations. 

In summary, the simulated SDGs tend to form their stellar populations slightly later and with an inside-out radial growth 
mode slightly more pronounced than the inferences obtained from the fossil record method applied to MaNGA ETGs.
This is in line with the fact that our SDGs are on average bluer and more star-forming than observed isolated ETGs (see previous section). We also note  that all the simulated SDGs present a disc component. 
The SDGs with  $\log(M_\mathrm{Star})\lesssim 10$ are gas-rich galaxies with a mean gas fraction of 34 \%
within the optical radius. More massive galaxies show gas fractions
around 18 \%. These gas fractions are more typical of LTGs (e.g. Calette et al. submitted). 
The larger gas fractions can account for a
more active SF activity. In a previous work,
\citet{DeRossi2013} find that the SN feedback adopted in this
simulation is able to regulate the SF activity in galaxies
with rotational velocity smaller than $\sim 100 \ \mathrm{km \ s^{-1}}$ so that
these systems reach an equilibrium point where the SN feedback produced by a mild SFR
is enough to keep the gas turbulent and warm and at the same time
allows the SF to be fed smoothly. Observational constraints on the
gas fraction and the SF history of ETGs galaxies in this stellar mass
range would be important to further improve the SN
feedback models.
  
\section{Conclusions}

We have analysed the properties of simulated galaxies dominated by the
spheroid components in a $\Lambda$-CDM universe with the aim to
investigate  to what extent these systems are consistent with the observations.
In previous works, the dynamical and chemical properties of disc-dominated galaxies have been studied in
great detail, finding very good agreement with observations
\citep{Pedrosa2015,Tissera2016,Tissera2016b, Tissera2017}. Since the
evolutionary paths of spheroid-dominated galaxies are expected to be different, 
it is of relevance to assess to what extent the same models and
simulations  reproduce the two kinds of systems. 

After identifying the velocity dispersion-dominated spheroid and the 
rotation-dominated disc by means of a dynamical criterion, we classify our simulated
(field) galaxies as SDGs  as those with $B/T>0.5$.  We notice
that  all our SDGs  actually have an extended disc component, 
  although all of them are found to be below the SDSS observability limit, and that an inner disc component coexists with a spheroid. 
It is important to keep in mind these particularities when explaining the differences between 
observed galaxies and our simulations. The main results from our analysis are as follows:

\begin{enumerate}[(i)]

  \item  The values of the spheroid Sersic index increases on average with the dynamical $B/T$, in
            agreement with observational results \citep[e.g.][]{Fisher2008}.
  
  \item The sizes of the simulated SDGs as a function of $M_{\rm Star}$ are consistent with observational determinations. 
    We measure both $R_{\rm hm}$ directly from the simulation 
    and $R_{\rm eff}$ from the Sersic fits to the whole galaxy. While
    the former radii are slightly larger ($\sim 0.2$ dex) than the 
    observational estimates, specially at low masses, the latter  agree with them. 
    
  \item The dynamical relations of our SDGs are in reasonable agreement with observational results
  for ETGs. The FJR defined as the correlation between $\sigma$ (central or at $R_{\rm hm}$) and the dynamical
  mass at  $R_{\rm hm}$ has a slope of $\sim 0.4$, in good agreement with the ATLAS$^{3\mathrm{D}}$ ETGs \citep{Cappellari2013}.
  SDGs also exhibit  a tilt in the FP consistent with these observations. The stellar and baryonic TFRs calculated from the disc components
  are consistent with the corresponding relations determined for a small subsample of ATLAS$^{3\mathrm{D}}$ ETGs, those that present an extended
  HI disc \citep{Heijer2015}. Interestingly enough, the TFRs of the SDGs are similar to the TFRs of the DDGs.  
  
  \item The  $F_{\rm dm}$ measured at $R_{\rm eff}$ increases for
    smaller SDGs, a trend also seen  in the
  observational inferences from the ATLAS$^{3\mathrm{D}}$ ETGs
  \citep{Cappellari2013, CappellariAtlasXX}. The simulated values of $F_{\rm dm}$ and their dependence on mass
  are consistent with the observational inferences, but none of  the simulated SDGs attains values below 0.15.

  \item Our SDGs are significantly bluer and with higher values of sSFR than the observed isolated ETGs in the same mass range. 
   This is partially due to the persistence of extended discs in the
   simulations, which tend to have young stellar populations. 
     Part of the disc coexists with the spheroidal component. Only a 
    few discs (5/18) have fractions of young stars ($<$ 3 Gyr) smaller than 5 \%, while this fraction is smaller than
   5 \%\  for most of the spheroids (16/18). The average mass-weighted stellar ages of the spheroids is $\sim
    8.5$ Gyr, while for discs the average is $\sim 6.7 $ Gyr, though the scatter is quite large. The colours and sSFR 
    values of  the spheroid components are then closer to those observed for isolated ETGs .

\item The archaeological radial MGHs of our SDGs are on average dominated by a moderate inside-out growth mode, though
some galaxies present periods of outside-in and inside-out modes, and two are dominated by the outside-in growth mode. 
Compared to the fossil record inferences applied to the observed MaNGA ETGs, the simulated SDGs form on average their stellar populations 
later and with an inside-out radial growth mode which is  slightly more prominent.  Larger
stellar-mass galaxies are predicted to assemble on average at earlier times than the less massive ones (downsizing), but
the differences are smaller than for observations.  

\end{enumerate}

We conclude that cosmological simulations in the context of the $\Lambda$-CDM hierarchical 
scenario are able to produce isolated galaxies dominated by spheroids with structural and dynamical
properties in good  agreement with observations. This is encouraging
since the subgrid parameters have not been fine-tuned to reproduce any
of them.
However, all our simulated field SDGs have a disc component that extends to much further away than the spheroid and with stellar populations younger 
than the spheroid. As a result, our SDGs are bluer and with higher sSFR values than the
observed isolated ETGs of similar masses. Even by not taking into account the disc components, our SDGs 
are on average slightly bluer than the observed ETGs.
We have shown that the extended discs in our simulations likely would not be detectable in observational surveys like SDSS; however, they might be detected by the LSST survey.

The above-mentioned  tension with the observations 
could be alleviated by introducing mechanisms able to avoid disc growth after major mergers and/or to quench SF efficiently.
Our simulations do not include the effect of feedback by AGNs. 
AGN feedback could prevent the formation of gaseous discs after major mergers, and consequently could eliminate the possibility 
of post starbursts. However, the presence of luminous AGNs in galaxies of low- and intermediate-masses such as the ones studied here, 
is not expected to be common.
 It is more feasible that our results point out  the necessity of
 more efficient feedback driven by both type-II and type-Ia SNe (see also \cite{Conroy2015} for an alternative heating source).  
It is also important to consider new observational results regarding
ETGs where discs and younger stellar populations are being identified \citep[e.g.][]{McIntosh2014, Schawinski2014}
  
There are still many open problems in the study of spheroid-dominated galactic
objects. It is therefore important to continue to investigate, and
to compare simulations with new observational results.  Fortunately, a
number of advances have been made in recent years. We expect to shed
light on these issues through our research.

\begin{acknowledgements}
We acknowledge Dr. H\'ector Hern\'andez-Toledo for making available in electronic form the UNAM-KIAS Catalog
of Isolated Galaxies. This work was partially supported by PICT 2011-0959 and PIP 2012-0396 (Mincyt, 
Argentina) and the Southern Astrophysics Network (SAN; Conicyt Chile). PBT acknowledges partial support from Nucleo UNAB 2015 of 
Universidad Andres Bello and  Fondecyt 1150334 (Conicyt). The Fenix simulation was run at the Barcelona Supercomputing Centre.
\end{acknowledgements}

\bibliographystyle{aa}

\def\apj{ApJ}
\def\apjl{ApJ}
\def\aj{AJ}
\def\mnras{MNRAS}
\def\aa{A\&A}
\def\nat{Nature}
\def\araa{ARA\&A}
\def\aap{A\&A}

\bibliography{fenix}

\begin{appendix} 

\section{Synthetic images of spheroid-dominated galaxies}
\label{app:image1}

We generated synthetic images of the SDGs
using the radiative transfer code SKIRT \citep{Baes2005}. The images are
generated using the stellar population
synthesis models of \citet{BruzualCharlot2003}
to assign a SED to each star particle in a given 
galaxy, based on their age, mass, and metallicity.  Then SKIRT
calculates the
propagations of photons towards a simulated imaging instrument  using a Monte Carlo technique. The photons
considered are emitted with wavelengths between 0.1 and 100 microns in a
logarithmic grid with 100 points. No gas or dust is considered in the
computations. The simulated imaging instrument corresponds to a $256 \times 256$ pixel
camera placed 10 Mpc away from the galaxy and with a spectral sensitivity equal
to the SDSS $u, g, r, i$, and $z$ broadband filters. Integrated magnitudes and
colours are computed from the resulting fully integrated SED of the galaxy.
Similar techniques have been used to compute mock images and colours in the
Illustris simulation \citep[e.g.][]{Torrey2015, Bignone2017}.

The left panels of Fig. \ref{fig:image1} show synthetic images of the 18 SDGs. We include synthetic
colour-composite images combining SDSS $g$, $r$, and $i$ mock images 
which show a
variety of systems from the morphological point of view.
By construction, all galaxies
have $B/T>0.5$; hence,  even if the disc components are clearly in place,
the dispersion-dominated component is more massive.
The distributions of $\epsilon$ parameters (middle panels) and the projected surface density for the spheroid
and disc components (right panels) are also shown in Fig. \ref{fig:image1}.

To determine the observability of our galactic discs,
we assume that the synthetic galaxies are at $z\sim 0.05$ and process the images to have the same pixel scale and similar PSF as SDSS. 
The limit radii, where the integrated surface brightness is less
than 23 mag arcsec$^{-2}$ in the r band, is estimated. 
This is the surface brightness limit for the main galaxy sample target selection in SDSS using the mean surface brightness within the Petrosian half-light radius \citep{Strauss2002}. 
The circles in the left panel of Fig. \ref{fig:image1} show the limit
radii. Therefore, we can appreciate that most of the disc components
are below the  SDSS observability.

\begin{figure*} 
    \includegraphics[width=0.283\textwidth]{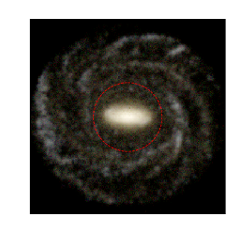}
    \includegraphics[width=0.33\textwidth]{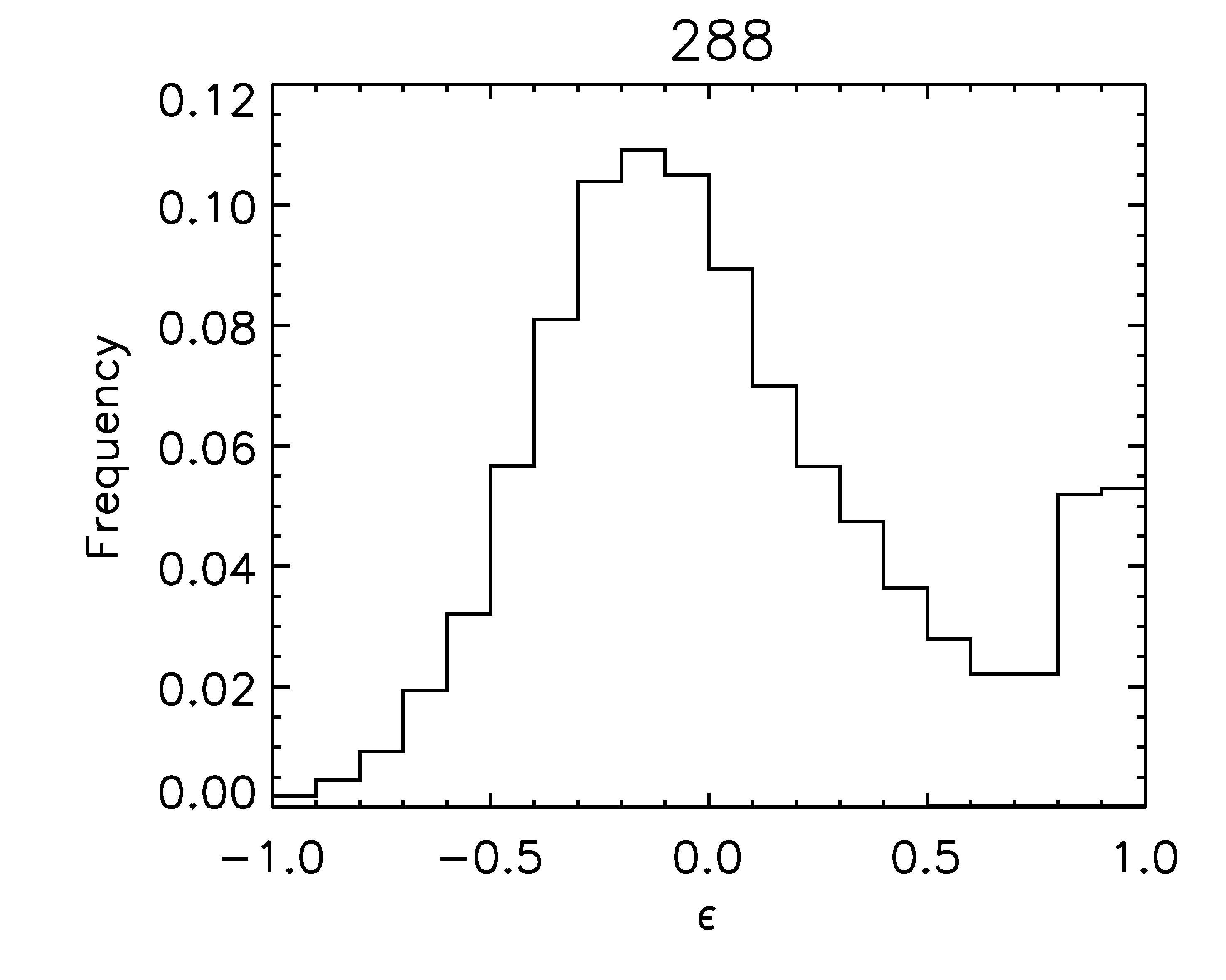} 
    \includegraphics[width=0.33\textwidth]{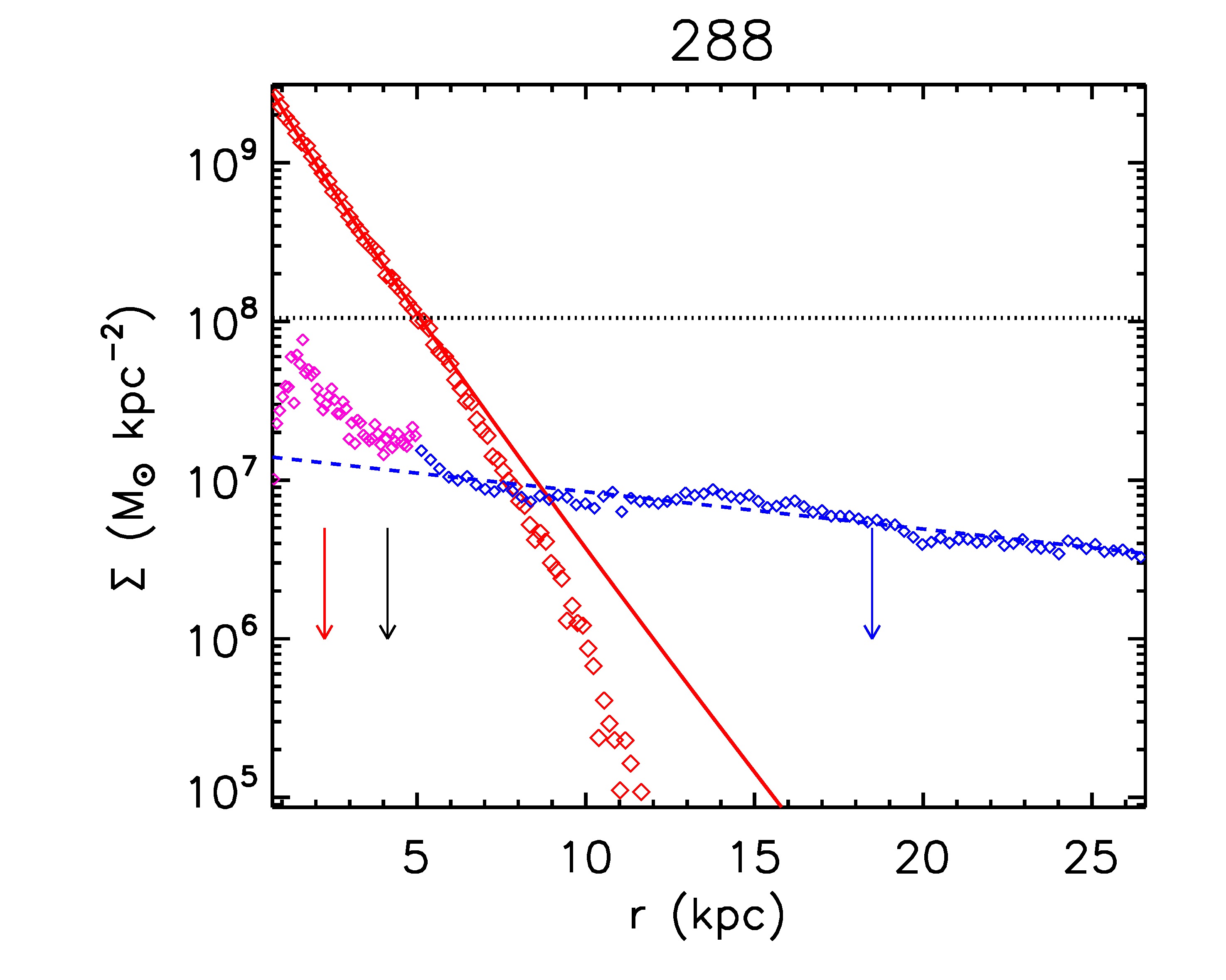} 
   \\
    \includegraphics[width=0.283\textwidth]{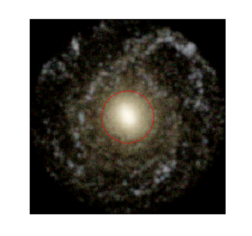}
    \includegraphics[width=0.33\textwidth]{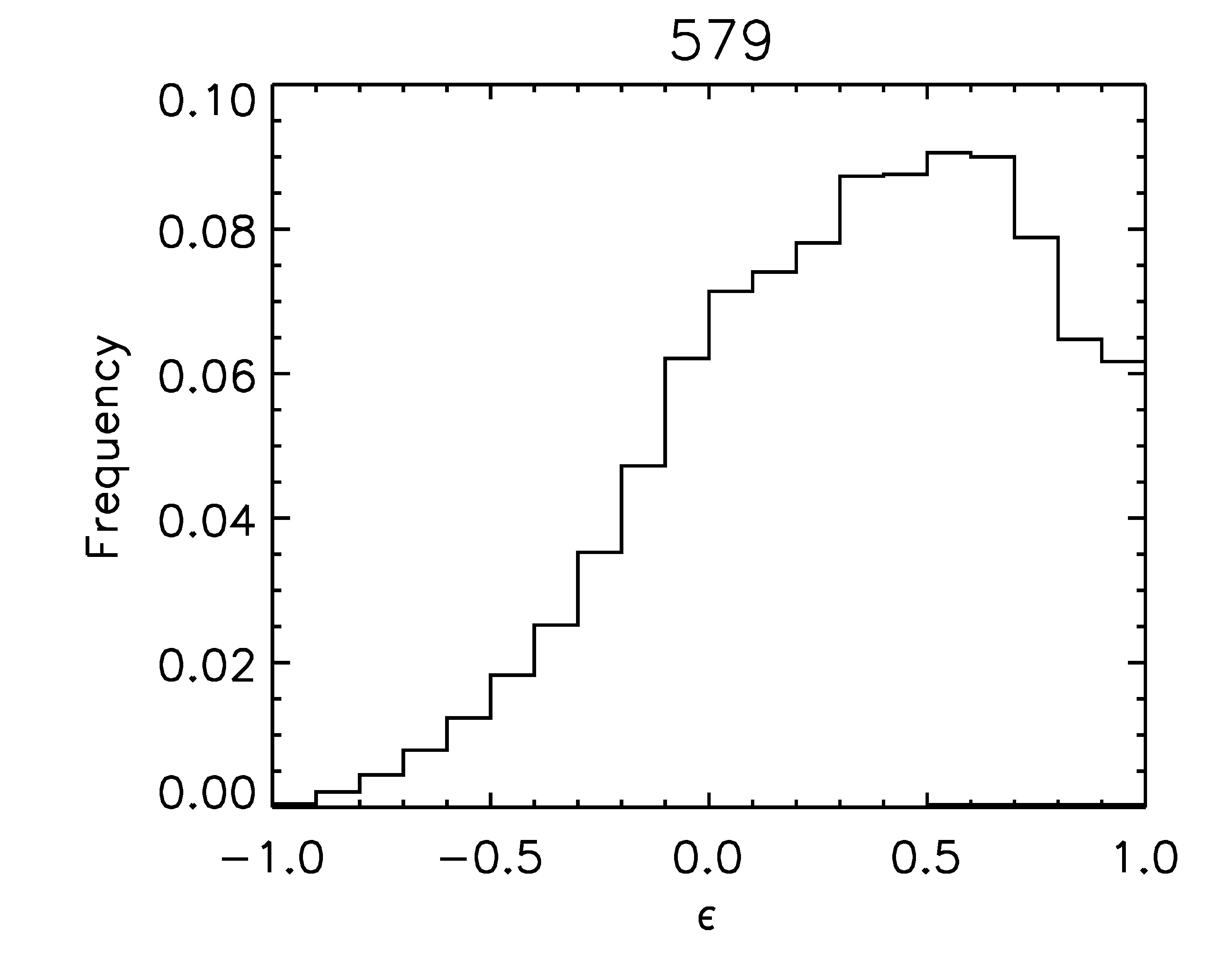}
    \includegraphics[width=0.33\textwidth]{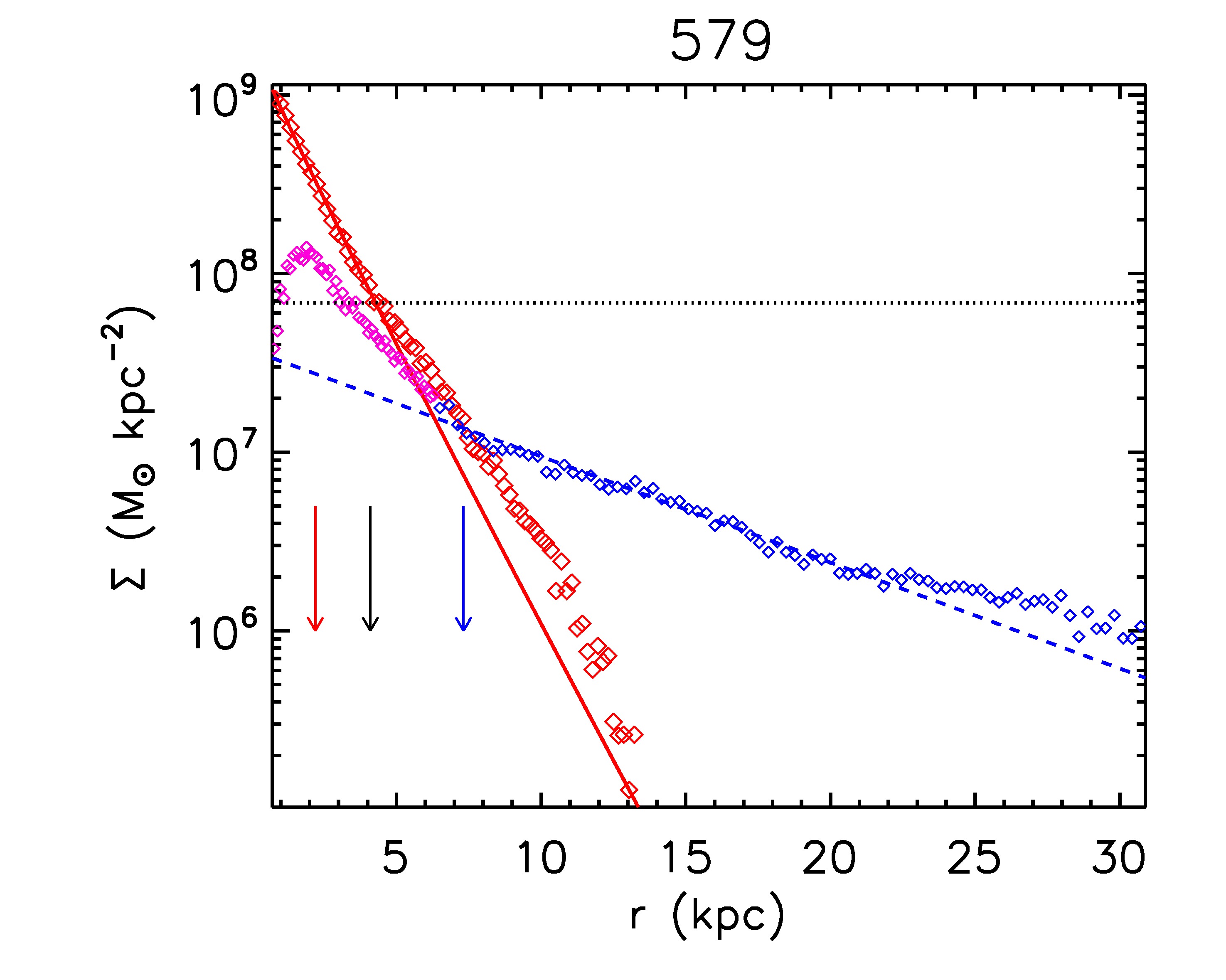}
    \\
    \includegraphics[width=0.283\textwidth]{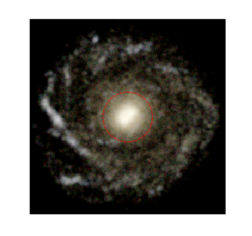}
    \includegraphics[width=0.33\textwidth]{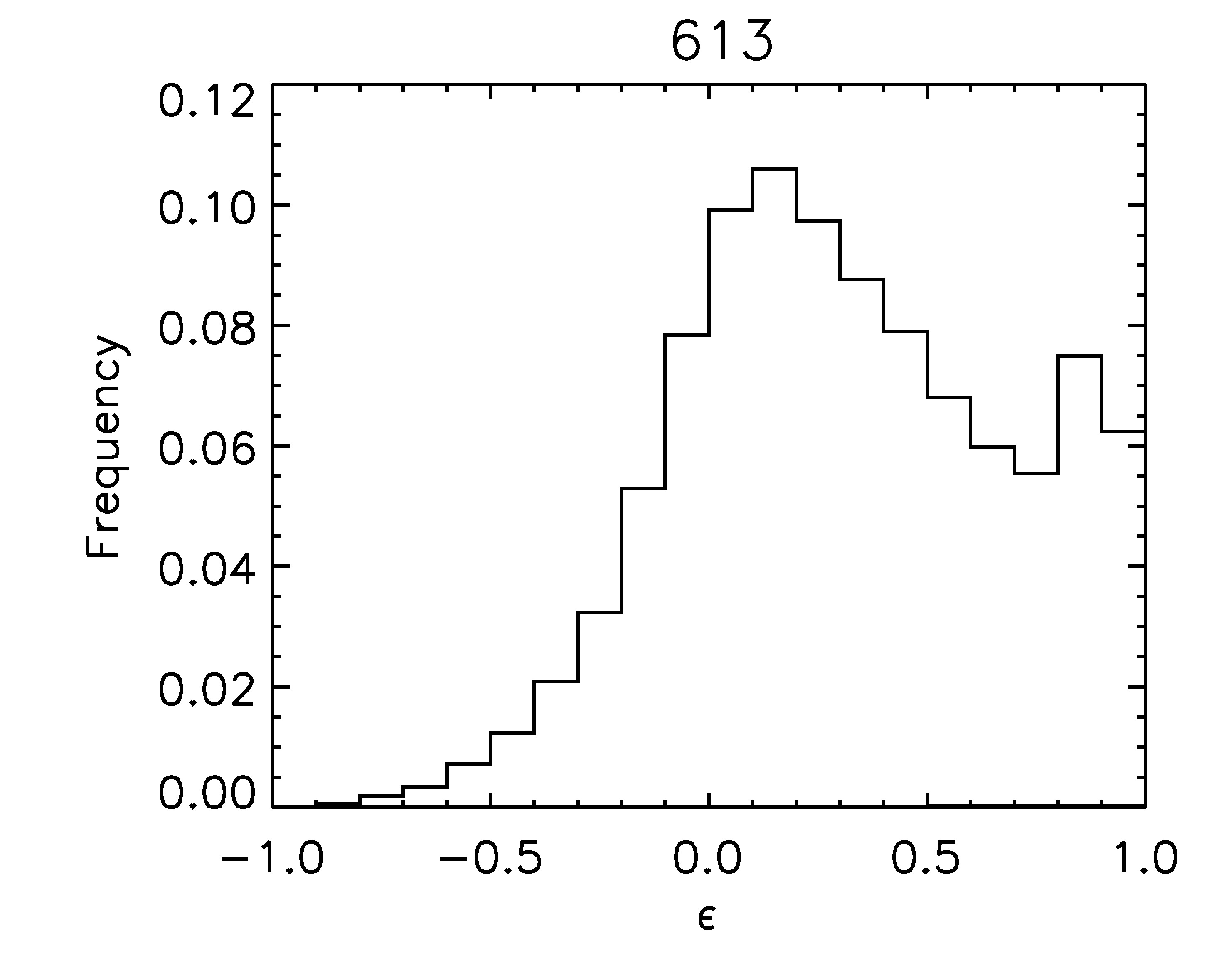}
    \includegraphics[width=0.33\textwidth]{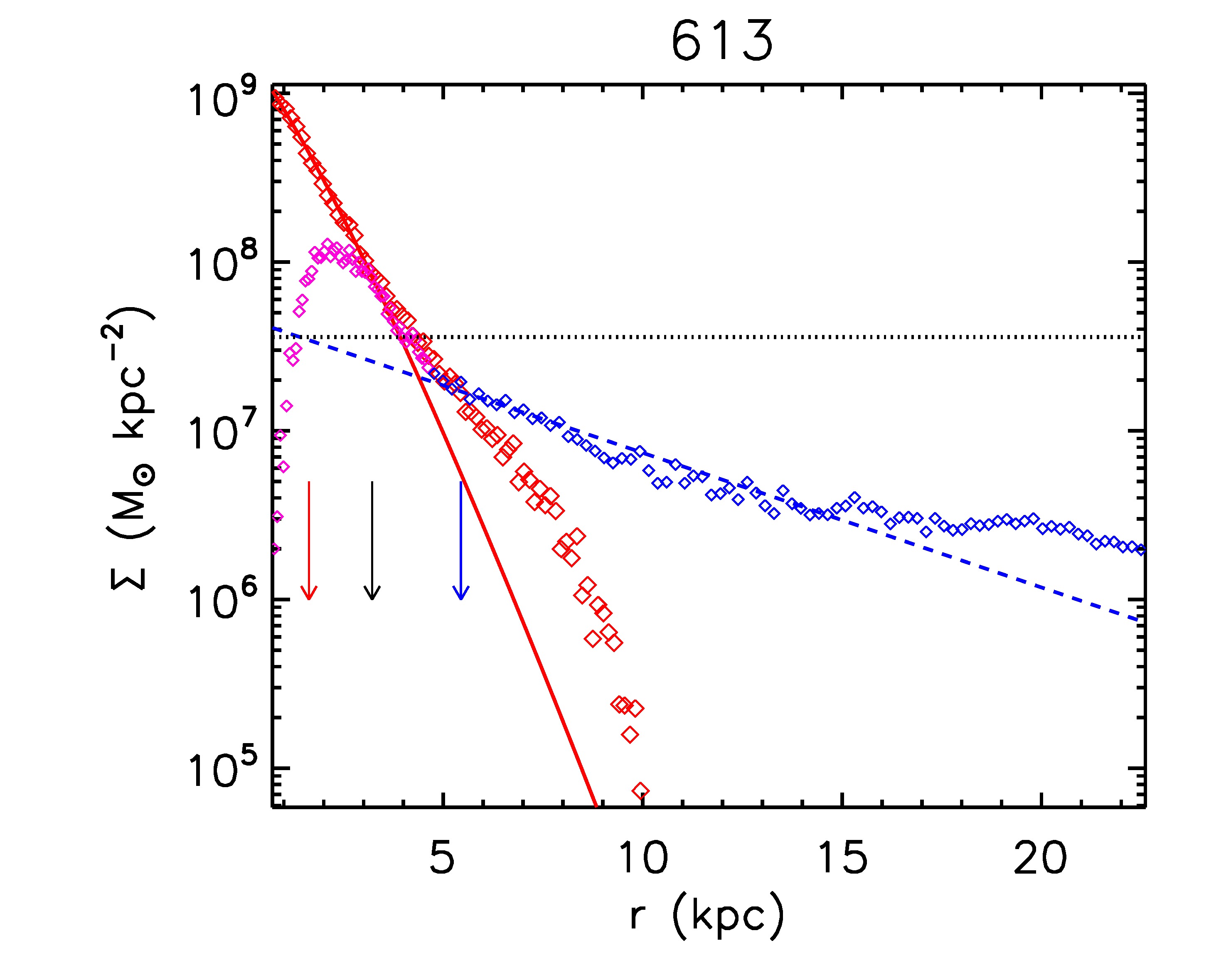}
    \\        
    \includegraphics[width=0.283\textwidth]{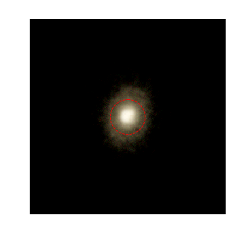}
    \includegraphics[width=0.33\textwidth]{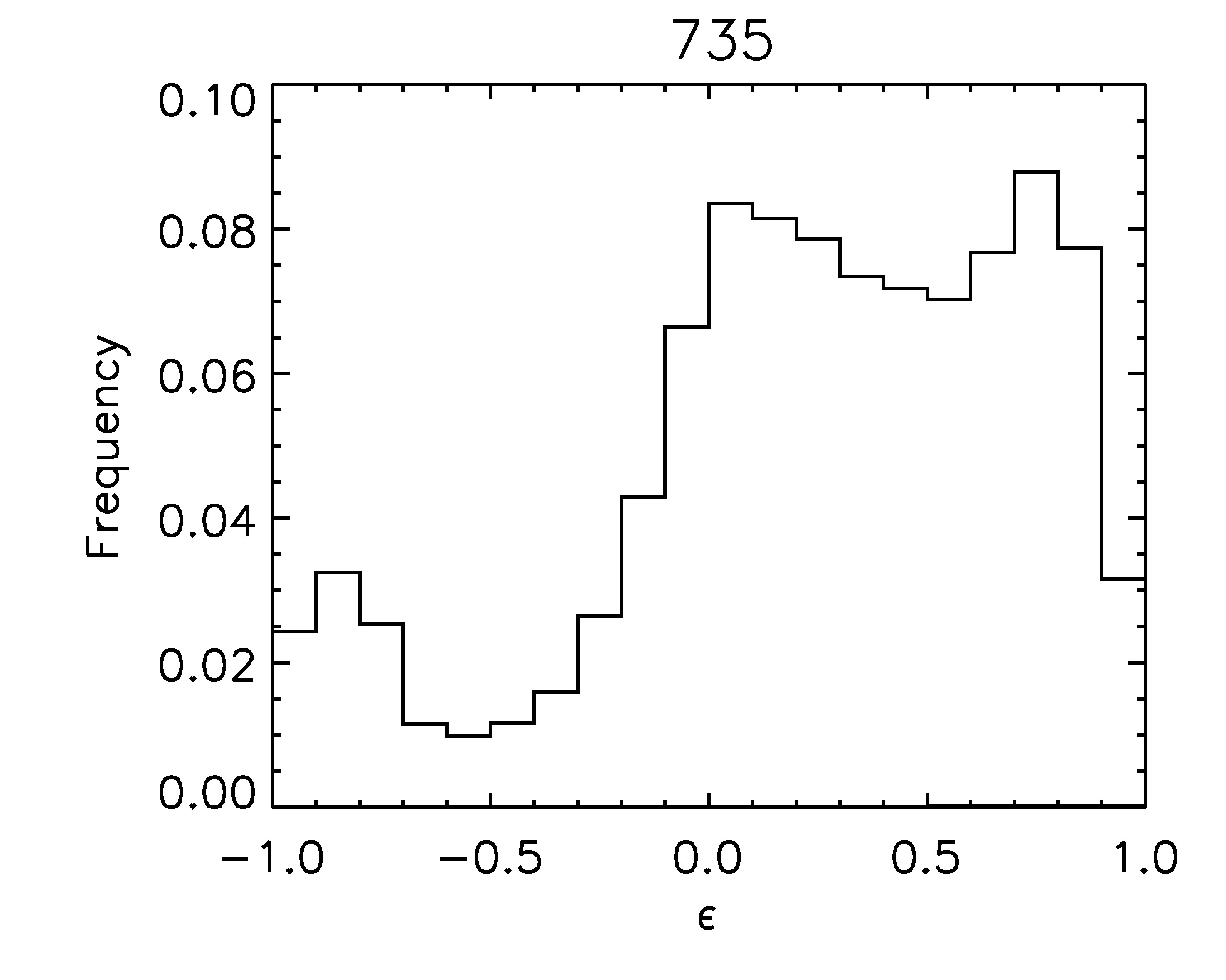}
    \includegraphics[width=0.33\textwidth]{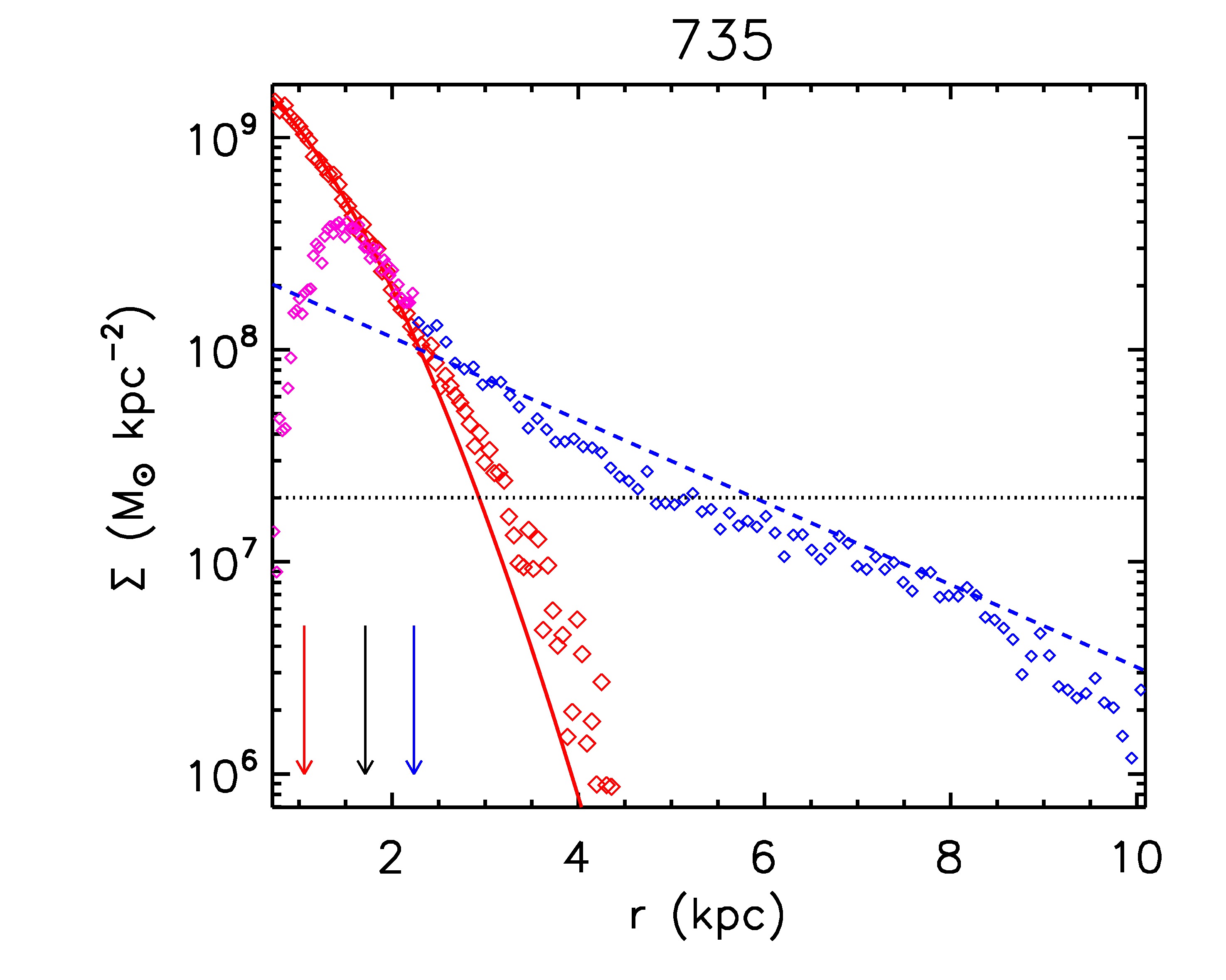}
    \\    
  \caption{Left panels: Synthetic images of
    18 SDGs
      obtained with the SKIRT code \citep{Baes2005} with a box side of 30 kpc. 
      The red circles show the limit radii for SDSS observability (see Appendix \ref{app:image1} for details).
      Middle panels: Distributions of the $\epsilon$
      parameter for star particles within the optical radius. 
      Right panels:  Projected
      stellar-mass surface profiles for the spheroids (red diamonds) and  the
      disc components (blue diamonds). Part of the disc components
      coexist spatially with the spheroid component (magenta diamonds). The best fit Sersic profile
      for the spheroid component (red solid lines) and the exponential
      profiles for the discs (blue dashed lines)
      are also included. We indicate with a red, black, and blue arrow
      the   $R_{\mathrm{Sersic}}$, $R_{\mathrm{hm}}$, and $R_d$  (see
      Table 1). 
      We also include the total surface brightness at the
        limit radius where the galaxy could be detected with the SDSS (black dotted line).
      The relations are shown out to $R_{\mathrm{opt}}$.
      The rows show galaxies in order of descending stellar mass. The galaxy ID is indicated above each middle and right panel.}
    \label{fig:image1}
\end{figure*}

\addtocounter{figure}{-1}

\begin{figure*}
    \includegraphics[width=0.283\textwidth]{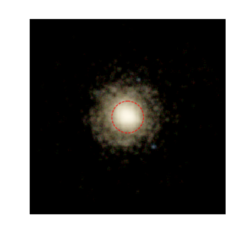}
    \includegraphics[width=0.33\textwidth]{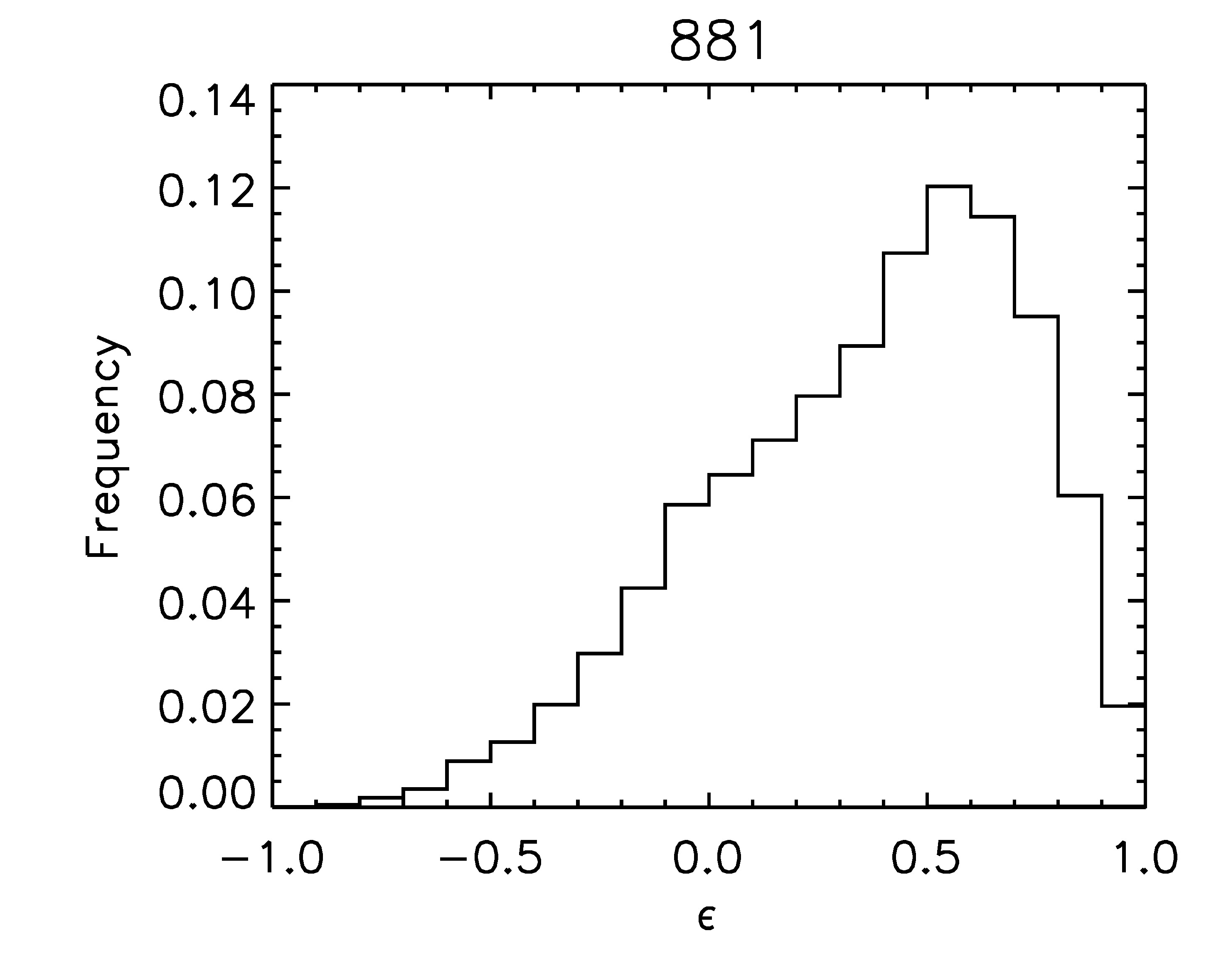}
    \includegraphics[width=0.33\textwidth]{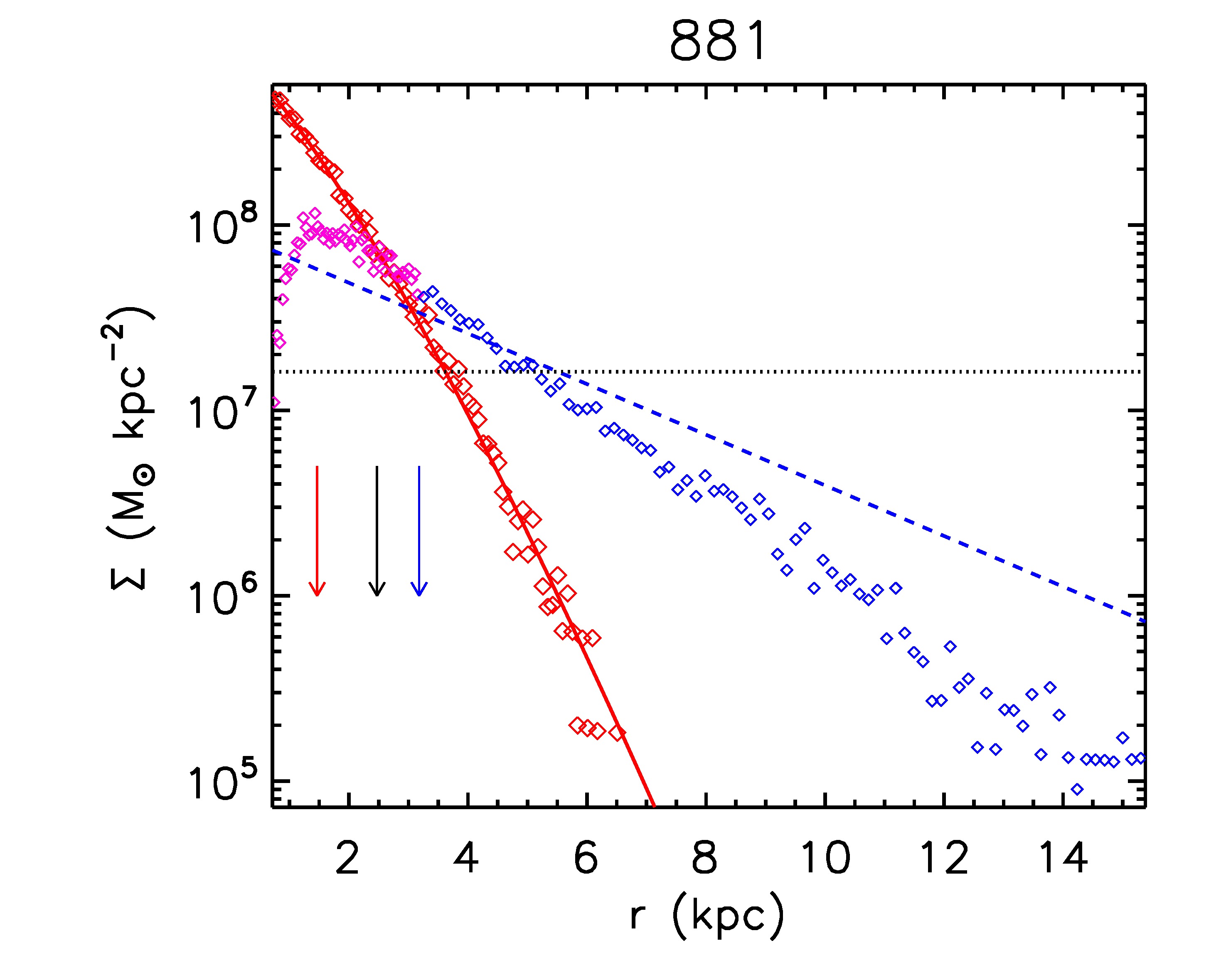} 
  \\
    \includegraphics[width=0.283\textwidth]{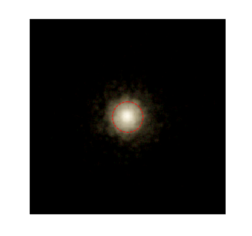}
    \includegraphics[width=0.33\textwidth]{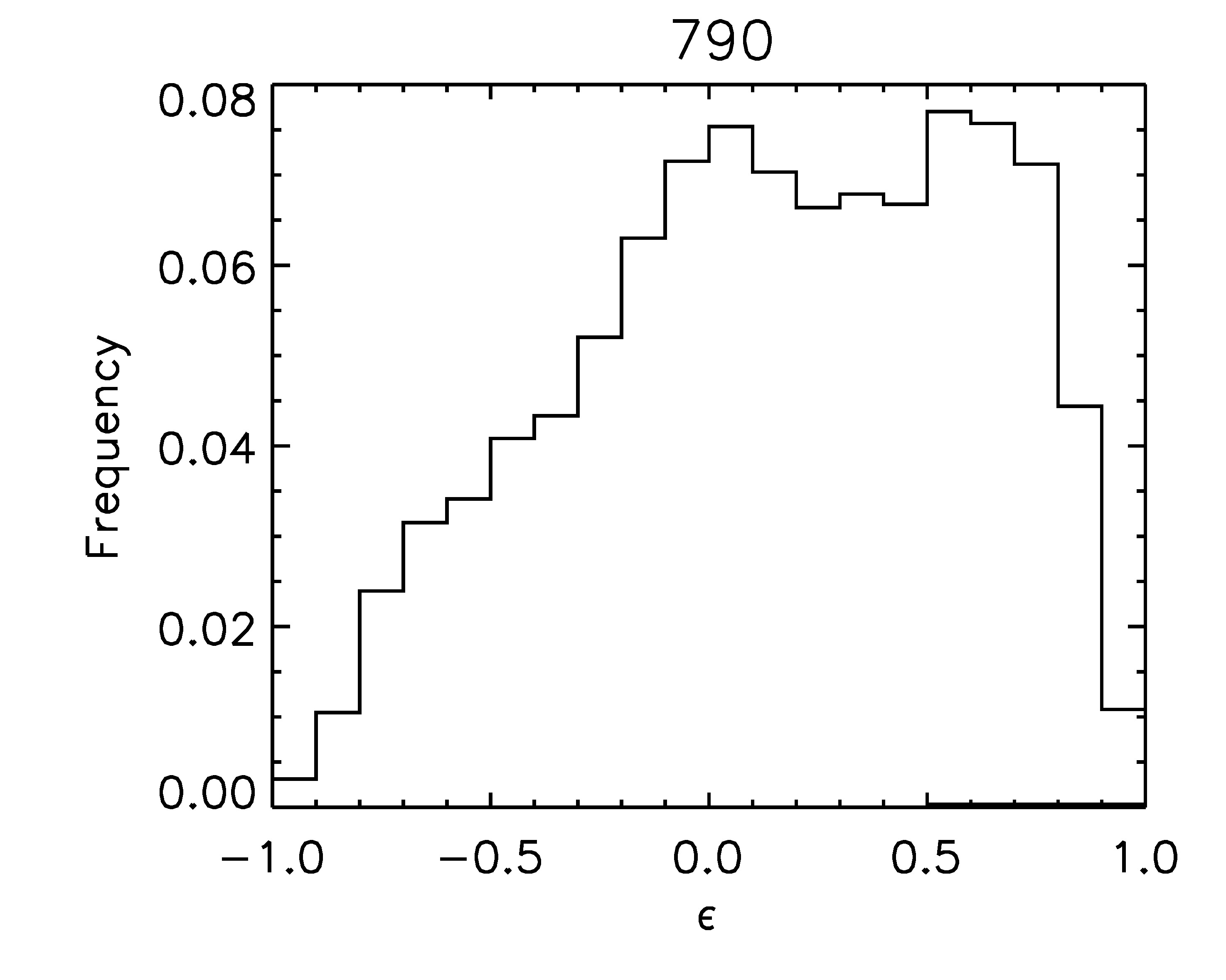}
    \includegraphics[width=0.33\textwidth]{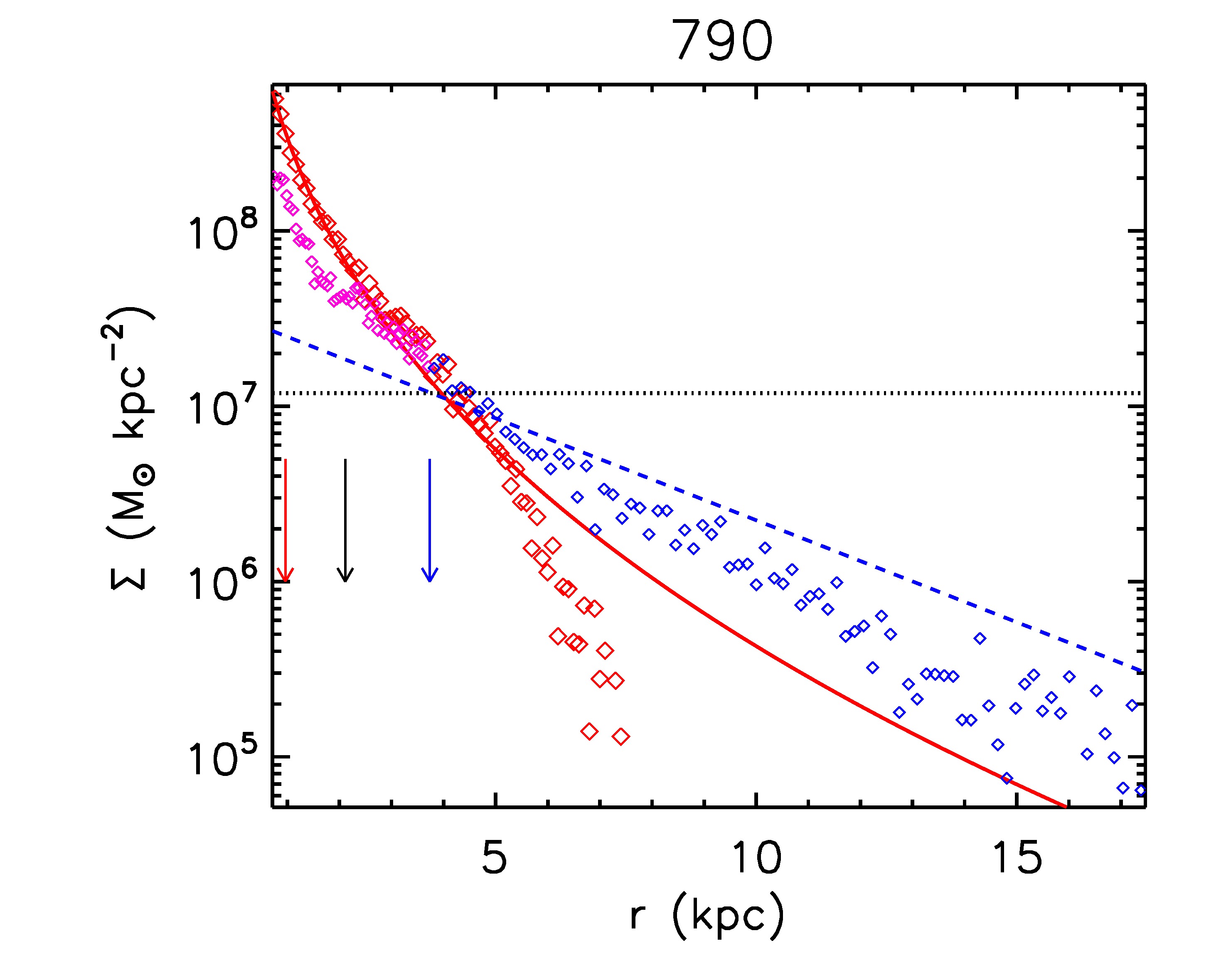}
\\
    \includegraphics[width=0.283\textwidth]{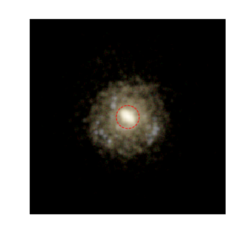}
 \includegraphics[width=0.33\textwidth]{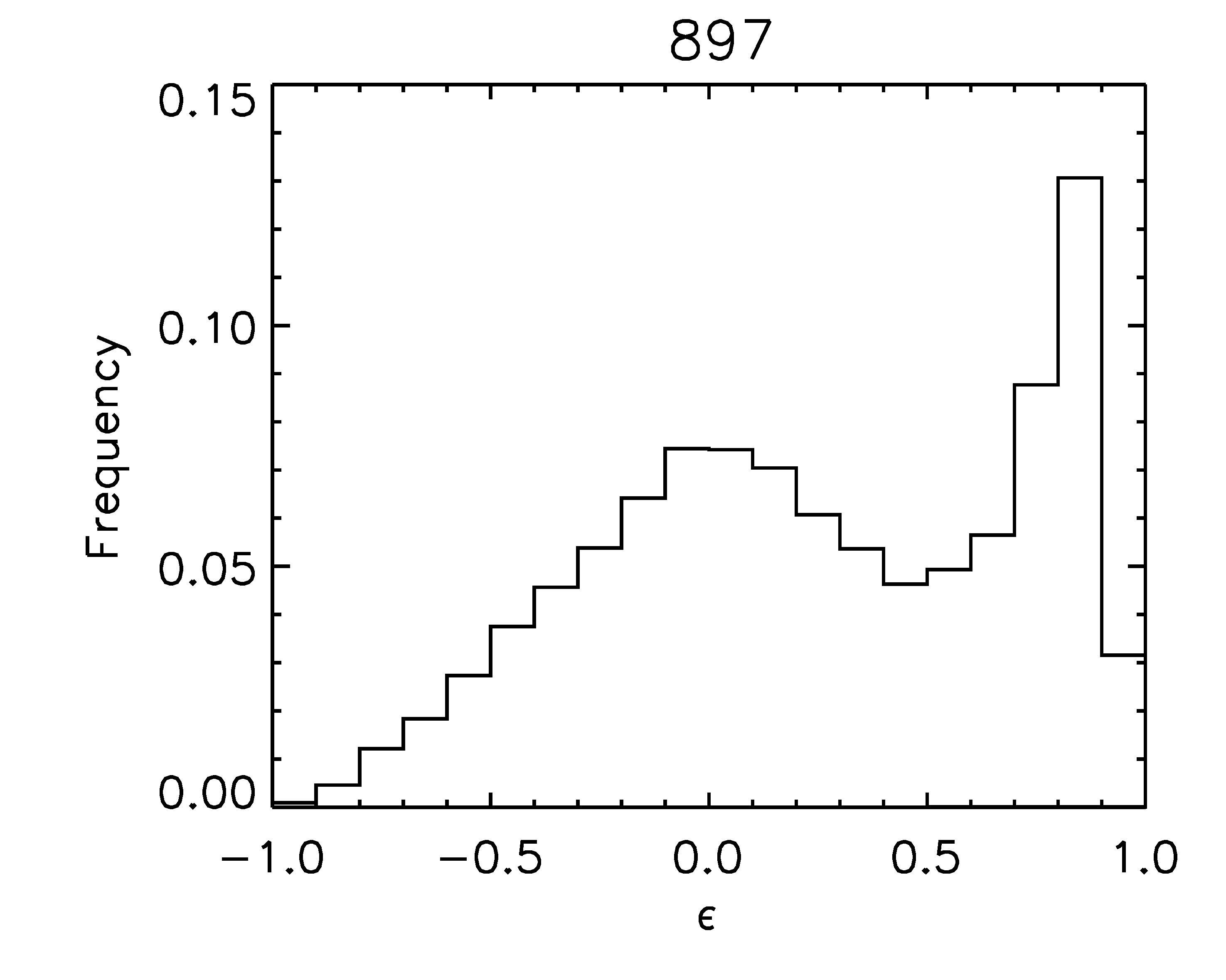}
    \includegraphics[width=0.33\textwidth]{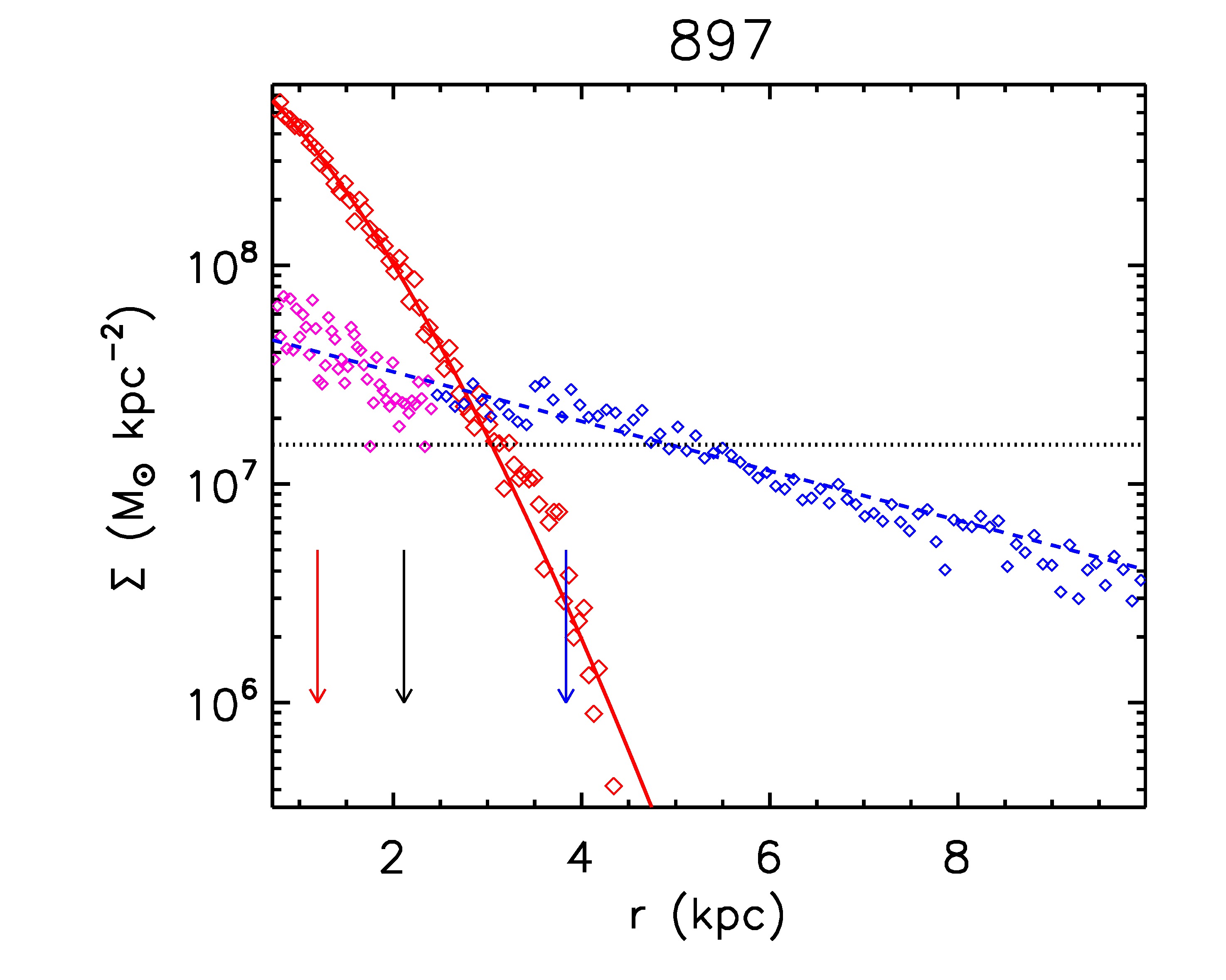} 
    \\
    \includegraphics[width=0.283\textwidth]{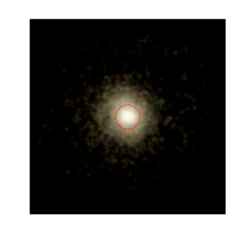}
 \includegraphics[width=0.33\textwidth]{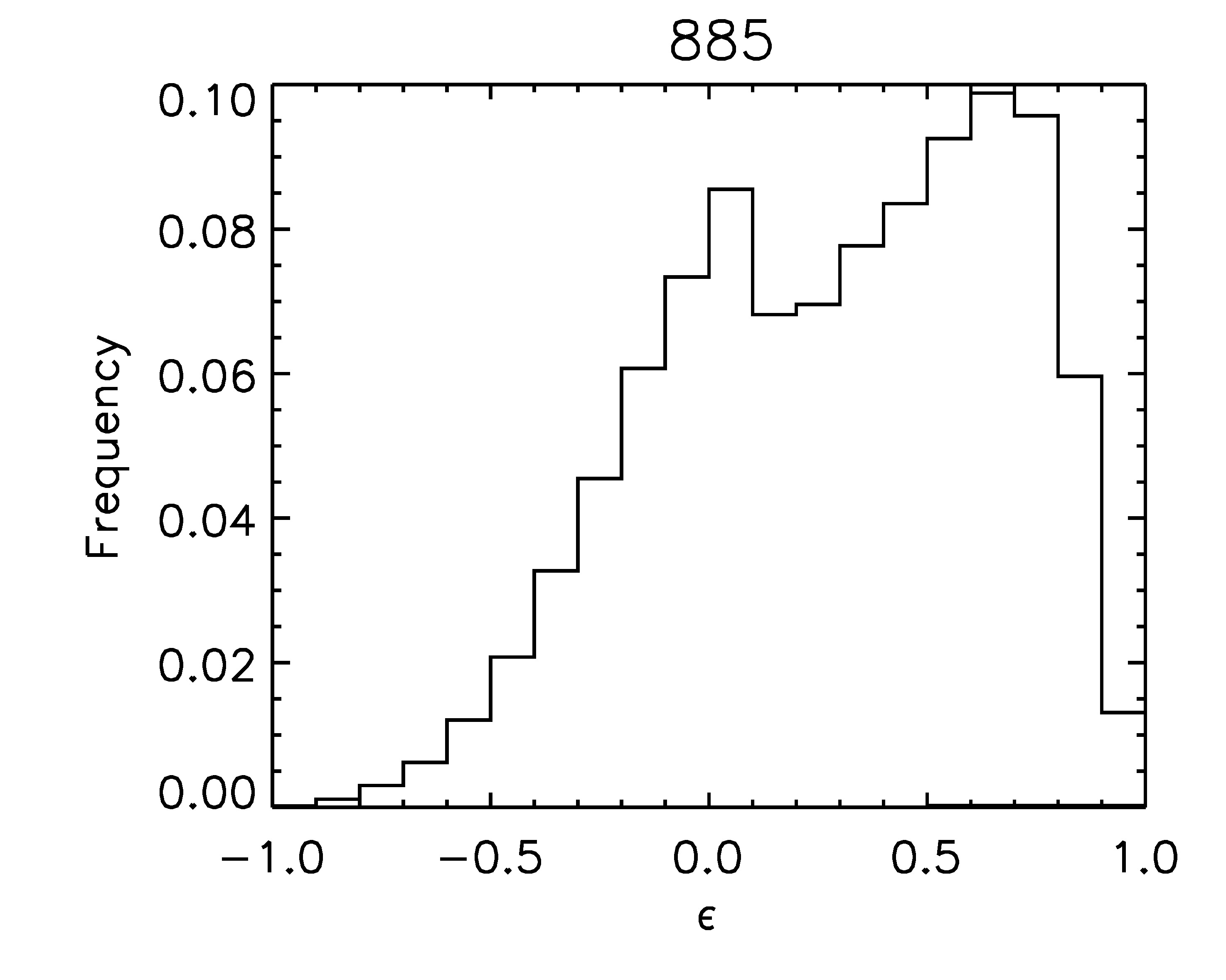}
    \includegraphics[width=0.33\textwidth]{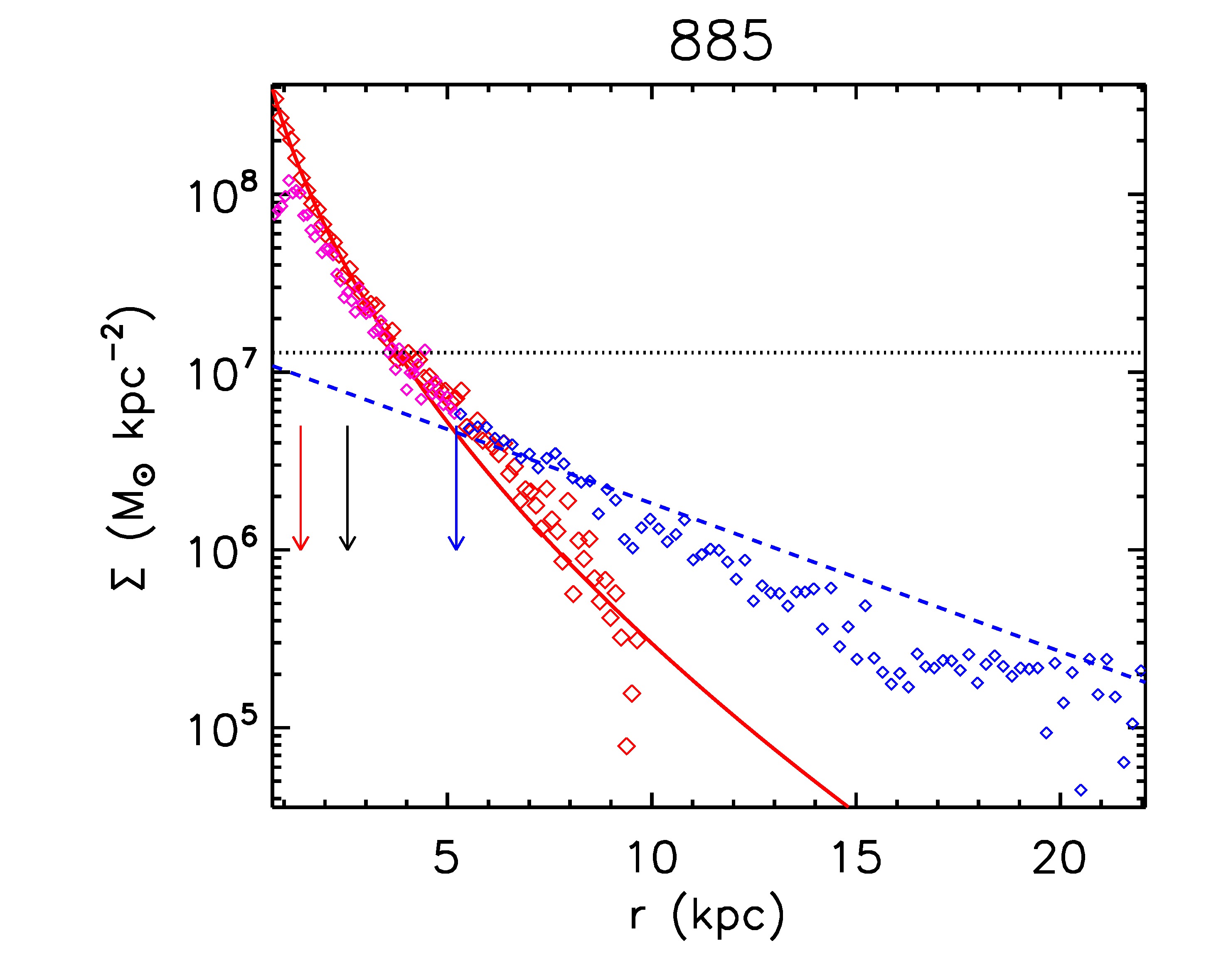} 
    \\
    \caption{(continued)}
\end{figure*}

\addtocounter{figure}{-1}

\begin{figure*}
    \includegraphics[width=0.283\textwidth]{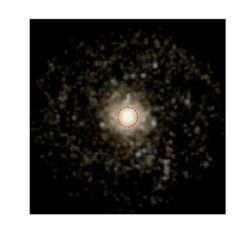}
    \includegraphics[width=0.33\textwidth]{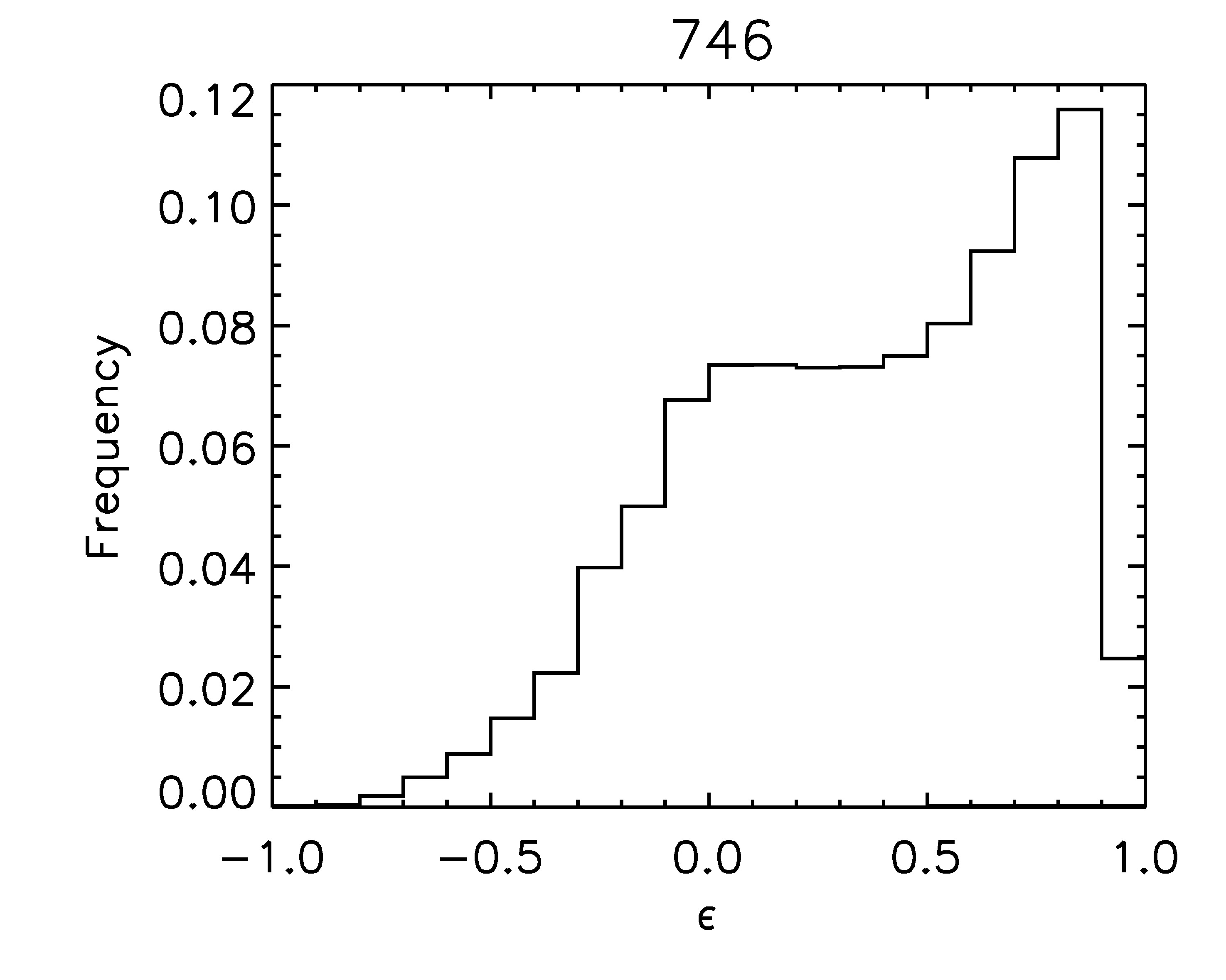}
    \includegraphics[width=0.33\textwidth]{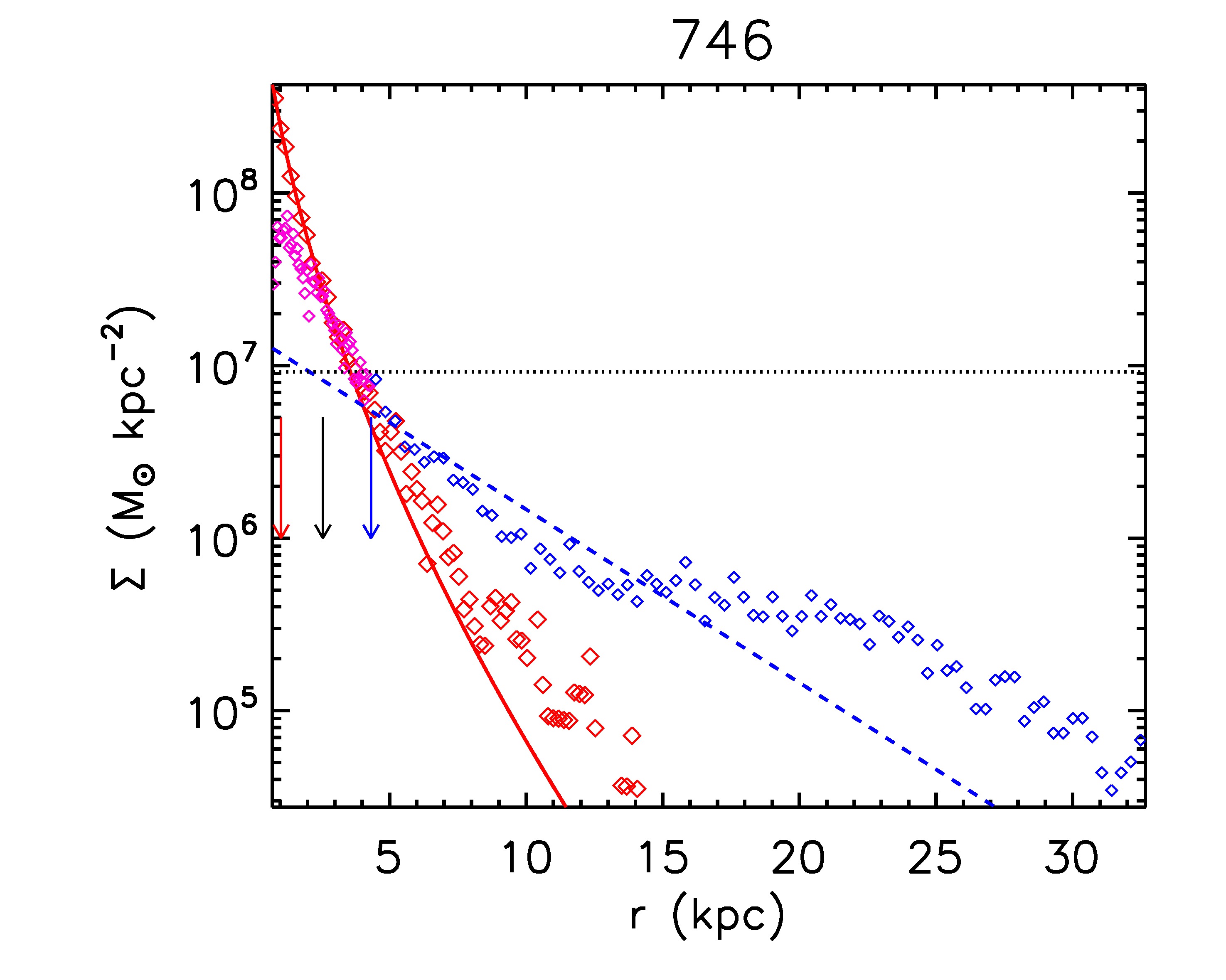} 
  \\
    \includegraphics[width=0.283\textwidth]{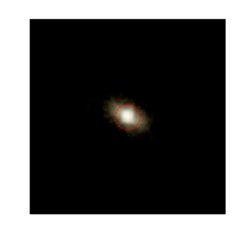}
    \includegraphics[width=0.33\textwidth]{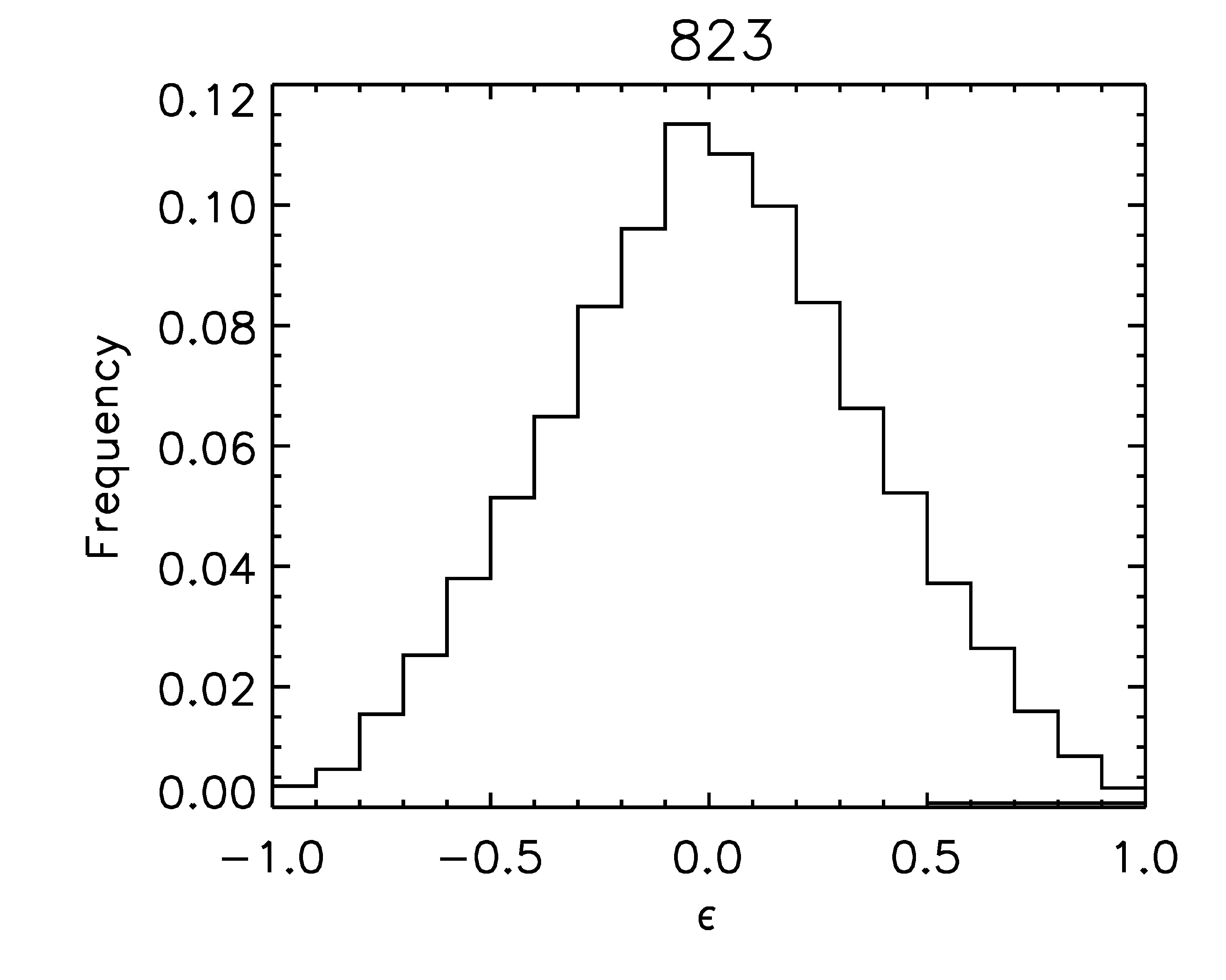}
    \includegraphics[width=0.33\textwidth]{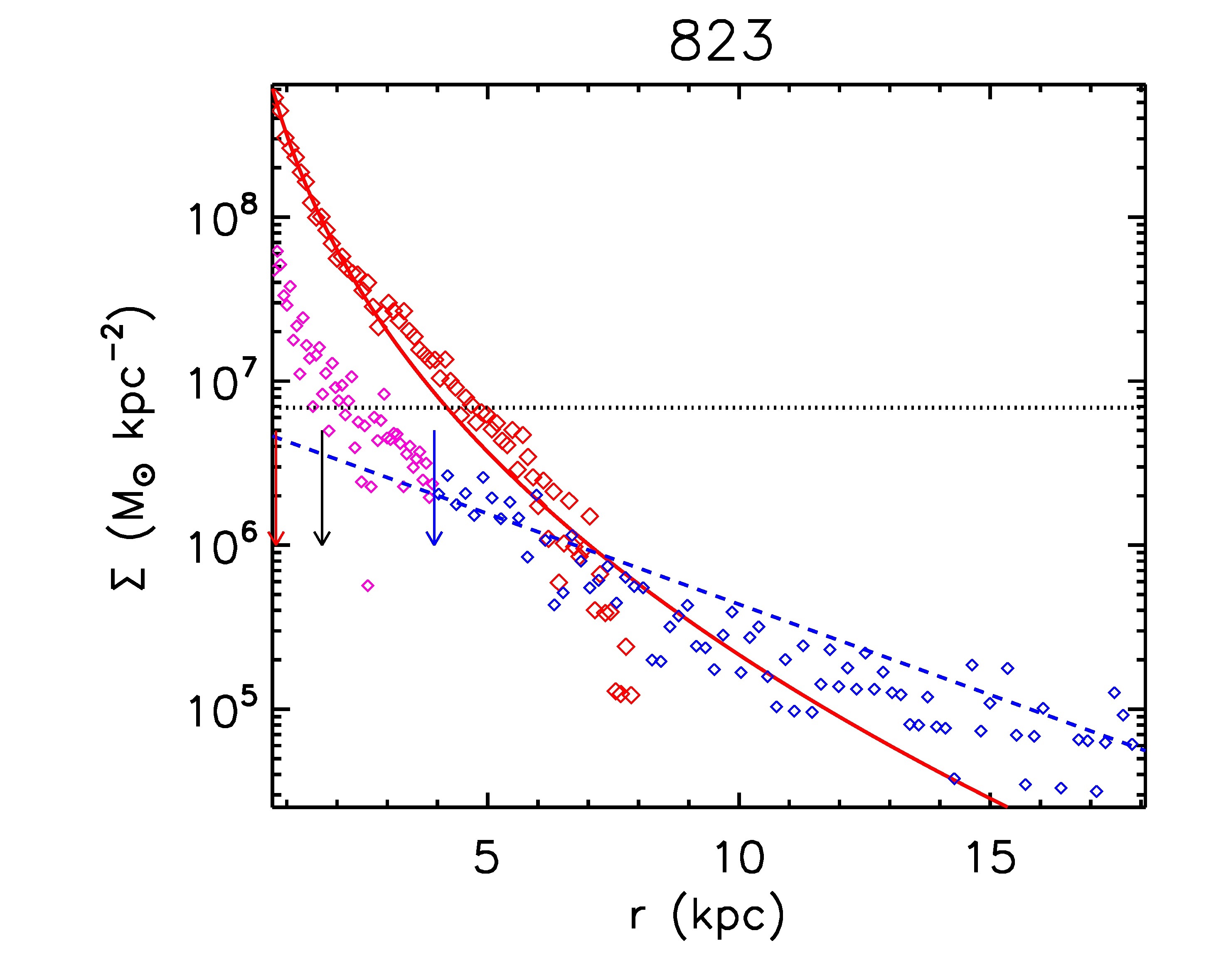}
\\
    \includegraphics[width=0.283\textwidth]{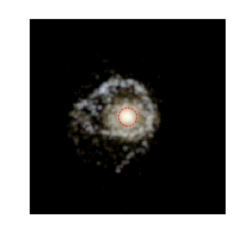}
 \includegraphics[width=0.33\textwidth]{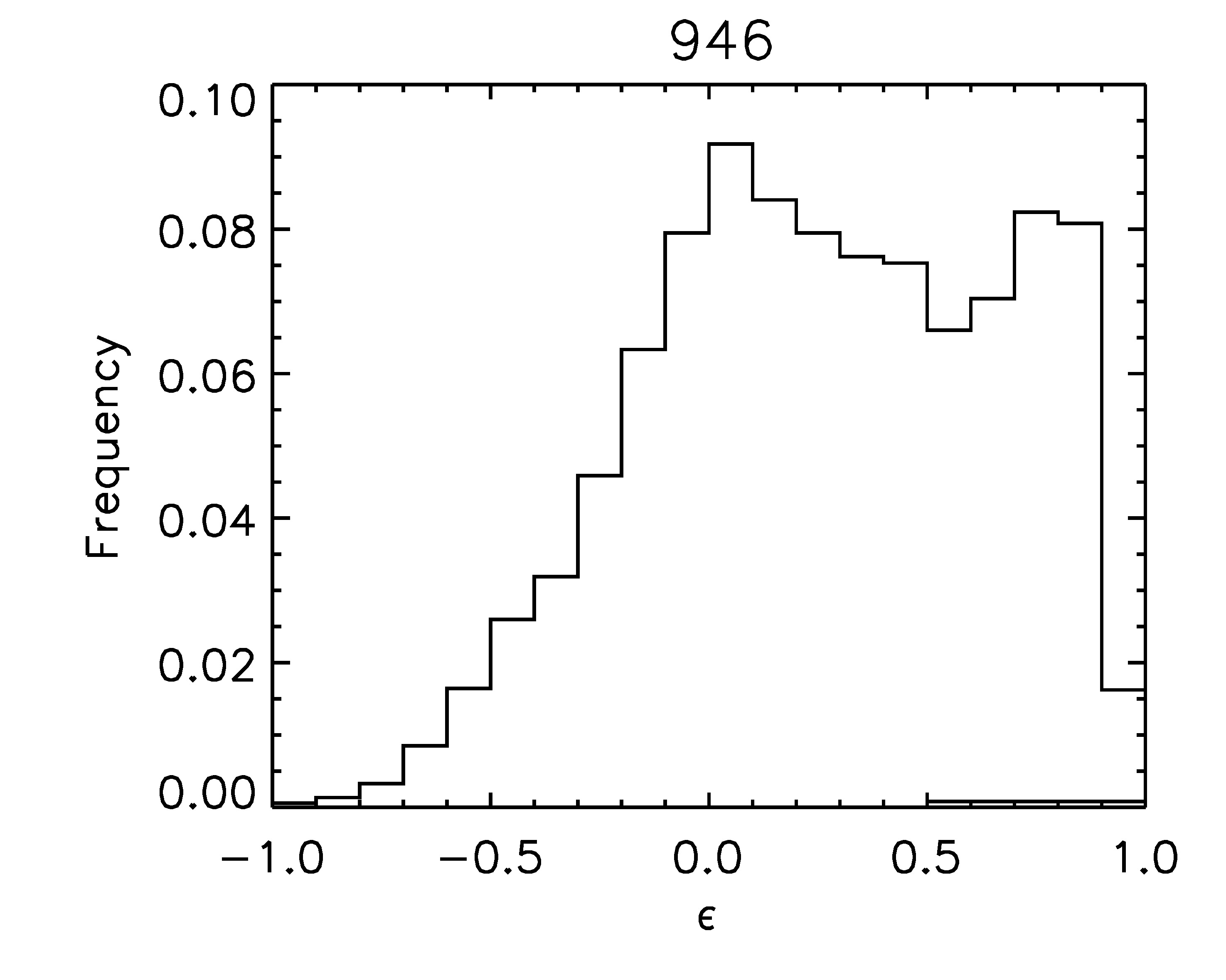}
    \includegraphics[width=0.33\textwidth]{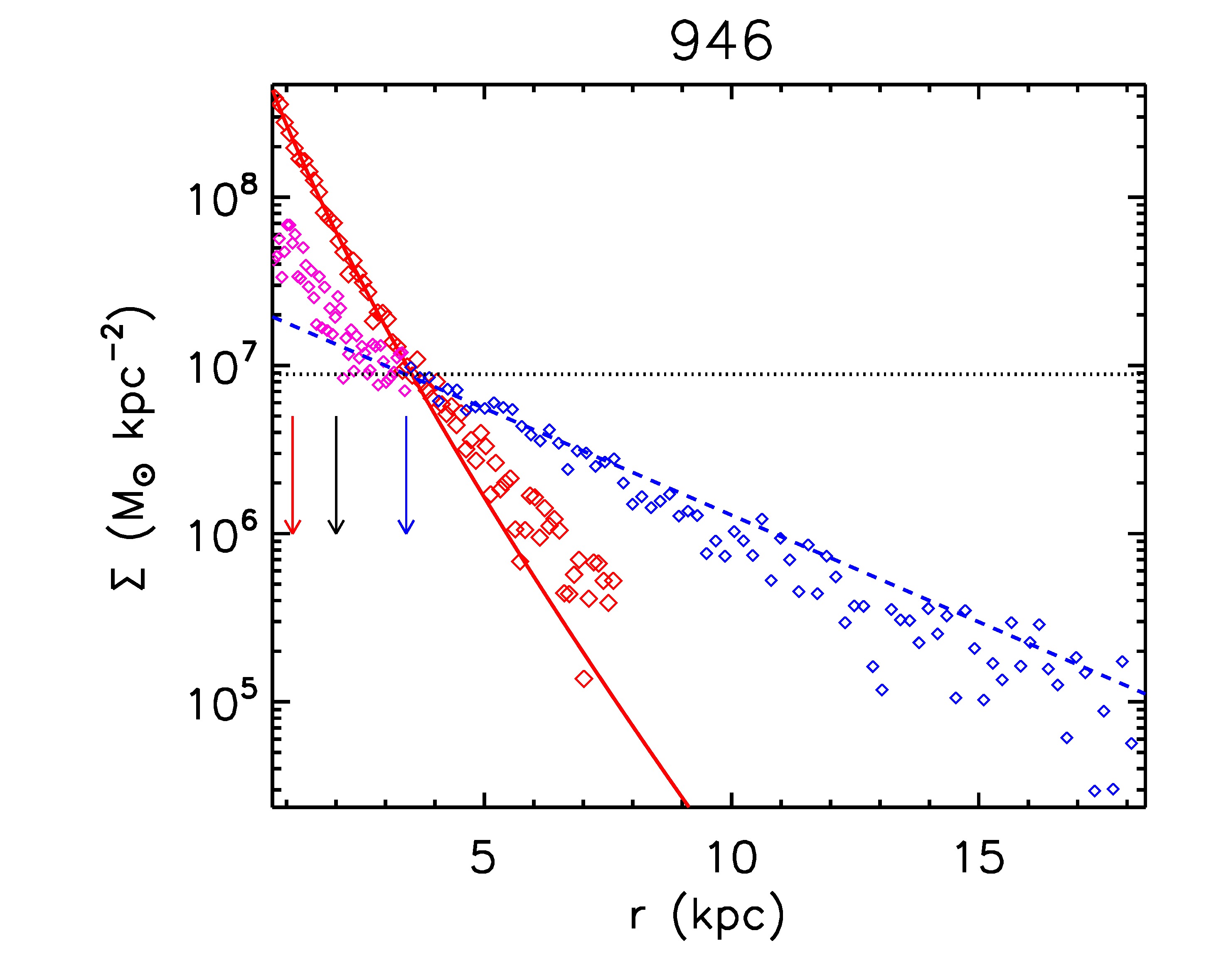} 
    \\
    \includegraphics[width=0.283\textwidth]{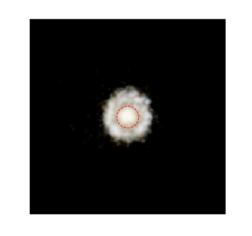}
 \includegraphics[width=0.33\textwidth]{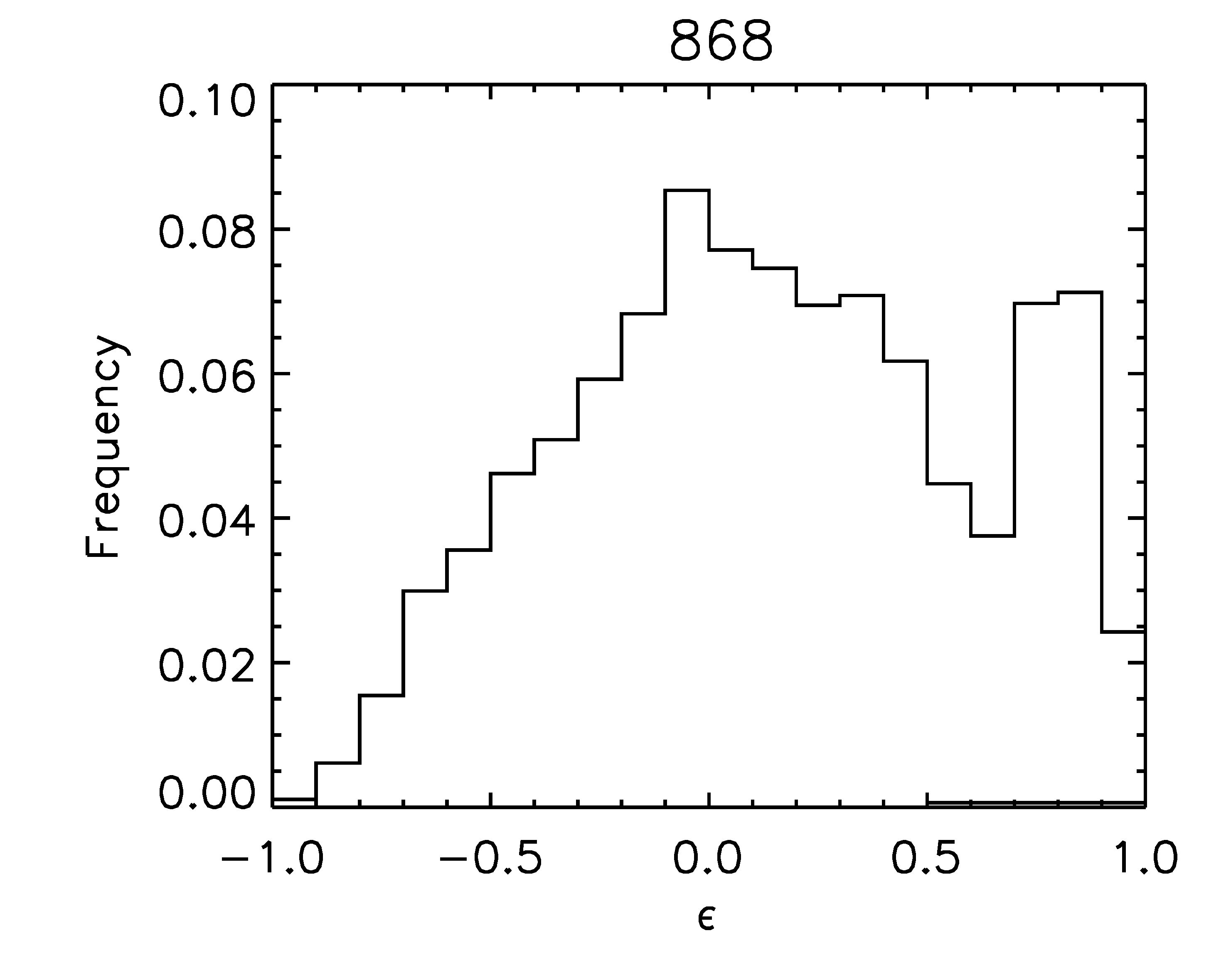}
    \includegraphics[width=0.33\textwidth]{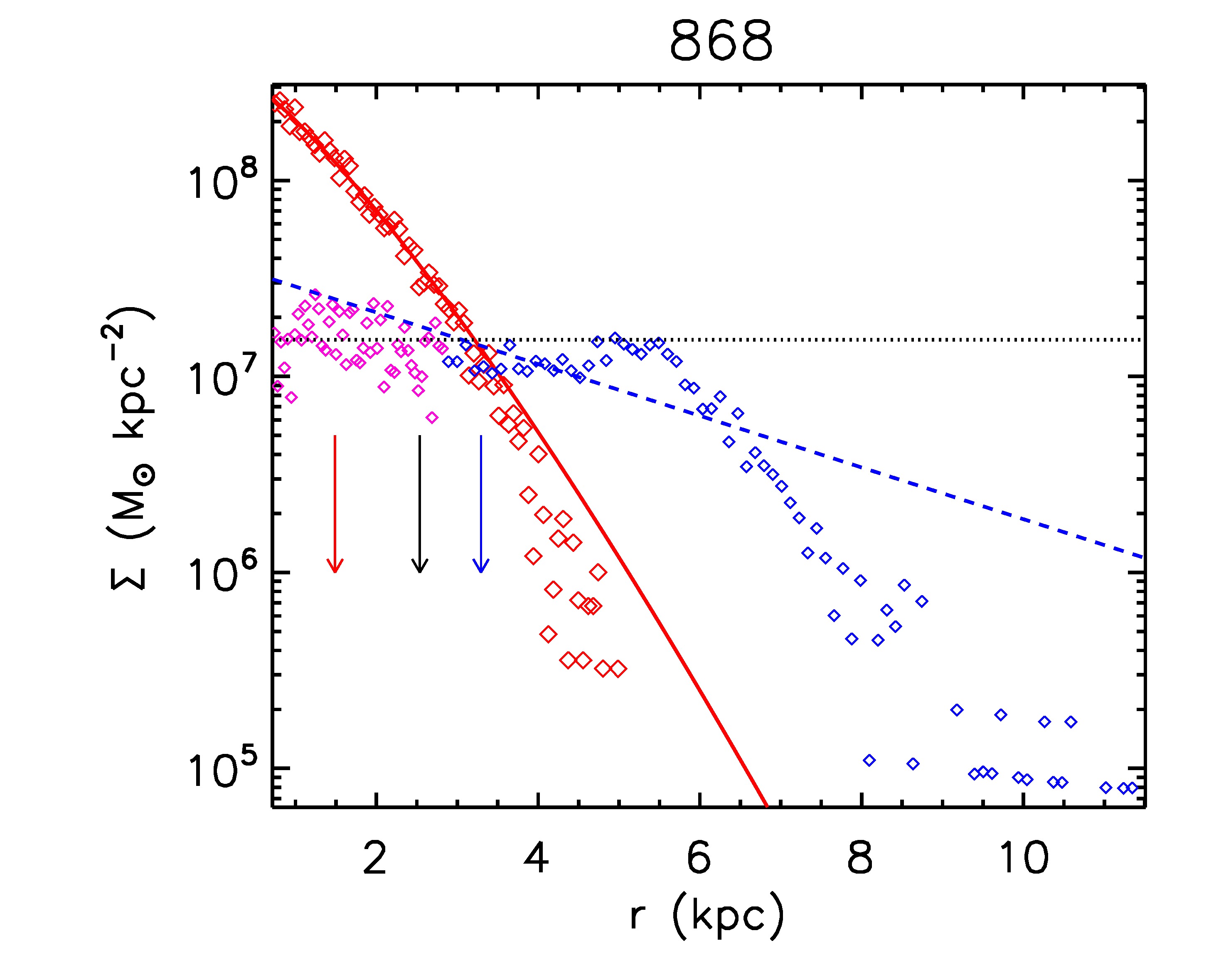} 
    \\
    \caption{(continued)}
\end{figure*}

\addtocounter{figure}{-1}

\begin{figure*}
    \includegraphics[width=0.283\textwidth]{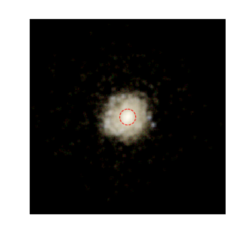}
    \includegraphics[width=0.33\textwidth]{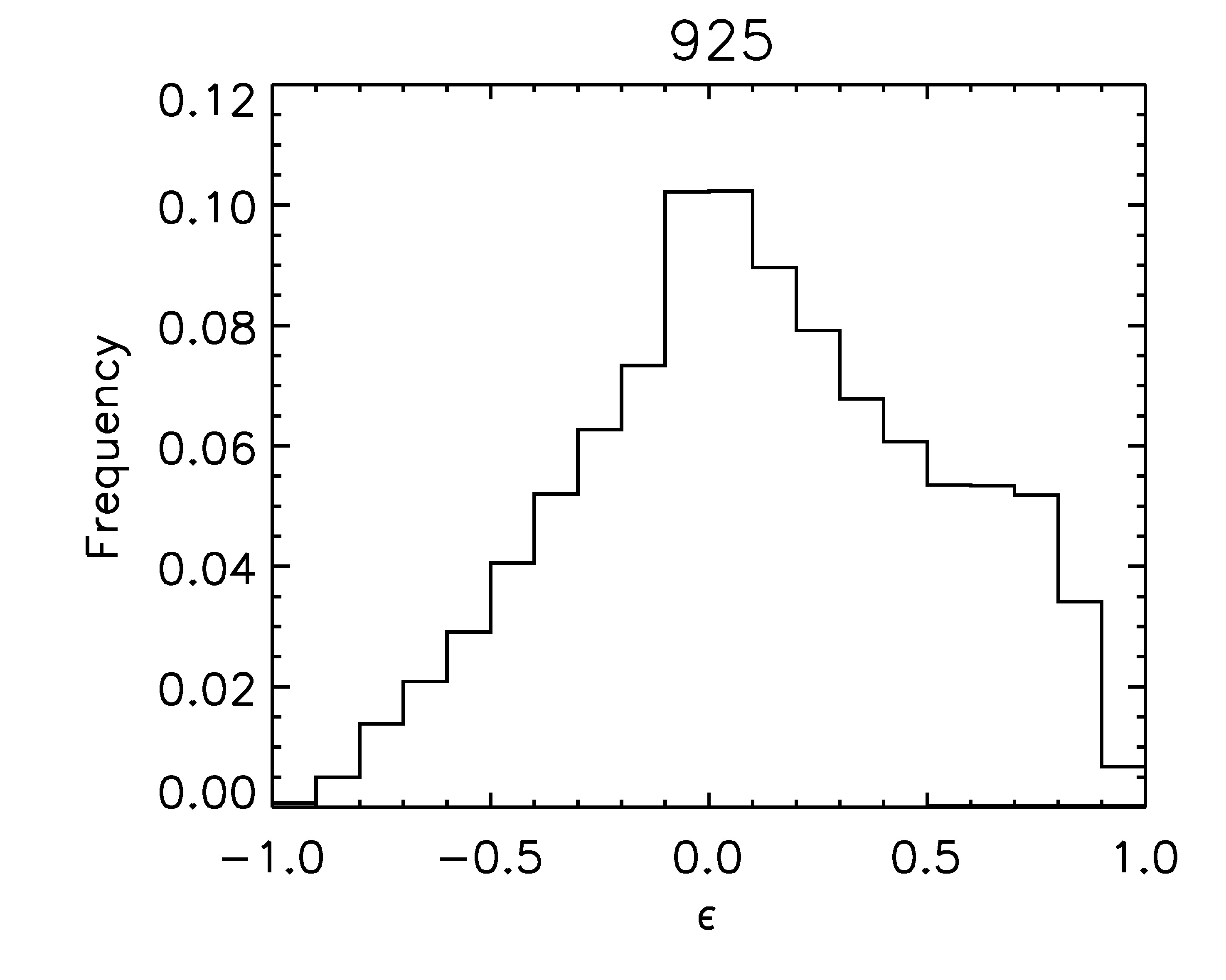}
    \includegraphics[width=0.33\textwidth]{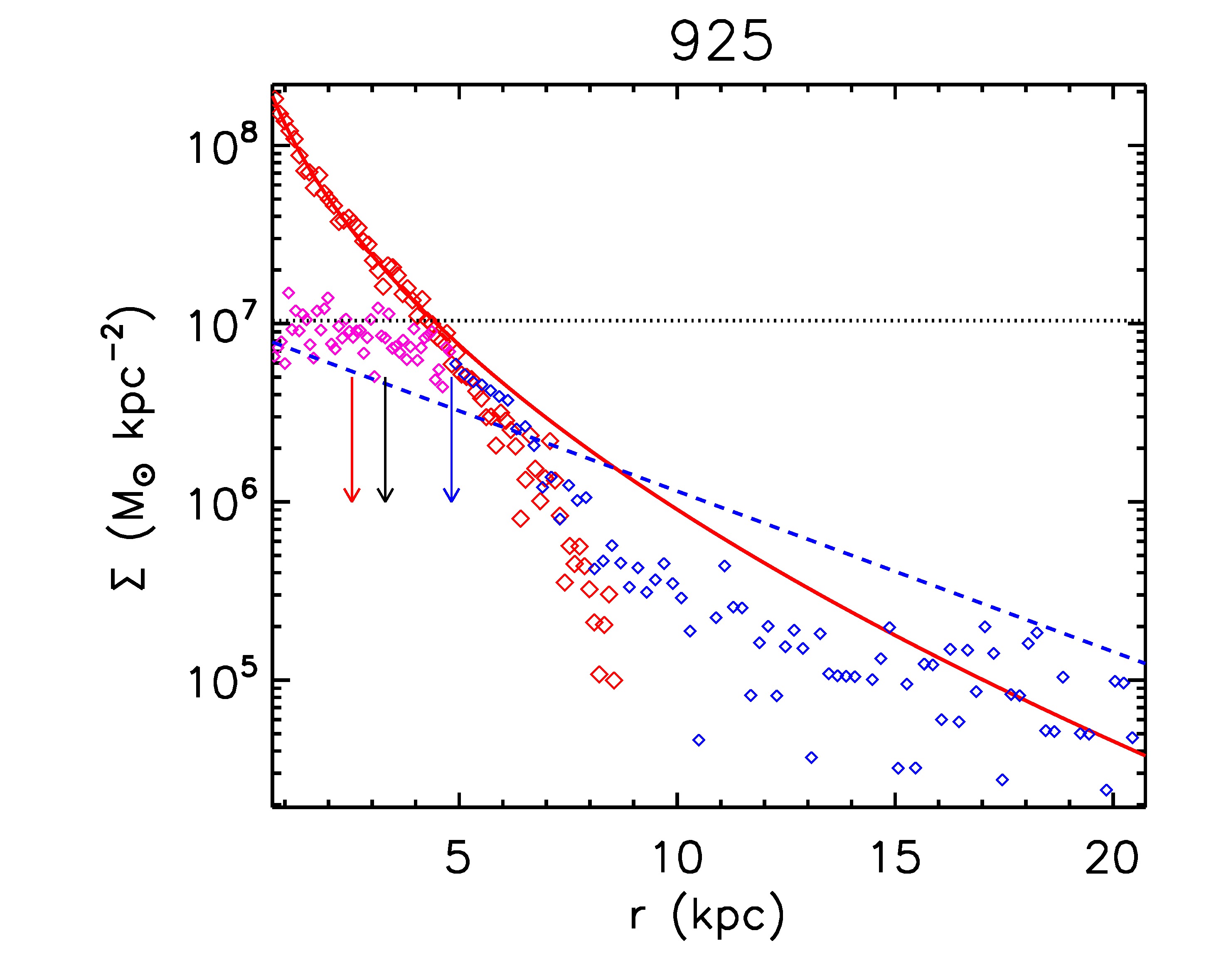} 
  \\
    \includegraphics[width=0.283\textwidth]{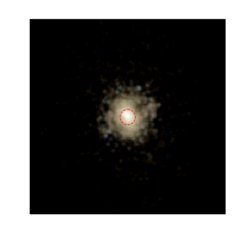}
    \includegraphics[width=0.33\textwidth]{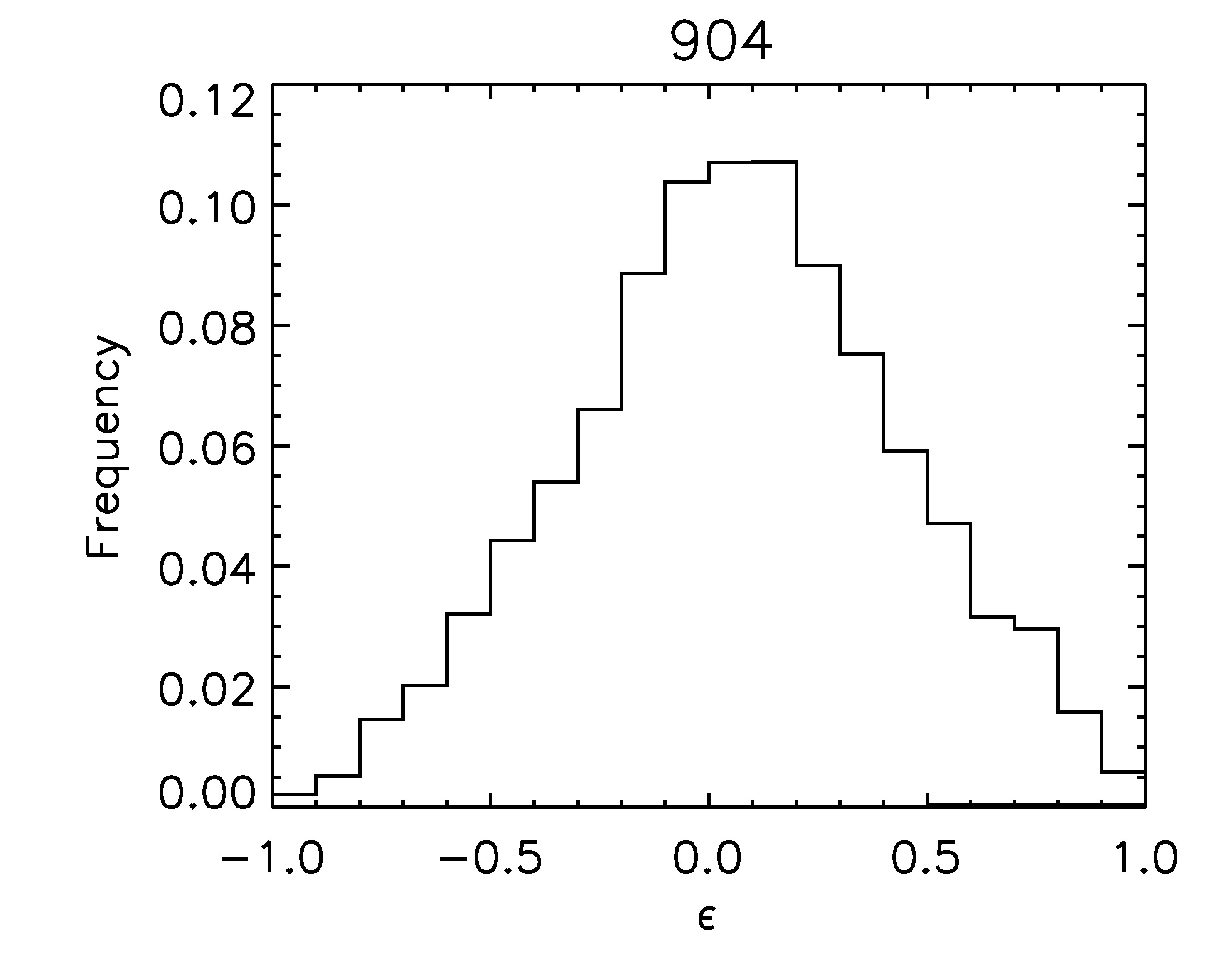}
    \includegraphics[width=0.33\textwidth]{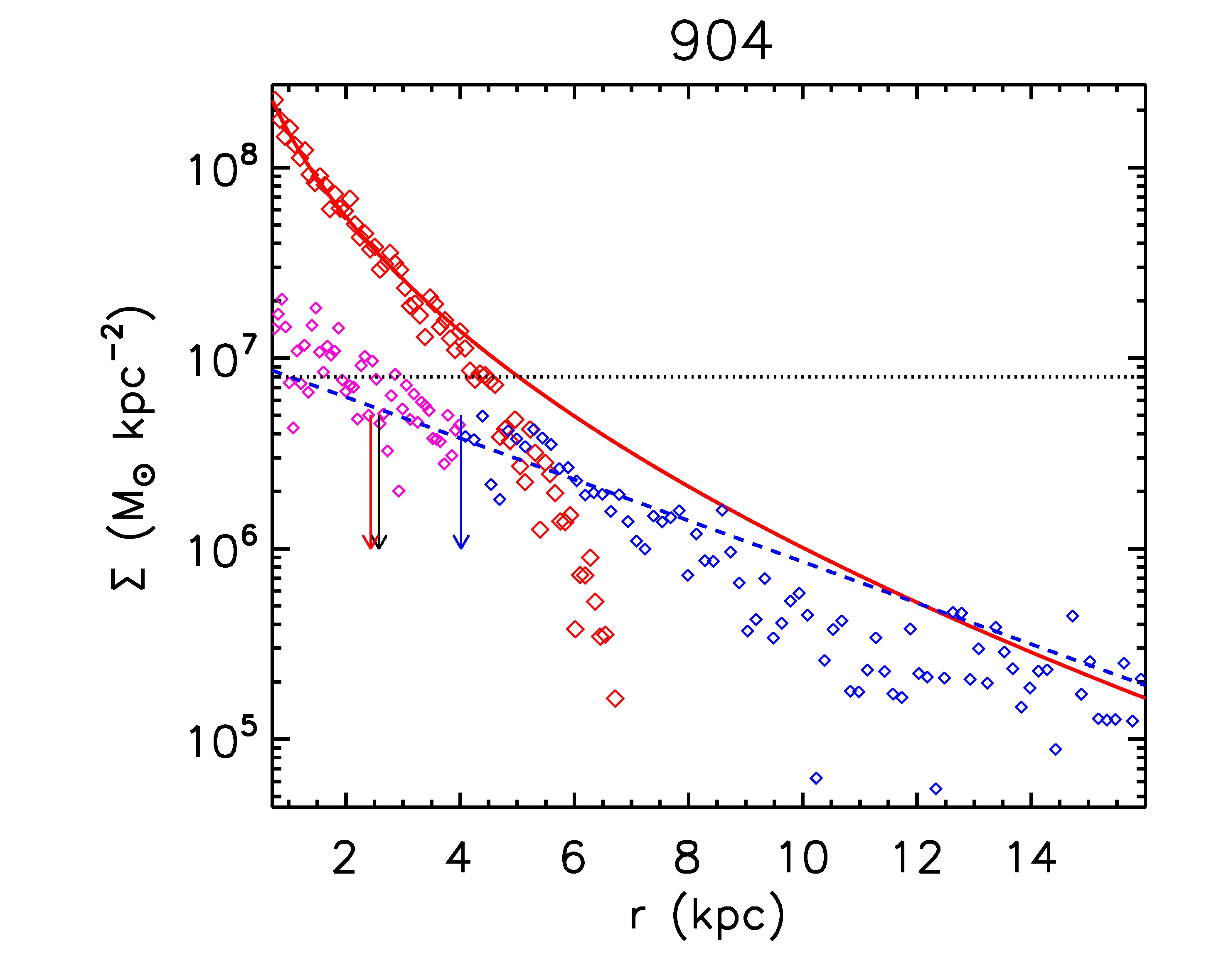}
\\
    \includegraphics[width=0.283\textwidth]{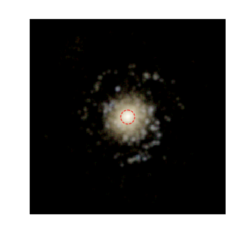}
 \includegraphics[width=0.33\textwidth]{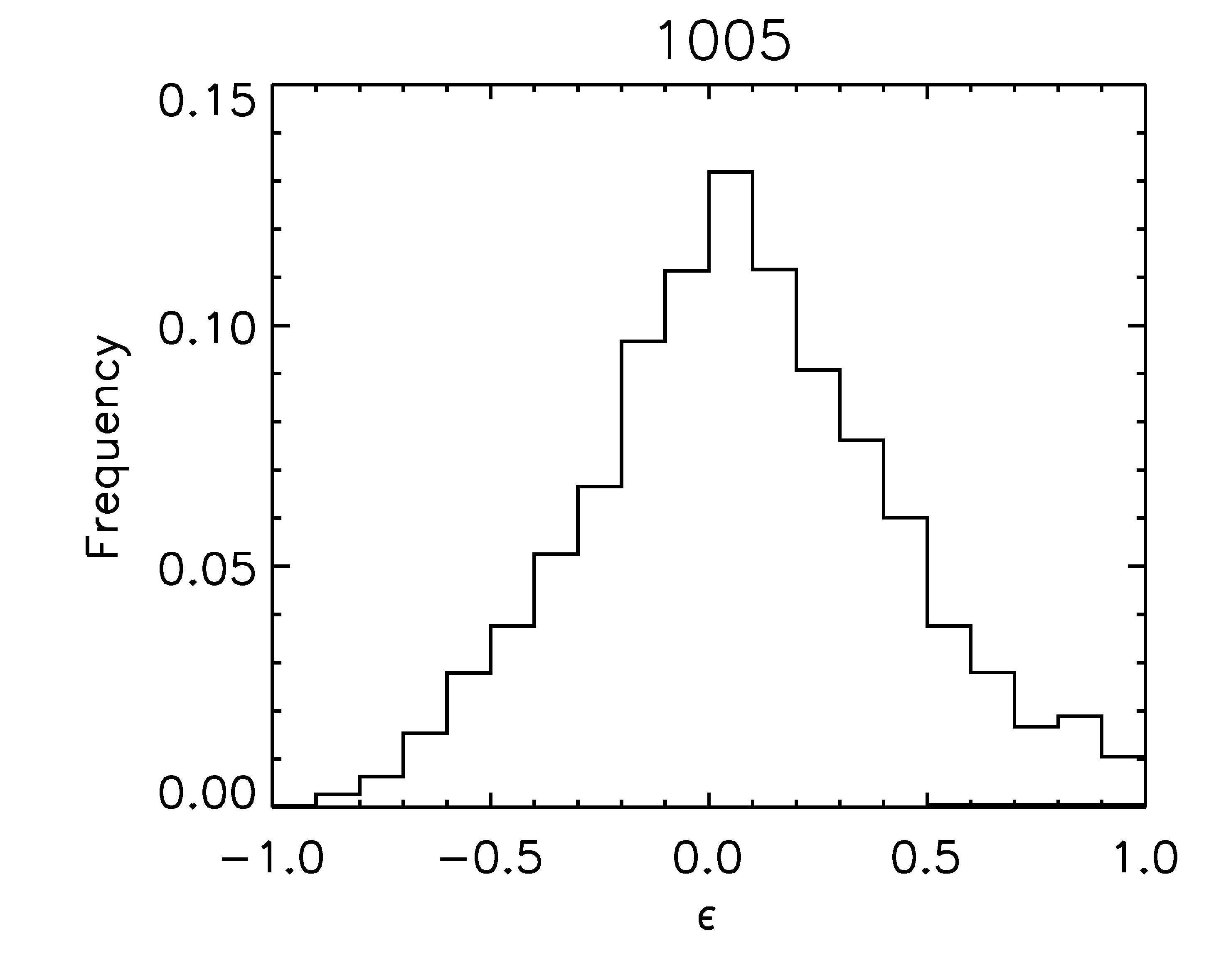}
    \includegraphics[width=0.33\textwidth]{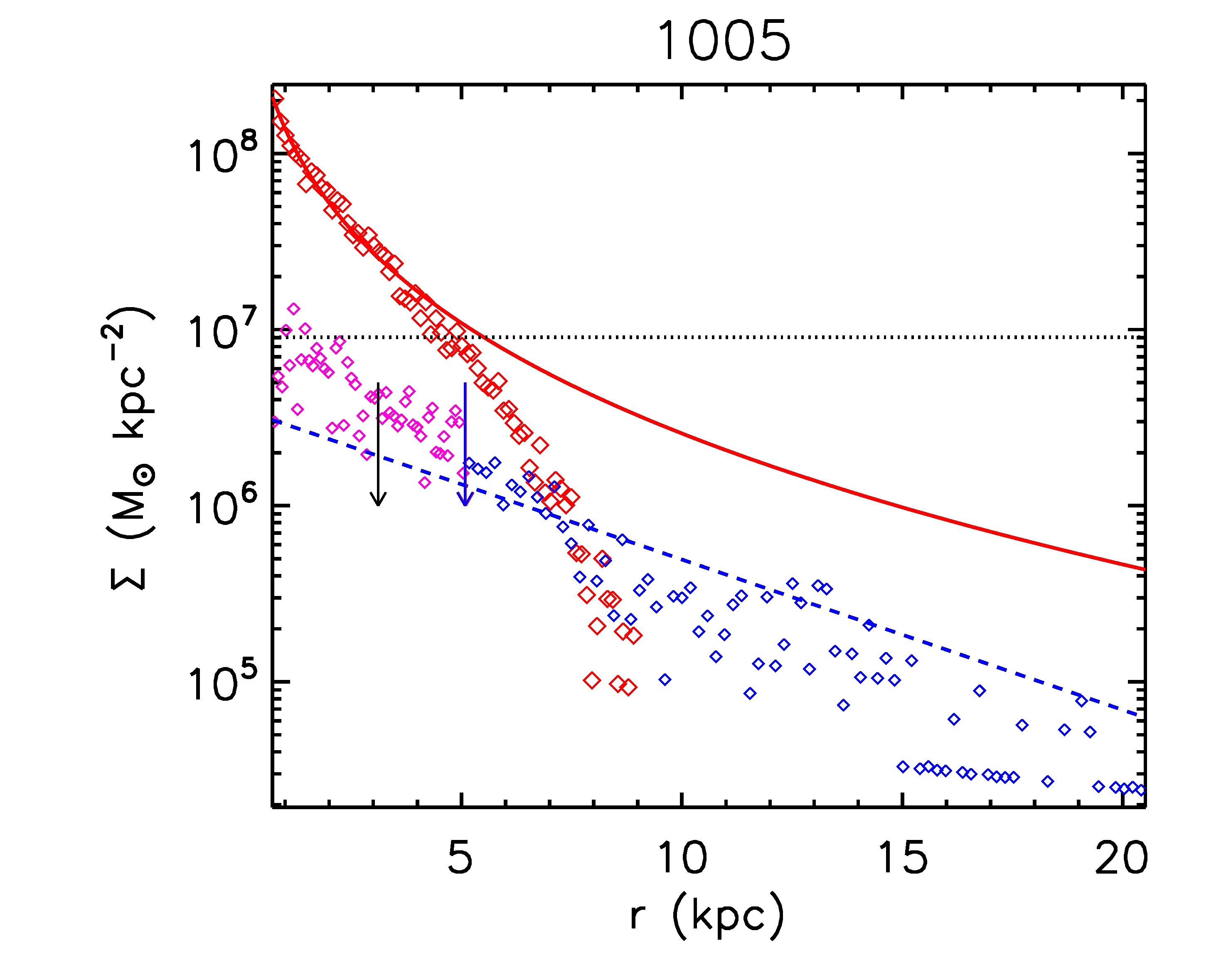} 
    \\
    \includegraphics[width=0.283\textwidth]{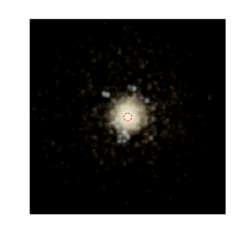}
    \includegraphics[width=0.33\textwidth]{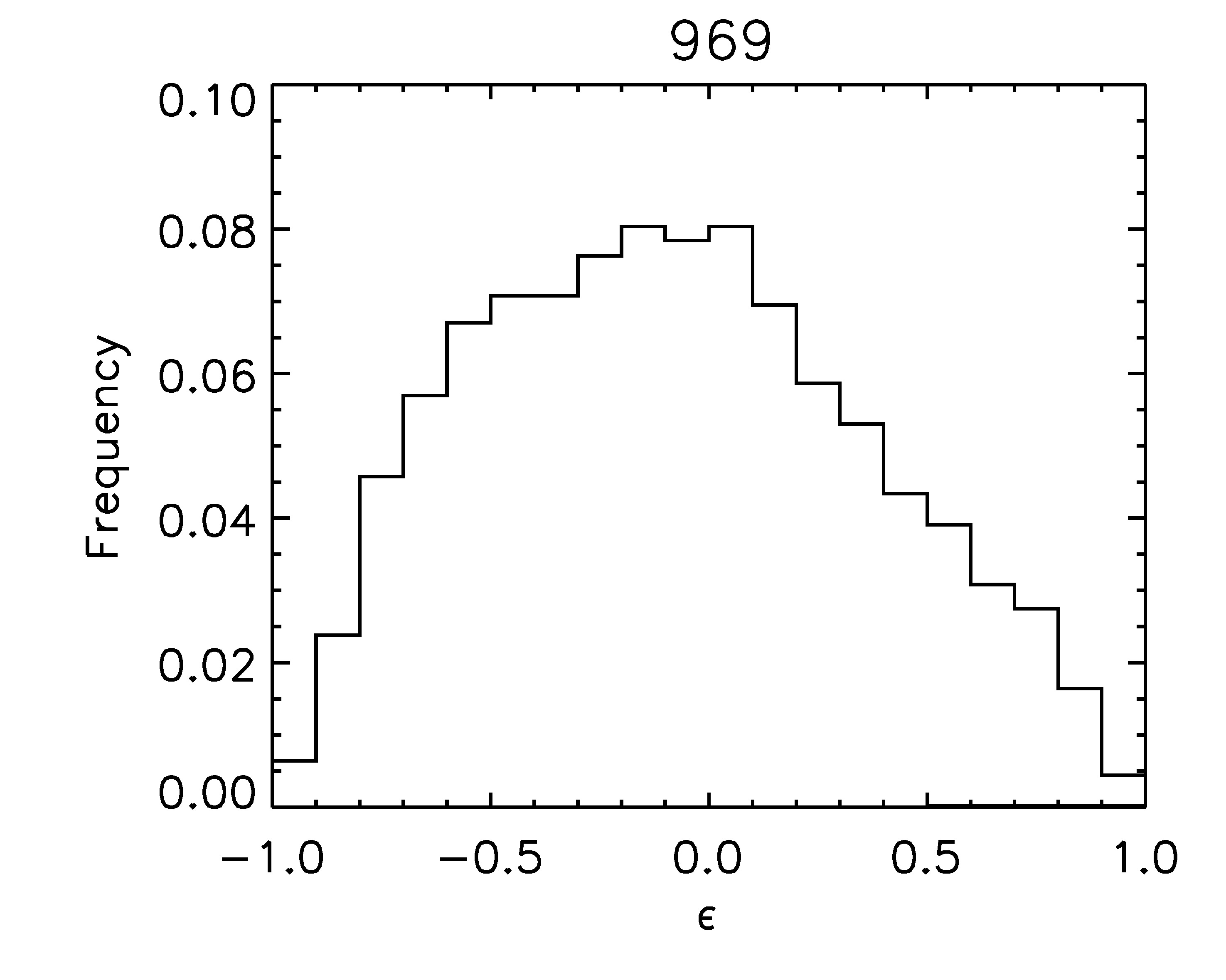}
    \includegraphics[width=0.33\textwidth]{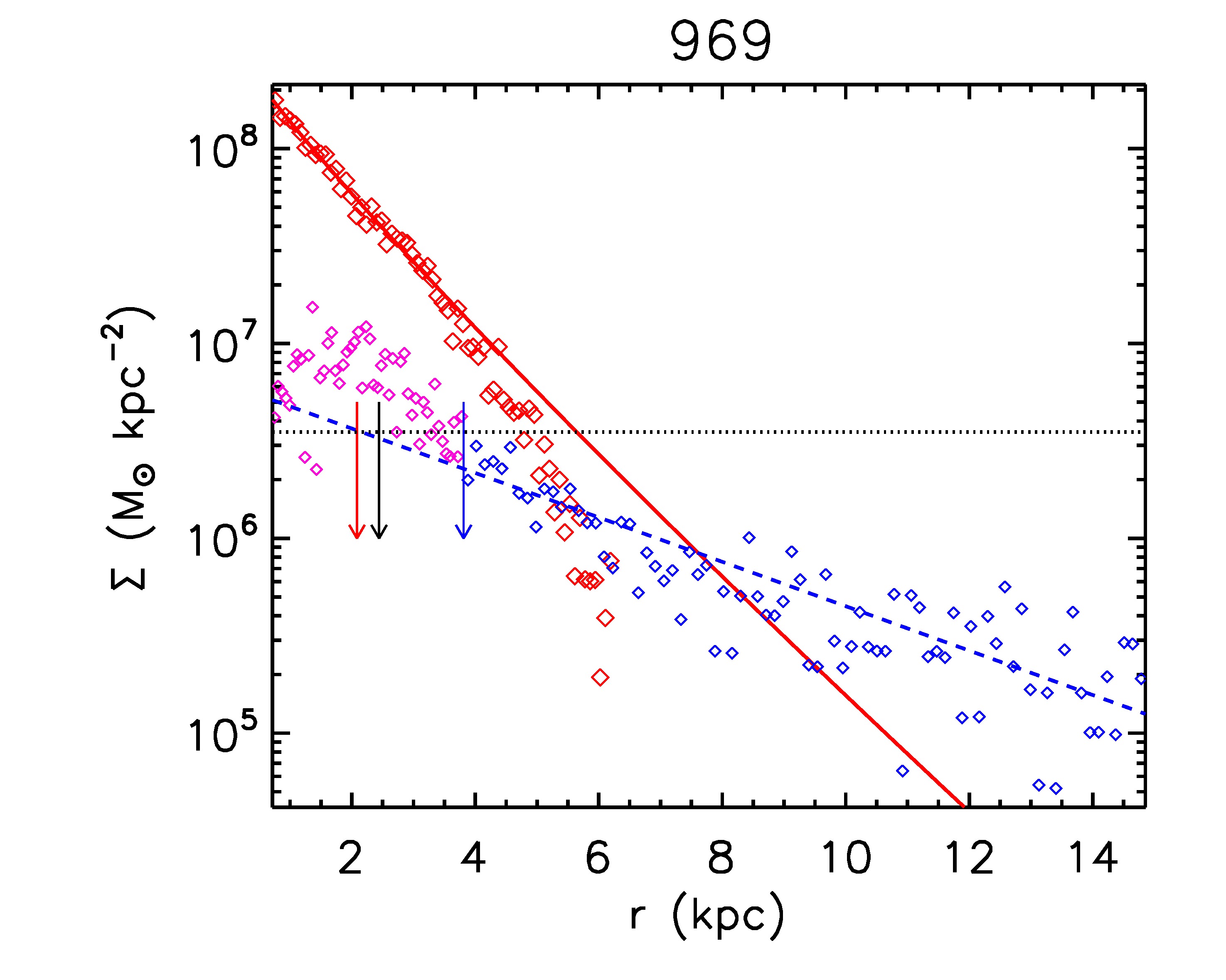} 
    \\
    \caption{(continued)}
\end{figure*}

\addtocounter{figure}{-1}

\begin{figure*}
    \includegraphics[width=0.283\textwidth]{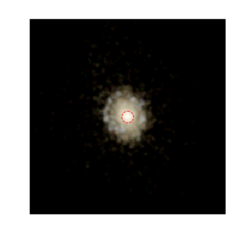}
    \includegraphics[width=0.33\textwidth]{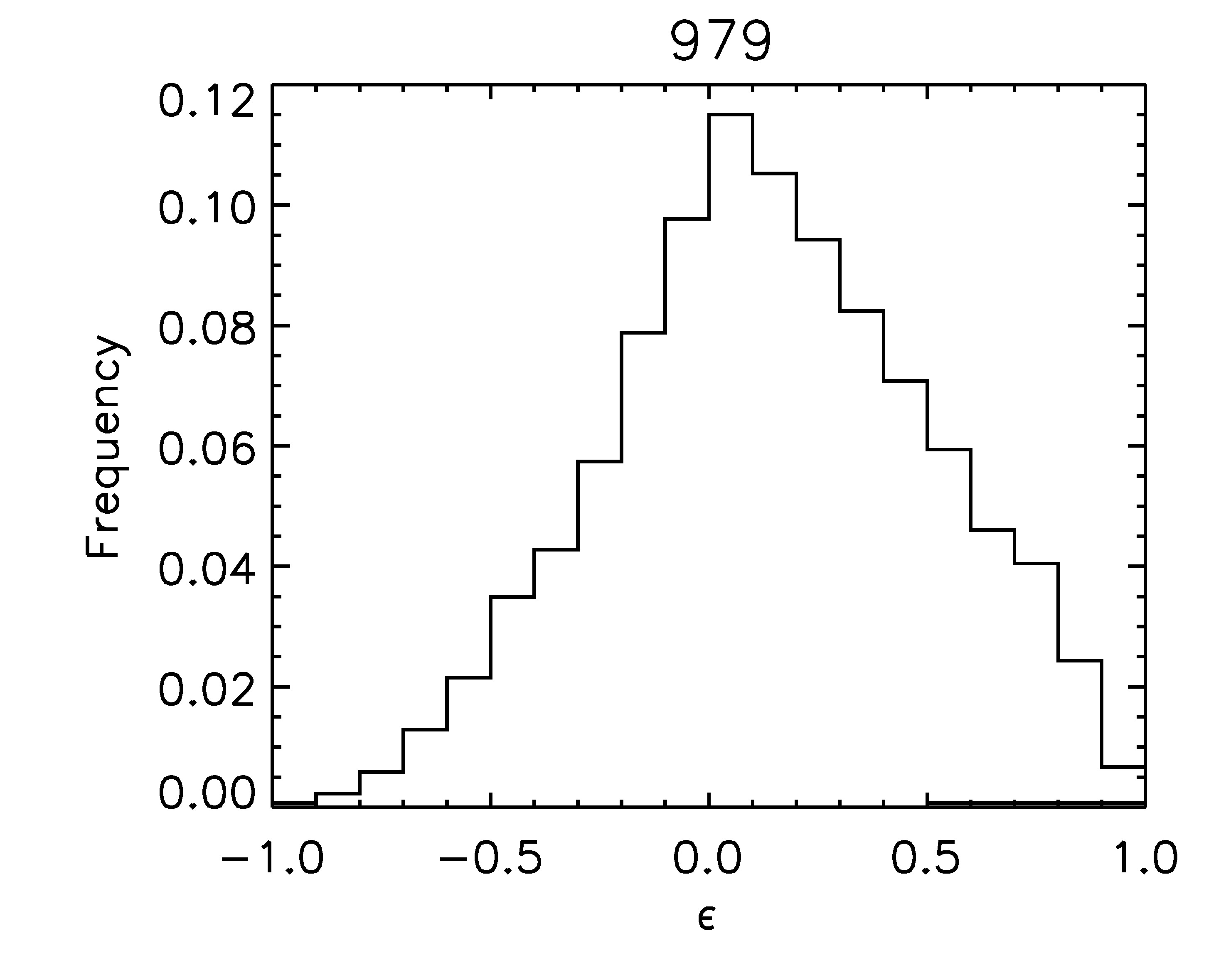}
    \includegraphics[width=0.33\textwidth]{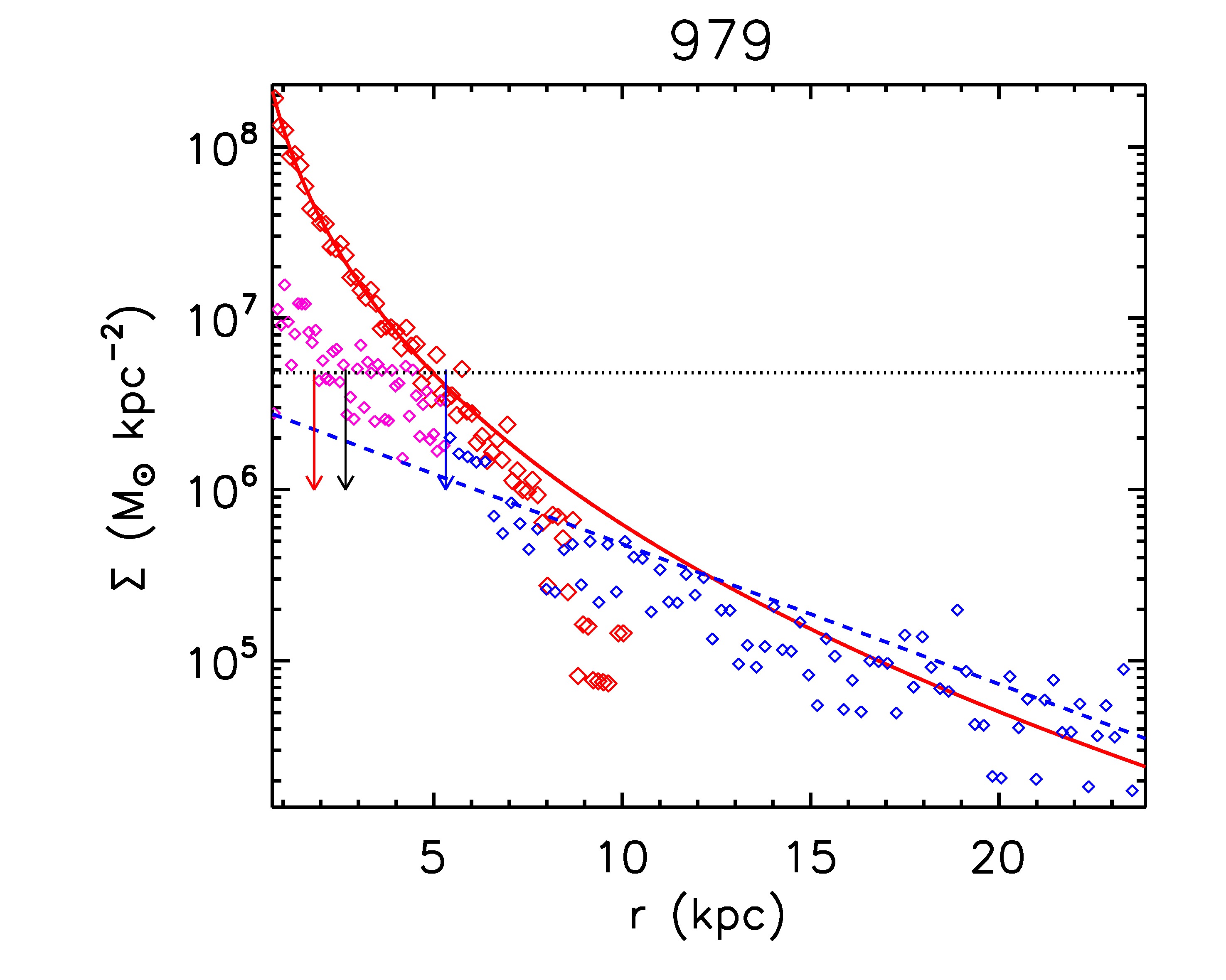} 
\\
    \includegraphics[width=0.283\textwidth]{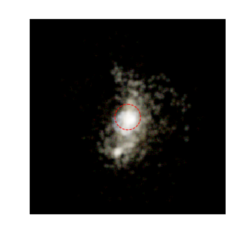}
 \includegraphics[width=0.33\textwidth]{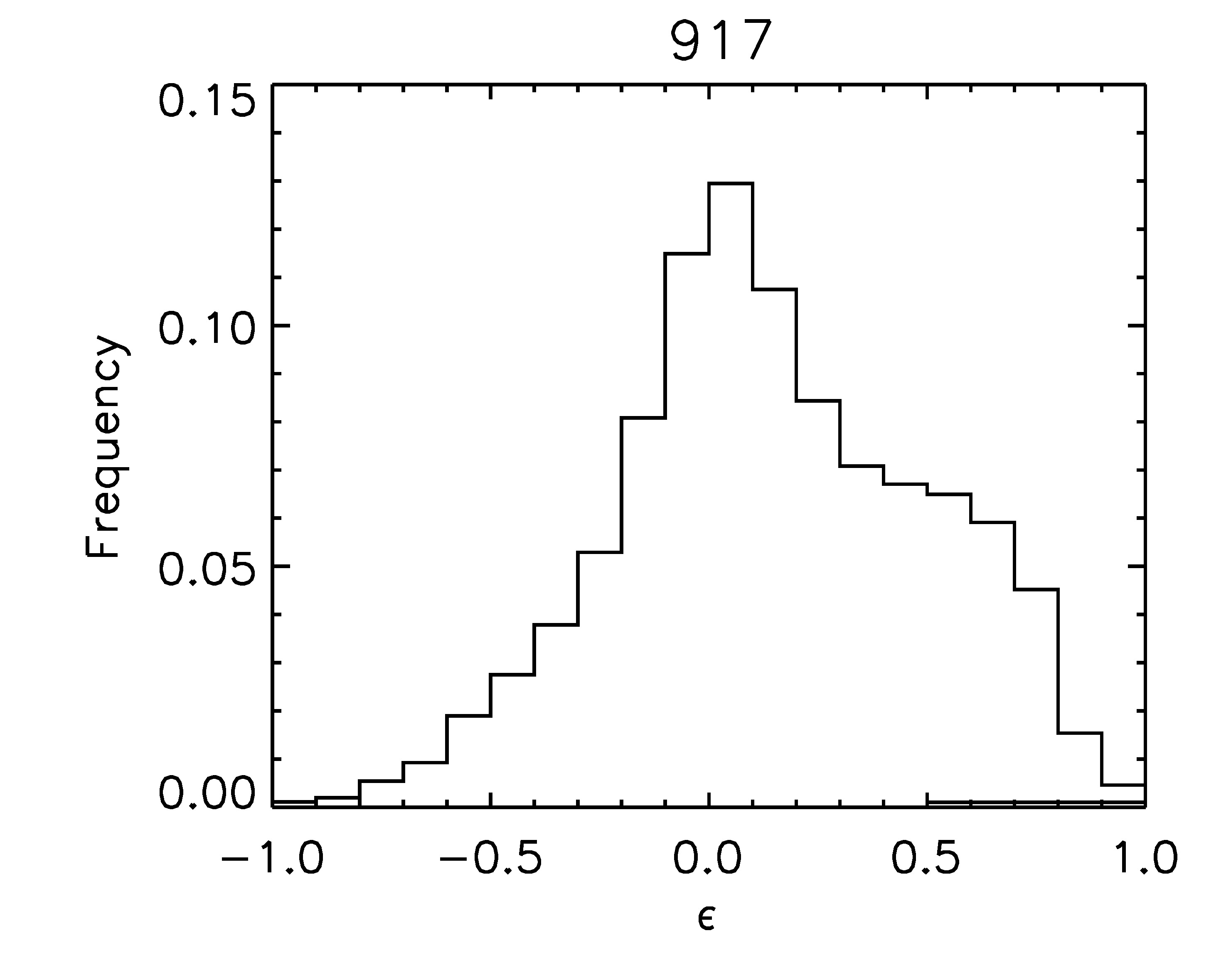}
    \includegraphics[width=0.33\textwidth]{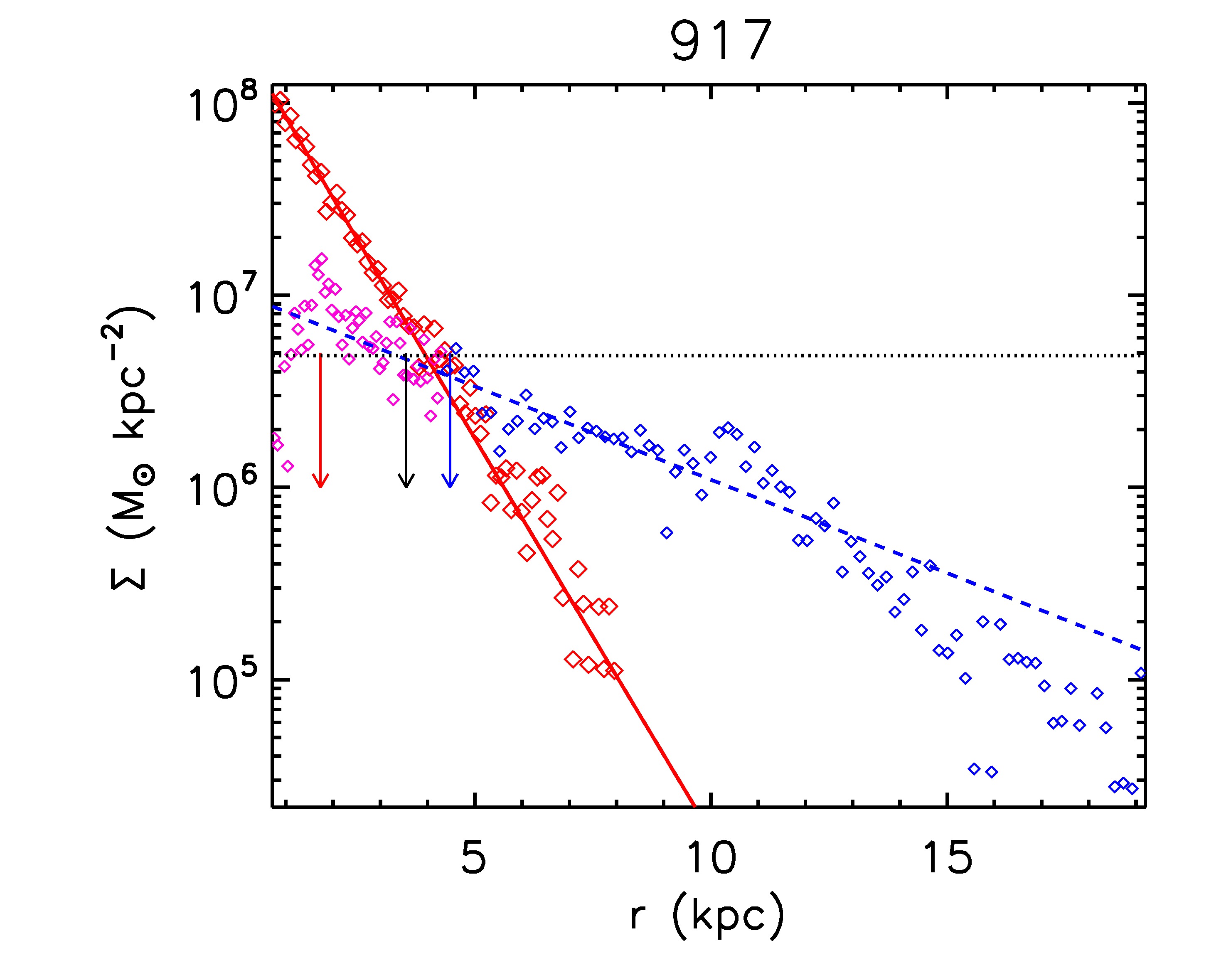} 
    \\
    \caption{(continued)}
\end{figure*}

\renewcommand{\thefigure}{\arabic{figure}}

\end{appendix}

\end{document}